\documentclass[twocolumn]{aastex61}

\shorttitle{ACTPol: Two-Season Cluster Catalog}
\shortauthors{Hilton et al. (ACT Collaboration)}

\begin{document}

\title{The Atacama Cosmology Telescope: The Two-Season ACTPol\\Sunyaev-Zel'dovich Effect Selected Cluster Catalog}

\correspondingauthor{Matt Hilton}
\email{hiltonm@ukzn.ac.za}

\author{Matt Hilton}
\affiliation{Astrophysics \& Cosmology Research Unit, School of Mathematics, Statistics \& Computer Science, 
University of KwaZulu-Natal,\\ Westville Campus, Durban 4041, South Africa}

\author{Matthew Hasselfield}
\affiliation{Institute for Gravitation and the Cosmos, The Pennsylvania State University, University Park, PA 16802, USA}
\affiliation{Department of Astronomy and Astrophysics, The Pennsylvania State University, University Park, PA 16802, USA}
\affiliation{Department of Astrophysical Sciences, Peyton Hall, Princeton University, Princeton, NJ 08544, USA}

\author{Crist\'{o}bal Sif\'{o}n}
\affiliation{Department of Astrophysical Sciences, Peyton Hall, Princeton University, Princeton, NJ 08544, USA}
\affiliation{Leiden Observatory, Leiden University, PO Box 9513, NL-2300 RA Leiden, Netherlands}

\author{Nicholas Battaglia}
\affiliation{Department of Astrophysical Sciences, Peyton Hall, Princeton University, Princeton, NJ 08544, USA}
\affiliation{Center for Computational Astrophysics, Flatiron Institute, 162 Fifth Avenue, New York, NY 10010, USA}

\author{Simone Aiola}
\affiliation{Joseph Henry Laboratories of Physics, Jadwin Hall, Princeton University, Princeton, NJ 08544, USA}

\author{V. Bharadwaj}
\affiliation{Astrophysics \& Cosmology Research Unit, School of Mathematics, Statistics \& Computer Science, 
University of KwaZulu-Natal,\\ Westville Campus, Durban 4041, South Africa}

\author{J. Richard Bond}
\affiliation{Canadian Institute for Theoretical Astrophysics, University of Toronto, Toronto, ON, M5S 3H8, Canada}

\author{Steve K. Choi}
\affiliation{Joseph Henry Laboratories of Physics, Jadwin Hall, Princeton University, Princeton, NJ 08544, USA}

\author{Devin Crichton}
\affiliation{Astrophysics \& Cosmology Research Unit, School of Mathematics, Statistics \& Computer Science, 
University of KwaZulu-Natal,\\ Westville Campus, Durban 4041, South Africa}
\affiliation{Department of Physics and Astronomy, The Johns Hopkins University, 3400 N. Charles St., Baltimore, MD 21218-2686, USA}

\author{Rahul Datta}
\affiliation{Department of Physics, University of Michigan, Ann Arbor, MI 48103, USA}
\affiliation{NASA/Goddard Space Flight Center, Greenbelt, MD 20771, USA}

\author{Mark J. Devlin}
\affiliation{Department of Physics and Astronomy, University of Pennsylvania, 209 South 33rd Street, Philadelphia, PA 19104, USA}

\author{Joanna Dunkley}
\affiliation{Department of Astrophysical Sciences, Peyton Hall, Princeton University, Princeton, NJ 08544, USA}
\affiliation{Joseph Henry Laboratories of Physics, Jadwin Hall, Princeton University, Princeton, NJ 08544, USA}

\author{Rolando D\"unner}
\affiliation{Instituto de Astrof\'isica and Centro de Astro-Ingenier\'ia, Facultad de F\'isica, Pontificia Universidad Cat\'olica de Chile, 
Av. Vicu\~na Mackenna 4860, 7820436 Macul, Santiago, Chile}

\author{Patricio A. Gallardo}
\affiliation{Department of Physics, Cornell University, Ithaca, NY 14853, USA}

\author{Megan Gralla}
\affiliation{Steward Observatory, University of Arizona, 933 N Cherry Avenue, Tucson, AZ 85721, USA}

\author{Adam D. Hincks}
\affiliation{Department of Physics, University of Rome ``La Sapienza'', Piazzale Aldo Moro 5, I-00185 Rome, Italy}

\author{Shuay-Pwu P. Ho}
\affiliation{Joseph Henry Laboratories of Physics, Jadwin Hall, Princeton University, Princeton, NJ 08544, USA}

\author{Johannes Hubmayr}
\affiliation{NIST Quantum Devices Group, 325 Broadway, Mailcode 817.03, Boulder, CO 80305, USA}

\author{Kevin M. Huffenberger}
\affiliation{Department of Physics, Florida State University, Tallahassee FL, 32306, USA}

\author{John P. Hughes}
\affiliation{Department of Physics and Astronomy, Rutgers University, 136 Frelinghuysen Road, Piscataway, NJ 08854-8019, USA}
\affiliation{Center for Computational Astrophysics, Flatiron Institute, 162 Fifth Avenue, New York, NY 10010, USA}

\author{Brian J. Koopman}
\affiliation{Department of Physics, Cornell University, Ithaca, NY 14853, USA}

\author{Arthur Kosowsky}
\affiliation{Department of Physics and Astronomy, University of Pittsburgh, Pittsburgh, PA 15260, USA}
\affiliation{Pittsburgh Particle Physics, Astrophysics, and Cosmology Center, University of Pittsburgh, Pittsburgh, PA 15260, USA}

\author{Thibaut Louis}
\affiliation{UPMC Univ Paris 06, UMR7095, Institut d’Astrophysique de Paris, F-75014, Paris, France}
\affiliation{Sub-Department of Astrophysics, University of Oxford, Keble Road, Oxford, OX1 3R, UK}

\author{Mathew S. Madhavacheril}
\affiliation{Department of Astrophysical Sciences, Peyton Hall, Princeton University, Princeton, NJ 08544, USA}

\author{Tobias A. Marriage}
\affiliation{Department of Physics and Astronomy, The Johns Hopkins University, 3400 N. Charles St., Baltimore, MD 21218-2686, USA}

\author{Lo\"ic Maurin}
\affiliation{Instituto de Astrof\'isica and Centro de Astro-Ingenier\'ia, Facultad de F\'isica, Pontificia Universidad Cat\'olica de Chile, 
Av. Vicu\~na Mackenna 4860, 7820436 Macul, Santiago, Chile}

\author{Jeff McMahon}
\affiliation{Department of Physics, University of Michigan, Ann Arbor, MI 48103, USA}

\author{Hironao Miyatake}
\affiliation{Jet Propulsion Laboratory, California Institute of Technology, Pasadena, CA 91109, USA}
\affiliation{Kavli Institute for the Physics and Mathematics of the Universe (Kavli IPMU, WPI), University of Tokyo, Chiba 277-8582, Japan}
\affiliation{Institute for Advanced Research, Nagoya University, Nagoya 464-8601, Aichi, Japan}
\affiliation{Division of Physics and Astrophysical Science, Graduate School of Science, Nagoya University, Nagoya 464-8602, Japan}

\author{Kavilan Moodley}
\affiliation{Astrophysics \& Cosmology Research Unit, School of Mathematics, Statistics \& Computer Science, 
University of KwaZulu-Natal,\\ Westville Campus, Durban 4041, South Africa}

\author{Sigurd N\ae ss}
\affiliation{Center for Computational Astrophysics, Flatiron Institute, 162 Fifth Avenue, New York, NY 10010, USA}

\author{Federico Nati}
\affiliation{Department of Physics and Astronomy, University of Pennsylvania, 209 South 33rd Street, Philadelphia, PA 19104, USA}

\author{Laura Newburgh}
\affiliation{Department of Physics, Yale University, New Haven, CT 06520, USA}

\author{Michael D. Niemack}
\affiliation{Department of Physics, Cornell University, Ithaca, NY 14853, USA}

\author{Masamune Oguri}
\affiliation{Research Center for the Early Universe, University of Tokyo, Tokyo 113-0033, Japan}
\affiliation{Department of Physics, University of Tokyo, Tokyo 113-0033, Japan}
\affiliation{Kavli Institute for the Physics and Mathematics of the Universe (Kavli IPMU, WPI), University of Tokyo, Chiba 277-8582, Japan}

\author{Lyman A. Page}
\affiliation{Joseph Henry Laboratories of Physics, Jadwin Hall, Princeton University, Princeton, NJ 08544, USA}

\author{Bruce Partridge}
\affiliation{Department of Physics and Astronomy, Haverford College, Haverford, PA 19041, USA}

\author{Benjamin L. Schmitt}
\affiliation{Department of Physics and Astronomy, University of Pennsylvania, 209 South 33rd Street, Philadelphia, PA 19104, USA}

\author{Jon Sievers}
\affiliation{Astrophysics \& Cosmology Research Unit, School of Chemistry \& Physics, 
University of KwaZulu-Natal,\\ Westville Campus, Durban 4041, South Africa}

\author{David N. Spergel}
\affiliation{Center for Computational Astrophysics, Flatiron Institute, 162 Fifth Avenue, New York, NY 10010, USA}
\affiliation{Department of Astrophysical Sciences, Peyton Hall, Princeton University, Princeton, NJ 08544, USA}

\author{Suzanne T. Staggs}
\affiliation{Joseph Henry Laboratories of Physics, Jadwin Hall, Princeton University, Princeton, NJ 08544, USA}

\author{Hy Trac}
\affiliation{McWilliams Center for Cosmology, Carnegie Mellon University, Department of Physics, 5000  Forbes Ave., Pittsburgh, PA 15213, USA}

\author{Alexander van Engelen}
\affiliation{Canadian Institute for Theoretical Astrophysics, University of Toronto, Toronto, ON, M5S 3H8, Canada}

\author{Eve M. Vavagiakis}
\affiliation{Department of Physics, Cornell University, Ithaca, NY 14853, USA}

\author{Edward J. Wollack}
\affiliation{NASA/Goddard Space Flight Center, Greenbelt, MD 20771, USA}



\begin{abstract}
We present a catalog of 182 galaxy clusters detected through the Sunyaev-Zel'dovich effect by the Atacama Cosmology Telescope
in a contiguous 987.5 deg$^{2}$ field. The clusters were detected as SZ decrements by applying a matched filter to 148 GHz
maps that combine the original ACT equatorial survey with data from the first two observing seasons using the ACTPol receiver. 
Optical/IR confirmation and redshift measurements come from a combination of large public surveys and our own follow-up observations. 
Where necessary, we measured photometric redshifts for clusters using a pipeline that achieves accuracy $\Delta z/(1 + z)=0.015$ 
when tested on SDSS data. Under the assumption that clusters can be described by the so-called Universal Pressure Profile and its 
associated mass-scaling law, the full signal-to-noise~$ > 4$ sample spans the mass range 
$1.6 < M^{\rm UPP}_{\rm 500c}/10^{14}{\rm M}_{\odot}<9.1$, with median $M^{\rm UPP}_{\rm 500c}=3.1 \times 10^{14}$ M$_{\odot}$. 
The sample covers the redshift range $0.1 < z < 1.4$ (median $z = 0.49$) and 28 clusters are new discoveries (median $z = 0.80$). 
We compare our catalog with other overlapping cluster samples selected using the SZ, optical,and X-ray wavelengths. 
We find the ratio of the UPP-based SZ mass to richness-based weak-lensing mass is
$\langle M^{\rm UPP}_{\rm 500c} \rangle / \langle M^{\rm \lambda WL}_{\rm 500c} \rangle = 0.68 \pm 0.11$. After 
applying this calibration, the mass distribution for clusters with $M_{\rm 500c} > 4 \times 10^{14}$ M$_{\odot}$ is 
consistent with the number of such clusters found in the South Pole Telescope SZ survey.

\end{abstract}





\defcitealias{Hasselfield_2013}{H13}
\defcitealias{Arnaud_2010}{A10}

\section{Introduction} 
\label{sec:intro}

Searching for clusters of galaxies using the thermal Sunyaev-Zel'dovich effect (SZ; \citealt{SZ_1972})
is now firmly established as a robust method for cluster detection \citep[e.g.,][]{Staniszewski_2009, Vanderlinde_2010, 
Marriage_2011, Hasselfield_2013, Bleem_2015, Planck2015_XXVII}. The SZ effect is the inverse Compton scattering of cosmic microwave background 
photons by the hot intracluster medium (ICM). The magnitude of the SZ signal is almost independent of redshift, and in 
principle this allows SZ surveys to track the evolution of the number density of massive clusters over 
most of the history of the Universe. Since the growth rate of these structures is dependent upon the energy density
of dark matter and dark energy, SZ surveys provide a method of measuring 
cosmological parameters that is complementary to studies using other probes \citep[e.g.,][]{Vanderlinde_2010,
Sehgal_2011, Hasselfield_2013, Reichardt_2013, Planck_XX_2013, Planck2015_XXIV, deHaan_2016}.

Although the SZ effect was first demonstrated in the late 1970s using pointed observations towards known clusters 
\citep[see the review by][]{Birkinshaw_1999}, the first blind detections were only made in the last decade,
initially using the South Pole Telescope \citep[SPT;][]{Staniszewski_2009}. The completed 2500\,deg$^2$
SPT survey SZ cluster catalog contains 516 confirmed clusters \citep{Bleem_2015} detected at signal-to-noise
$> 4.5$. Large area cluster searches have also been conducted using the Atacama Cosmology Telescope \citep[ACT;][]{Swetz_2011} 
and the \textit{Planck} satellite \citep[e.g.,][]{Planck2015_XXIV}. At the time of writing, more than 1000
clusters have been detected in blind SZ searches.

The initial ACT cluster search is described in \citet{Marriage_2011}. A total of 23 clusters were found in a
survey area of 455\,deg$^2$, centered on $-55\deg$ declination, after applying a matched filter to a 
map of the 148\,GHz sky. Optical confirmation and redshifts were obtained using 4\,m class telescopes 
\citep{Menanteau_2010}. From 2009--2010, ACT observations were concentrated on an equatorial field covering 504\,deg$^2$, with complete
coverage by the SDSS Stripe 82 optical survey \citep[S82 hereafter;][]{Annis_2014}. The final cluster sample
extracted from the ACT survey contains 91 confirmed clusters with redshifts, in a total area of 959\,deg$^2$
\citep{Hasselfield_2013, Menanteau_2013}. The sample is 90 per cent complete for 
$M_{\rm 500c} \gtrsim 5 \times 10^{14}$\,M$_{\sun}$ at $z < 1.4$ (assuming a mass-scaling relation based on \citealt{Arnaud_2010},
as described in \citealt{Hasselfield_2013}; note that $M_{\rm 500c}$ is the mass within the radius $R_{\rm 500c}$ that encloses a mean density 
500 times that of the critical density at the cluster redshift).

\begin{figure*}
\includegraphics[width=\textwidth]{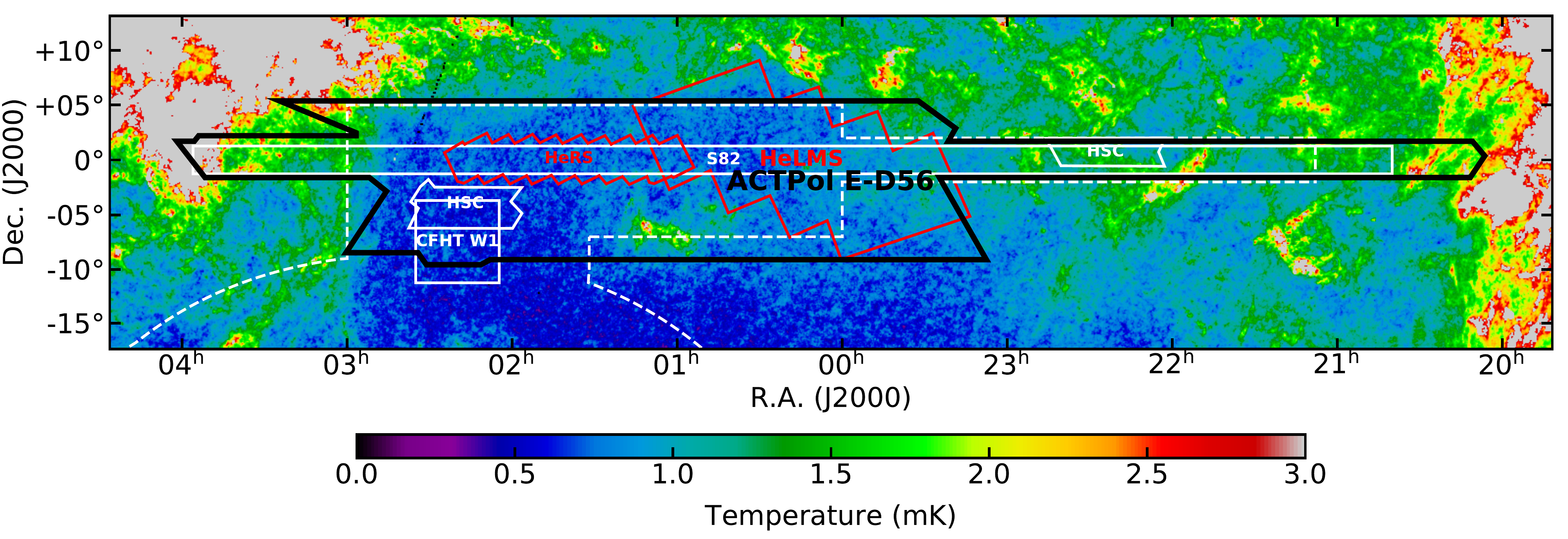}
\caption{The location of the combined ACT equatorial and ACTPol D56 field (E-D56; covering area 987.5\,deg$^{2}$ after masking) overlaid 
on the \textit{Planck} 353\,GHz map, which is sensitive to thermal emission by dust.  The locations of Herschel surveys 
(HeLMS [part of HerMES; \citealt{Oliver_2012}], HeRS [\citealt{Viero_2014}]) and deep optical surveys (CFHTLS W1, HSC [ongoing, current coverage marked; \citealt{HSC_2017}], SDSS S82 [\citealt{Annis_2014}]) are also shown. 
The expected final footprint of the Dark Energy Survey \citep[DES;][]{Diehl_2016} is shown as the dashed white line. Almost the entire E-D56 field is covered by the SDSS legacy survey.}
\label{fig:E-D56}
\end{figure*}

In this paper, we present the first SZ cluster sample derived from observations by the Atacama Cosmology 
Telescope Polarization experiment (ACTPol). The ACTPol receiver \citep{Thornton_2016} is a significant upgrade to
the Millimeter Bolometer Array Camera \citep[MBAC;][]{Swetz_2011}, which was used for the initial ACT survey. The
two 148\,GHz ACTPol bolometer arrays are both roughly a factor of three times more sensitive than MBAC. This
allows ACTPol to detect clusters with smaller SZ signals that have lower masses than those detected by ACT.
In this work, we combine the ACTPol maps of the D56 field \citep{Naess_2014, Louis_2016} with the ACT
equatorial survey maps \citep{Dunner_2013}, and search for clusters in a combined survey area of 987.5\,deg$^2$, which
we will refer to as the ``E-D56'' field. We find a total of 182 confirmed clusters detected with signal-to-noise ratio 
(SNR)~$> 4$ in this survey area. This is double the number of clusters detected in the original ACT survey, in a similar
sized survey region. Tables~\ref{tab:detections}, \ref{tab:redshifts}, and \ref{tab:masses} in the 
Appendix list the coordinates and detected properties, redshifts, and derived masses of the clusters respectively.

The structure of this paper is as follows. We begin by describing the processing of the ACT 148\,GHz data
and the SZ cluster candidate selection and characterization in Section~\ref{sec:ACTSZ}.  In Sections~\ref{sec:OpticalPublic} and 
\ref{sec:OpticalPrivate}, we describe the confirmation of candidates as galaxy clusters using optical/IR data
and the measurement of their redshifts -- this is a crucial first step needed to allow the sample to be used to 
obtain cosmological constraints. In Section~\ref{sec:Sample}, we present the ACTPol E-D56 field cluster sample 
and its properties. We discuss the sample in the context of other work in Section~\ref{sec:Discussion},
in particular applying a richness-based weak-lensing mass calibration to re-scale the SZ cluster masses.
Finally, we summarize our findings in Section~\ref{sec:Summary}.

We assume a flat cosmology with $\Omega_{\rm m}=0.3$, $\Omega_\Lambda=0.7$, and $H_0=70$~km~s$^{-1}$~Mpc$^{-1}$ 
throughout. All magnitudes are on the AB system \citep{Oke_1974}, unless otherwise stated.

\section{ACT Observations and SZ Cluster Candidate Selection}
\label{sec:ACTSZ}

\subsection{148 GHz Observations and Maps}
\label{sec:ACTObs}

\begin{figure*}
\includegraphics[width=\textwidth]{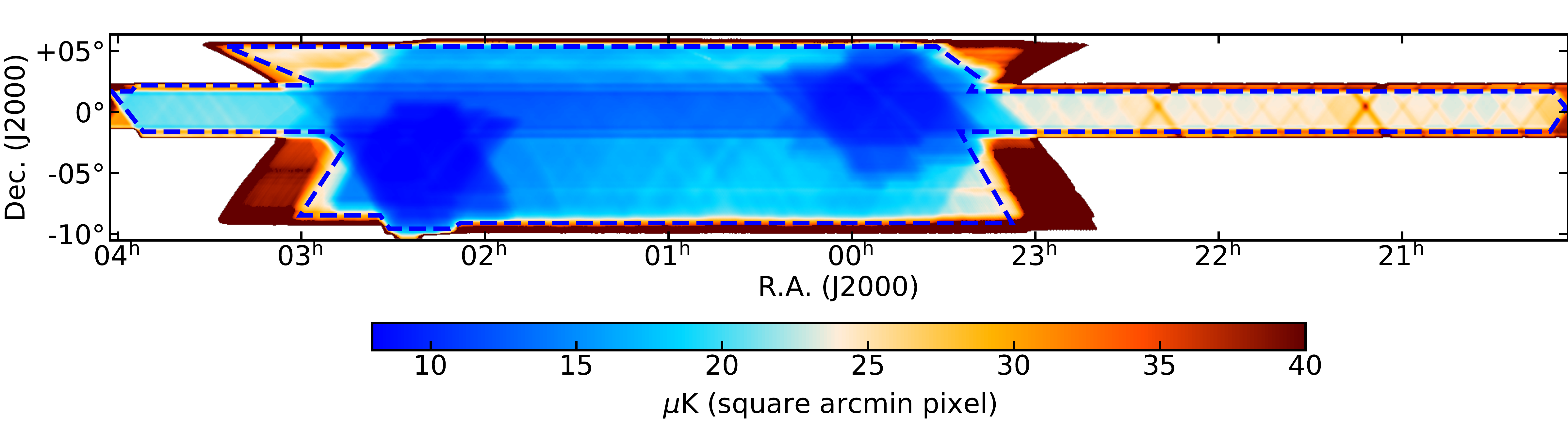}
\caption{The white noise level ($\mu$K per square arcmin pixel) across the inverse-variance weighted 
combination of the ACT equatorial and ACTPol maps (E-D56). The variation in the noise level in this map 
reflects the scan strategy. The cluster search was conducted within the area
bounded by the blue dashed line. The deepest regions are the D5 and D6 fields \citep{Naess_2014, Louis_2016},
located at approximately 23$^{\rm h}$30$^{\rm m}$ and 02$^{\rm h}$30$^{\rm m}$ respectively.}
\label{fig:noiseMap}
\end{figure*}

A description of the ACTPol maps used in this work can be found in \citet{Naess_2014} and 
\citet{Louis_2016}. ACTPol observed two deep fields on the celestial equator, referred to as D5 and D6, 
from 2013 September 11 to 2013 December 14 (Season 13), using a single 148\,GHz detector array (PA1). 
Each of the D5 and D6 fields covers an area of roughly 70\,deg$^2$. In Season 14 (2014 August 20 -- 
2014 December 31), an additional 148\,GHz detector array was added to the ACTPol receiver (PA2), and we obtained 
observations of a wider, approximately 700\,deg$^2$ field, referred to as D56, in which the deeper D5 and D6
fields are embedded. We use only ACTPol night-time observations for this analysis, as the beam for this subset
is well characterized and known to be stable. We made maps from the ACTPol data using
similar methods to those applied to ACT MBAC data, as described in \citet{Dunner_2013}. \citet{Louis_2016}
gives details of some changes and improvements in the data processing pipeline.

The ACTPol D56 field also overlaps with the previous ACT survey of the celestial equator, conducted using
the MBAC receiver \citep{Swetz_2011} at a frequency of 148\,GHz. These observations took place during 2009-2010, and
covered the entire 270\,deg$^2$ SDSS S82 optical survey region \citep{Annis_2014} to a white noise level of 
22\,$\mu$K per square arcmin 
\citep[when filtered on a 5.9$\arcmin$ filter scale;][\citetalias{Hasselfield_2013} hereafter]{Hasselfield_2013}.

In this work, we combine the 148\,GHz observations obtained by ACT using both the MBAC and ACTPol receivers,
in order to maximize our sensitivity for cluster detection using the SZ effect. The resulting survey area,
which we refer to as the E-D56 field, is shown in Fig.~\ref{fig:E-D56}, overplotted on the 2015 
\textit{Planck} 353\,GHz map \citep{Planck2015Overview_2016}, which is sensitive to dust emission. As shown, this region has significant
overlap with several large optical and IR public surveys. We combine a total of six maps, all now publicly available from 
LAMBDA\footnote{\url{https://lambda.gsfc.nasa.gov/product/act/}}, inverse-variance weighted by their white noise 
level. Fig.~\ref{fig:noiseMap} shows the resulting variation of the white noise level across the E-D56 survey 
region. The D5 and D6 regions, observed in 2013 with ACTPol, are easily identified by eye as the lowest noise 
regions. A common area of 296\,deg$^{2}$ within the E-D56 field is covered by both ACT and ACTPol observations.
The boundary of the E-D56 cluster search region itself is shown as the black polygon in Fig.~\ref{fig:E-D56}.
The survey boundary was chosen to enclose the area with a maximum white noise level of approximately 
30\,$\mu$K per square arcmin.

\begin{figure}
\includegraphics[width=\columnwidth]{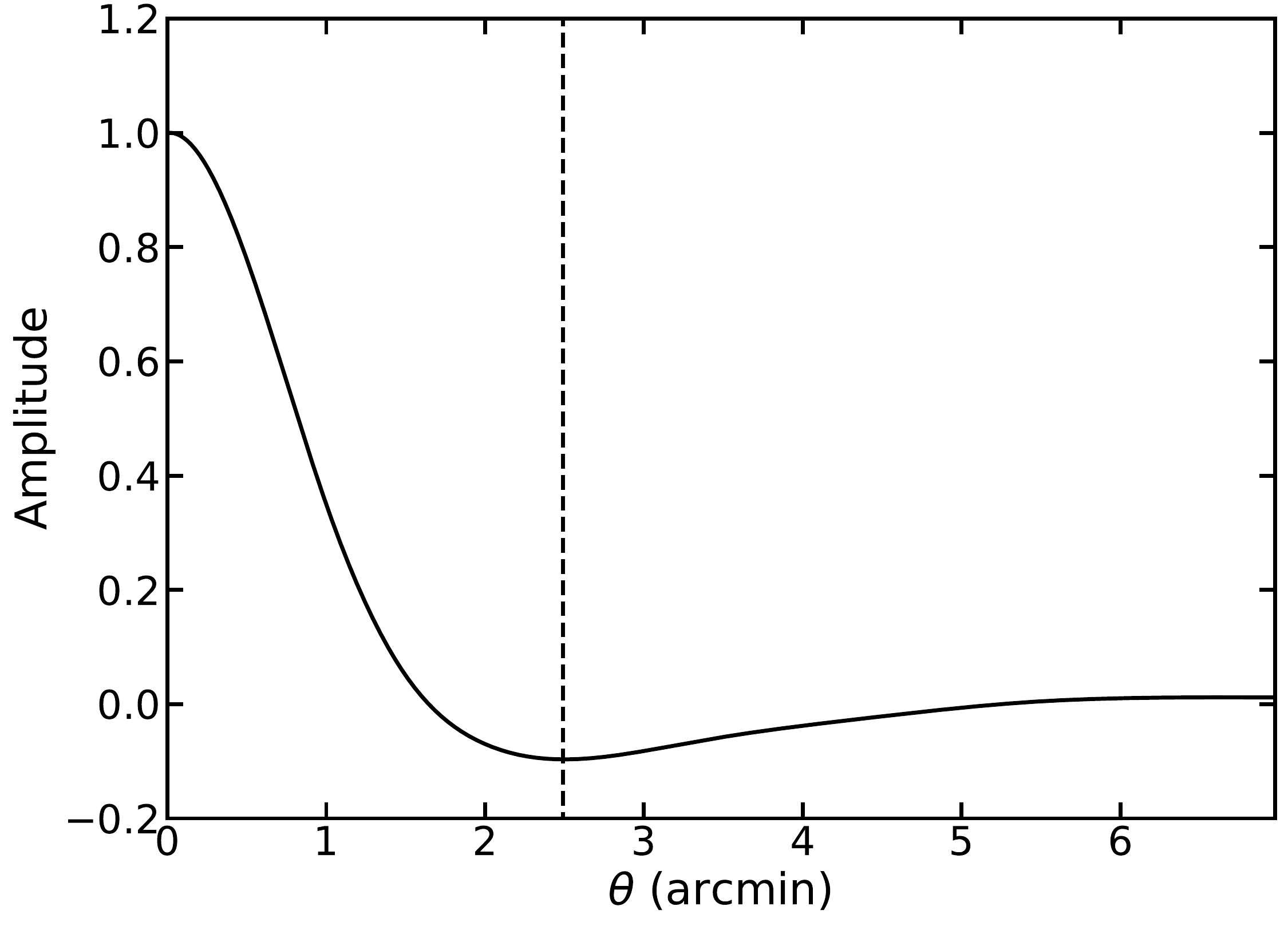}
\caption{The matched filter profile, for the $\theta_{\rm 500c} = 2.4\arcmin$ ($M_{\rm 500c} = 2 \times 10^{14}$\,M$_\sun$ 
at $z = 0.4$) filter scale. This is the reference scale used to characterize cluster masses and the survey 
completeness (see Sections~\ref{sec:SZMass} and \ref{sec:selFn}). The vertical dashed line marks the scale on 
which the map is additionally high-pass filtered. For comparison, the beam FWHM is 1.4$\arcmin$, and the ACT maps have
0.5$\arcmin$ pixel scale.}
\label{fig:filterProfile}
\end{figure}

\begin{figure*}
\includegraphics[width=\textwidth]{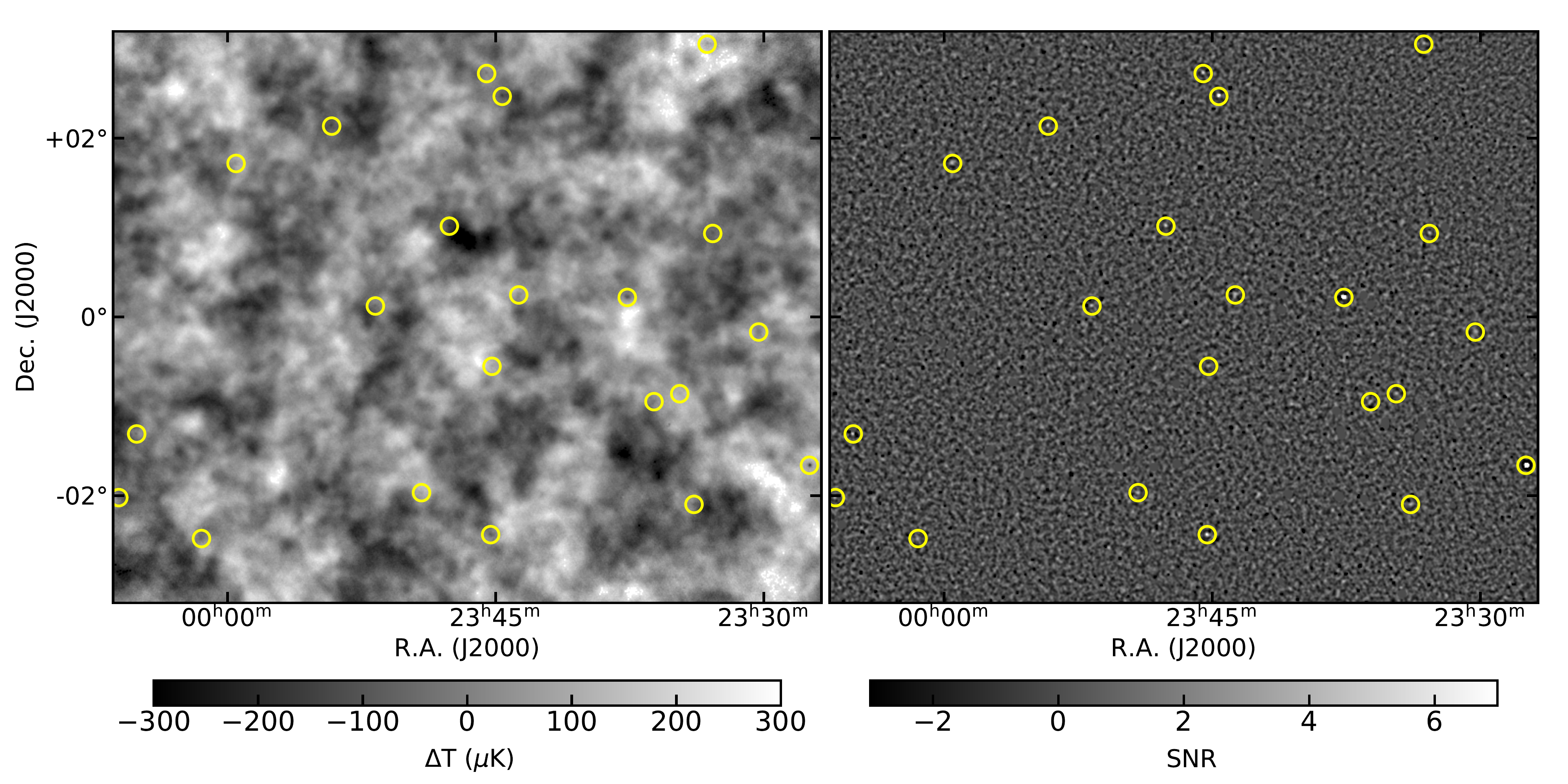}
\caption{Zoom-in on a 79\,deg$^2$ section of the E-D56 map, to show the comparison between the unfiltered (left) and 
filtered (right) maps. The filtered map is the result of convolution with the real-space matched filter kernel
(described in Section~\ref{sec:SZDetection}) with $\theta_{\rm 500c} = 2.4\arcmin$, corresponding to an UPP-model
cluster with $M_{\rm 500c} = 2 \times 10^{14}$\,M$_\sun$ at $z = 0.4$. The positions of detected clusters are highlighted
with yellow circles. The highest SNR cluster detected, ACT-CL J2327.4-0204 ($z = 0.70$; SNR = 23.7), is clearly visible near 
the lower right edge of both maps (in the unfiltered map, it appears as a decrement).}
\label{fig:filteredMap}
\end{figure*}

We masked the locations of point sources in the E-D56 map before searching for clusters, as high-pass filtering
of the maps leads to negative rings around point sources, which can then be falsely flagged as cluster candidates. 
Although sources have already been subtracted from the ACT and ACTPol maps we used in this work, in some cases this is not perfect,
and residuals left in the maps can also result in the detection of spurious cluster candidates after high pass
filtering (Section~\ref{sec:SZDetection}). We masked sources with fluxes in the range 0.015--0.1\,Jy, 0.1--1\,Jy, and 
$> 1$\,Jy with circles of radius 2.4$\arcmin$, 3.6$\arcmin$, and 7.2$\arcmin$ respectively. We also masked
the locations of three artifacts in the map, arising from the construction of the weighted-average map from
the individual ACT and ACTPol maps, with circles of radius 3.6$\arcmin$. The masking process reduced the available
sky area by 1.3\%, resulting in 987.5\,deg$^2$ being available for the cluster search. The median 
white noise level in the cluster search area is 16.8\,$\mu$K per square arcmin.

\subsection{SZ Cluster Candidate Detection}
\label{sec:SZDetection}

In previous ACT cluster searches \citep[][\citetalias{Hasselfield_2013}]{Marriage_2011}, clusters were detected using a matched
filter, applied in Fourier space, which amplifies the signal from cluster scales and in turn suppresses 
large scale noise fluctuations in the map, whether due to the CMB or the atmosphere.
The use of only 148\,GHz data in the previous and current analysis restricts us to using only spatial rather than 
spectral information for SZ cluster detection.

In this work, we take a slightly different approach to spatial filtering for cluster detection to \citetalias{Hasselfield_2013}. 
We begin by constructing
a matched filter in Fourier space, using a small section of the E-D56 map, chosen to coincide with the
D6 field at 02$^{\rm h}30^{\rm m}$ RA (see Fig.~\ref{fig:noiseMap}). The noise power spectrum used in the matched filter 
construction is that of the map itself; this is a good approximation, as the maps are dominated by the CMB on large scales,
and white noise on small scales, rather than cluster signal. As in \citetalias{Hasselfield_2013}, throughout this work we use the 
Universal Pressure Profile \citep[UPP;][\citetalias{Arnaud_2010} hereafter]{Arnaud_2010} and associated mass-scaling relation
to model the SZ signal from galaxy clusters (Section~\ref{sec:SZMass}). This is used as the signal template in the matched filter, 
after convolution with the ACT beam. To maximize the efficiency of detection of clusters at different 
scales, we create a family of 24 such matched filters, corresponding to 
$M_{\rm 500c} = (1, 2, 4, 8) \times 10^{14}$\,M$_\sun$ over the redshift range $0.2 \leq z \leq 1.2$, in steps of 
$\Delta z = 0.2$. Note that there is some degeneracy between lower mass and higher redshift.

In \citetalias{Hasselfield_2013}, each matched filter was applied to the map as a multiplication in Fourier space. However,
since the signal from clusters exists only at arcminute scales, it is feasible to construct a real-space
filter kernel from the matched filter, and apply it to the maps by convolution. One advantage of this 
latter approach is that it simplifies the analysis of maps with arbitrary boundaries, and does not require
the edges of the map to be tapered to avoid ringing in the Fourier transform. It also makes it straightforward
to split a large map into sections that can be analysed separately, using the exact same filter kernel.
This is useful for parallelizing both cluster detection in very large maps, as will be provided by 
Advanced ACTPol \citep{DeBernardis_2016}, and for computation of the survey selection function 
(Section~\ref{sec:selFn}). We therefore
constructed real-space kernels from the family of matched filters, truncating them at $7\arcmin$ 
radius, which results in a kernel with a footprint of $28 \times 28$ pixels. Fig.~\ref{fig:filterProfile} 
shows an example one-dimensional kernel profile. 

Having truncated the filter profile, we need to apply an additional high-pass filter to the maps, in order to
remove noise on scales larger than $7\arcmin$ and reduce contamination from erroneously classifying larger scale
noise features as cluster candidates. We do this by subtracting a Gaussian-smoothed version
of the unfiltered map from itself, with the smoothing scale set according to the location of the minimum of 
the matched filter kernel. This is typically $\sigma = 2.5\arcmin$, as in the example shown in 
Fig.~\ref{fig:filterProfile}. After high-pass filtering the maps in this way, we convolve them with the 
real-space matched filter kernel, which is normalized such that it returns the cluster central decrement 
$\Delta T$ in each pixel in the filtered map.

To detect clusters, we construct a signal-to-noise ratio (SNR) map for each filtered map, and in turn make
a segmentation map that identifies peaks (cluster candidates) with SNR~$ > 4$. We estimate the noise in each
filtered map by dividing it up into square 20$\arcmin$ cells and measuring the 3$\sigma$-clipped standard 
deviation in each cell, taking into account masked regions. This accounts for the significant variations in
depth seen across the map (Fig.~\ref{fig:noiseMap}). Finally, we apply the survey mask shown in 
Fig.~\ref{fig:noiseMap} to reject the noisiest regions at the edges of the E-D56 map. 
Fig.~\ref{fig:filteredMap} shows a side-by-side comparison of a section of the unfiltered 148\,GHz E-D56 map 
with the corresponding filtered map (in units of SNR), after application of the survey and point source masks. 

To construct the catalog of cluster candidates, we first make catalogs of candidates at each filter scale, from
each SNR map. We use a minimum detection threshold of a single pixel with SNR~$ > 4$ in any filtered map. We adopt the
location of the center-of-mass of the SNR~$ > 4$ pixels in each detected object in the filtered map as the 
coordinates of the cluster candidate. We then create a final master candidate list by cross-matching the catalogs 
assembled at each cluster scale using a 1.4$\arcmin$ matching radius. We adopt the maximum SNR across all filter
scales for each candidate as the `optimal' SNR detection. However, as in \citetalias{Hasselfield_2013}, and discussed
in Section~\ref{sec:SZMass}, we also adopt a single reference filter scale (chosen to be $\theta_{\rm 500c} = 2.4\arcmin$) 
at which we also measure the signal-to-noise ratio. Throughout this work we use SNR to refer to the `optimal' signal-to-noise ratio 
(maximized over all filter scales), and SNR$_{2.4}$ for the signal-to-noise ratio measured at the fixed 2.4$\arcmin$ filter scale. 

\begin{figure}
\includegraphics[width=\columnwidth]{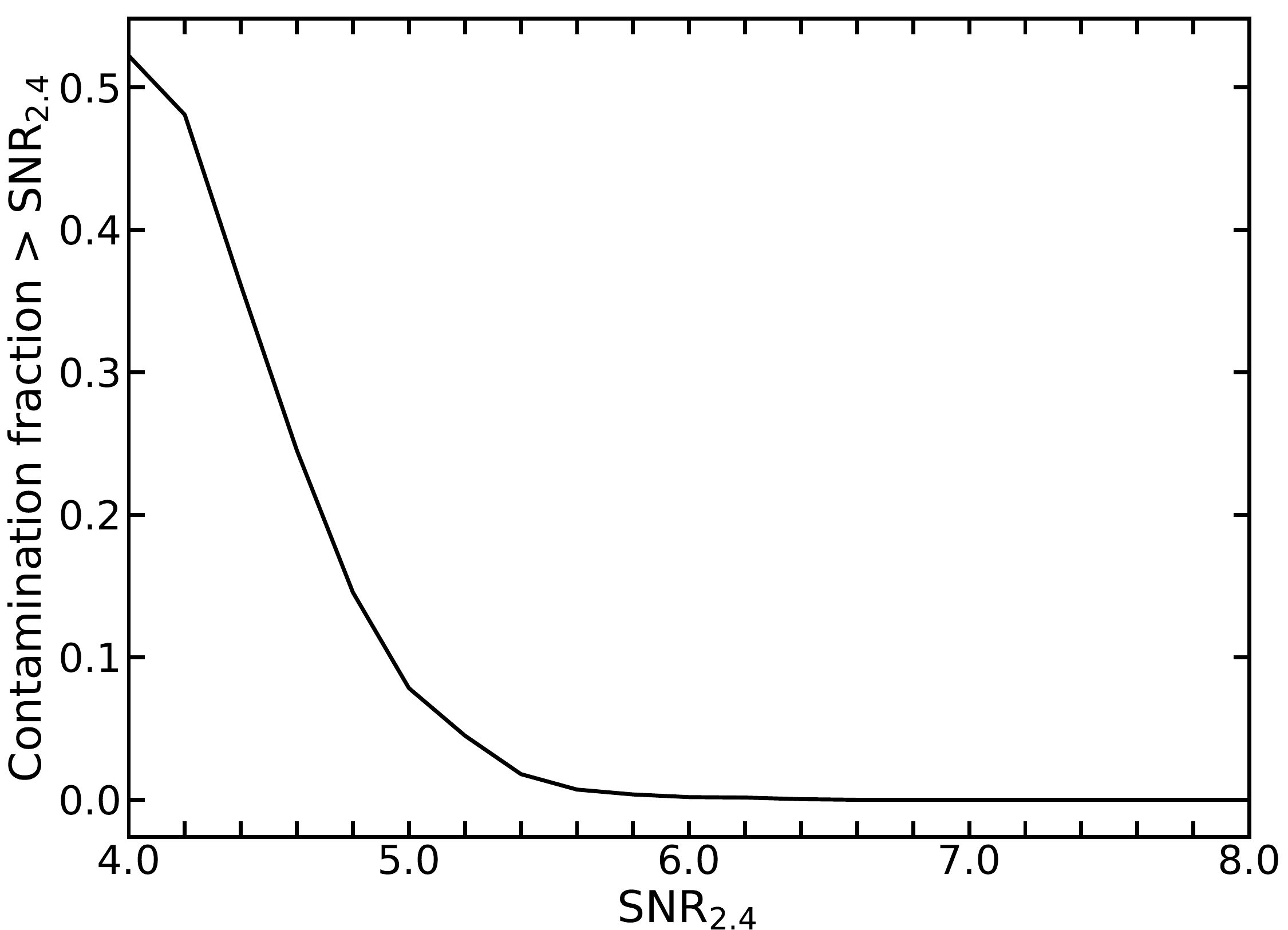}
\caption{Estimated contamination fraction (i.e., false positive detection rate) versus SNR$_{2.4}$. This was estimated by applying the
matched filter at the 2.4$\arcmin$ reference scale to simulated sky maps that contained no cluster signal, 
and averaging the results (see Section~\ref{sec:SZDetection}).}
\label{fig:contamination}
\end{figure}

We assess the fraction of false positive detections above a given SNR$_{2.4}$ cut by running the cluster detection algorithm over sky simulations
that are free of cluster signal. We generate 100 random realizations of the CMB sky using a CAMB \citep{Lewis_2000} power spectrum model with 
parameters consistent with \textit{Planck} 2015 results \citep{Planck2015_XIII}. To these we add white noise, varying across the E-D56 field 
according to the ACT scan strategy, and scaled
to match the noise level seen in the real data. We apply the same survey boundary and point source mask to these simulations as were applied
to the real data, in order to match the real survey area. Fig.~\ref{fig:contamination} shows the result after averaging over 100 simulated sky
maps: at SNR$_{2.4}$~$ > 4.0$, the false positive rate is 52\%, which falls to 30\% for SNR$_{2.4}$~$ > 4.5$,  8\% for SNR$_{2.4}$~$>5.0$, 
and 0.7\% for SNR$_{2.4}$~$ > 5.6$. The fraction
of cluster candidates that have been optically confirmed as clusters in the final catalog (see Section~\ref{sec:Sample}) shows 
that Fig.~\ref{fig:contamination} gives a reasonable estimate of the false positive rate.

\begin{figure*}
\includegraphics[width=\textwidth]{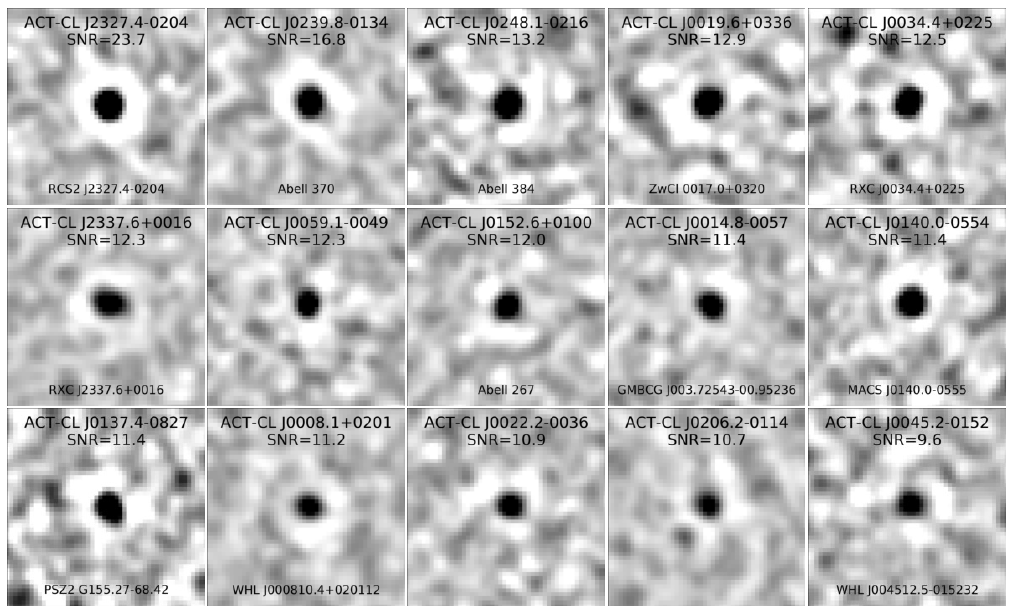}
\caption{Postage stamp images (25$\arcmin$ on a side;  0.5$\arcmin$ pixels; North is at the top, and East is to the left) 
for the 15 highest SNR detections in the catalog (see 
Table~\ref{tab:detections}), taken from the filtered ACT maps. The clusters are ordered by detection SNR,
optimized over all filter scales, from top left to bottom right. They cover the range $9.6 < {\rm SNR} < 23.5$, 
and the minimum SNR here is higher than all but two of the detections in the previous ACT equatorial survey \citep{Hasselfield_2013}. 
None of these are new discoveries.
The greyscale is linear and runs from -150\,$\mu$K (black) to +50\,$\mu$K (white). ACT-CL\,J0034.9+0233,
which is at the same redshift as ACT-CL J0034.4+0225, is clearly visible (detected at SNR = 5.1) towards the
northeast in the image of the latter. Similarly, ACT-CL\,J0206.4$-$0118 ($z = 0.195$, detected at SNR = 5.1) is seen to the
southeast of ACT-CL\,J0206.2$-$0114 ($z = 0.676$, detected at SNR = 10.7).
}
\label{fig:SZMontage}
\end{figure*}

Fig.~\ref{fig:SZMontage} presents postage stamp images of the fifteen highest SNR
candidates detected in the E-D56 field, which cover the range $9.6 < {\rm SNR} < 23.5$. None of them are new cluster discoveries.  
Ten of these were previously detected by ACT (three of which were entirely new systems: ACT-CL\,J0059.1$-$0049, 
ACT-CL\,J0022.2$-$0036, and ACT-CL\,J0206.2$-$0114) and the remainder were known before the era of modern SZ surveys. 
For comparison, only 2/68 objects in the H13 equatorial ACT survey were detected with SNR higher than the lowest SNR 
cluster shown in Fig.~\ref{fig:SZMontage}, which reflects the greater depth and larger area coverage of the ACTPol maps.

The final candidate list contains a total of 
517 cluster candidates detected with SNR~$ > 4$ (110 candidates with SNR~$ > 5$). As described in Sections~\ref{sec:OpticalPublic}
and \ref{sec:OpticalPrivate}, 182/517 candidates have been optically confirmed as clusters and have 
redshift measurements at the time of writing. We discuss the redshift completeness and purity of the sample in Section~\ref{sec:Sample}.
Table~\ref{tab:detections} presents the SZ properties of the 182 candidates detected with SNR~$ > 4$ that are optically
confirmed as clusters.

\subsection{Cluster Characterization}
\label{sec:SZMass}

Although we select cluster candidates using a suite of matched filters in order to maximize the cluster yield,
we follow \citetalias{Hasselfield_2013} by choosing to characterize the cluster signal and its relation to mass using a single 
fixed filter scale. This approach is called Profile Based Amplitude Analysis (PBAA), and has the advantage
that it avoids the complication of inter-filter noise bias (see the discussion in \citetalias{Hasselfield_2013}, where this method
was introduced) and in turn simplifies the survey selection function (see Section~\ref{sec:selFn}). 
However, we note that the cluster finder still maximizes SNR over location in the sky, which results in a small positive bias
in the recovered SNR values \citep[at most $\approx 7$\% at SNR$_{2.4} = 4.0$; see, e.g.,][]{Vanderlinde_2010}.

We use the UPP to model the cluster signal, and we relate mass to the SZ signal using the \citetalias{Arnaud_2010} scaling relation, applying the methods described
in \citetalias{Hasselfield_2013}. For a map filtered at a fixed scale, the cluster central Compton parameter $\tilde{y}_{0}$ is related to mass
through
\begin{equation}
\tilde{y}_{0} = 10^{A_0} E(z)^2 \left( \frac{M_{\rm 500c}}{M_{\rm pivot}} \right)^{1+B_0} Q(M_{\rm 500c}, z) f_{\rm rel} (M_{\rm 500c}, z) \, ,
\label{eq:y0}
\end{equation}
where $10^{A_0} = 4.95 \times 10^{-5}$ is the normalization, $B_0 = 0.08$, $M_{\rm pivot} = 3 \times 10^{14}$\,M$_{\sun}$ (these
values are equivalent to the \citetalias{Arnaud_2010} scaling relation; see \citetalias{Hasselfield_2013}).
We describe the cluster--filter scale mismatch function, $Q (M_{\rm 500c}, z)$, and the relativistic correction, $f_{\rm rel}$, below. 

\begin{figure}
\includegraphics[width=\columnwidth]{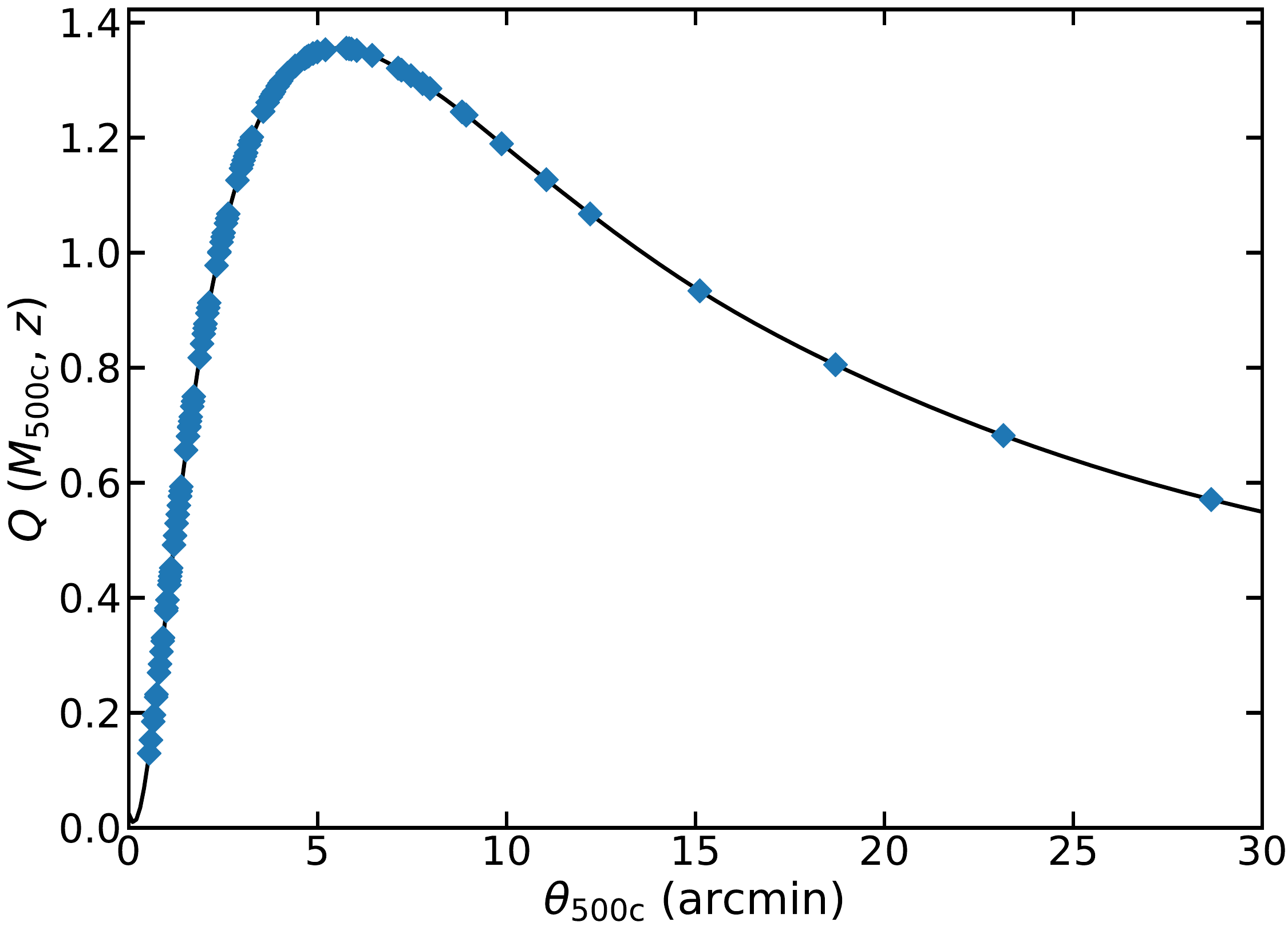}
\caption{The filter mismatch function, $Q$, which is used to reconstruct cluster central Compton parameters
and in turn infer cluster masses (see Section~\ref{sec:SZMass}), under the assumption that clusters are 
described by the UPP and \citetalias{Arnaud_2010} scaling relation. In this work, we use a matched filter constructed from a UPP
model with $M_{\rm 500c} = 2 \times 10^{14}$\,M$_{\sun}$ at $z = 0.4$ ($\theta_{\rm 500c} = 2.4\arcmin$) as our 
reference. The blue diamonds mark scales at which the value of $Q$ was evaluated numerically, over wide ranges in
mass ($13.5 < \log M_{\rm 500c} < 16$) and redshift ($0.1 < z < 1.7$), while the 
solid line is a spline fit.}
\label{fig:QFit}
\end{figure}

The function $Q(M_{\rm 500c}, z)$, shown in Fig.~\ref{fig:QFit}, accounts for the mismatch between the size of a cluster with a different mass and 
redshift to the reference model used to define the matched filter (including the effect of the beam) and in turn $\tilde{y}_{0}$ 
(see Section~3.1 of \citetalias{Hasselfield_2013}).
In this work, we use a UPP-model cluster with $M_{\rm 500c} = 2 \times 10^{14}$\,M$_{\sun}$ at $z = 0.4$ to define the 
reference filter scale. This has an angular scale of $\theta_{\rm 500c} = 2.4\arcmin$, which is smaller than the 
$\theta_{\rm 500c} = 5.9\arcmin$ scale
adopted in \citetalias{Hasselfield_2013}; this is motivated by the fact that this scale is better matched to the majority of the clusters
in our sample, and results in higher SNR $\tilde{y}_{0}$ measurements than would be achieved by
filtering on a larger scale. Our cluster observable $\tilde{y}_{0}$ is therefore extracted from the map filtered 
at the $\theta_{\rm 500c} = 2.4\arcmin$ scale at each detected cluster position. We also define an equivalent 
signal-to-noise ratio at this fixed filter scale, which we will refer to as SNR$_{2.4}$. 

The relativistic correction $f_{\rm rel}$ in equation~(\ref{eq:y0}) is implemented in the same way as in \citetalias{Hasselfield_2013}, i.e., we use the 
\citet{Arnaud_2005} mass--temperature relation in order to convert $M_{\rm 500c}$ to temperature at a given
cluster redshift, and then apply the formulae of \citet{Itoh_1998} to calculate $f_{\rm rel}$. These corrections are
at the $<10$\% level for the ACTPol sample.

For cosmological applications, the quantity of interest in equation~(\ref{eq:y0}) is $M_{\rm 500c}$, but to extract a mass
for each cluster in the sample, we must also take into account the intrinsic scatter in the SZ signal--mass scaling
relation, and also the fact that the average recovered mass will be biased high due to the steepness of the cluster
mass function. To extract a mass estimate for each cluster with a redshift measurement, we calculate the 
posterior probability
\begin{equation}
P( M_{\rm 500c} | \tilde{y}_{0}, z) \propto P ( \tilde{y}_{0} | M_{\rm 500c}, z) P(M_{\rm 500c} | z) \, ,
\label{eq:PM500}
\end{equation}
assuming that there is intrinsic log normal scatter $\sigma_{\rm int}$ in $\tilde{y}_{0}$ about the mean 
relation defined in equation~(\ref{eq:y0}), in addition to the effect of the measurement error on $\tilde{y}_{0}$. 
Following \citetalias{Hasselfield_2013}, we take $\sigma_{\rm int} = 0.2$ throughout this work. \citetalias{Hasselfield_2013}
showed that this level of scatter is seen in both numerical simulations \citep[taken from][]{Bode_2012} and dynamical mass
measurements of ACT clusters \citep[taken from][]{Sifon_2013}. Here, $P(M_{\rm 500c} | z)$ is the halo
mass function at redshift $z$, for which we use the results of the calculation by \citet{Tinker_2008}, as implemented in the 
\texttt{hmf}\footnote{\url{https://pypi.python.org/pypi/hmf/2.0.5}} \texttt{python} package 
\citep*{Murray_2013}. We assume $\sigma_8 = 0.80$ for such calculations throughout this work. 
Where we use photometric redshifts, we also marginalize over the redshift uncertainty. We adopt the maximum of the
$P( M_{\rm 500c} | \tilde{y}_{0}, z)$ distribution as the cluster $M_{\rm 500c}$ estimate, and the uncertainties quoted on these 
masses are $1\sigma$ error bars that do not take into account any uncertainty on the scaling relation parameters. The mass estimates
obtained through equations~(\ref{eq:y0}) and (\ref{eq:PM500}) are referred to as $M^{\rm UPP}_{\rm 500c}$ throughout this work.

It is the inclusion of the $P(M_{\rm 500c} | z)$ term that corrects the derived cluster masses for the effect of the 
steep halo mass function on cluster selection. For the ACT UPP-based masses, and assuming the \citet{Tinker_2008} mass function, 
this leads to an $\approx 16$\% correction,  \citep{Battaglia_2016}. For some comparisons to other samples, and for the calculation of mass limits 
based on the survey selection function (Section~\ref{sec:selFn}), it is necessary to omit this correction.
We list such ``uncorrected'' mass estimates as $M_{\rm 500c}^{\rm Unc}$ in Table~\ref{tab:masses}.

\begin{figure}
\includegraphics[width=\columnwidth]{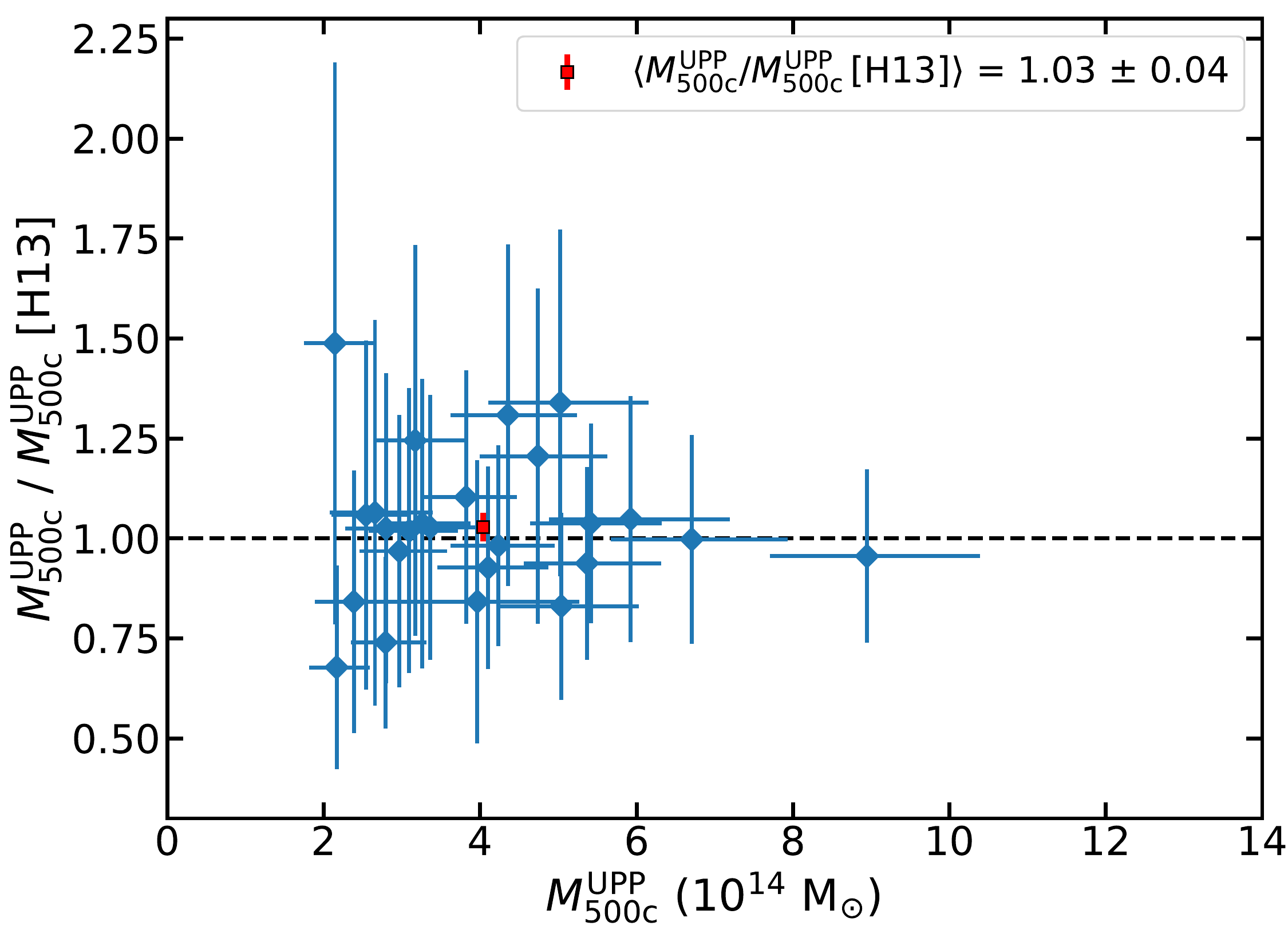}
\caption{End-to-end test of $M_{\rm 500c}$ recovery, comparing clusters cross-matched with \citetalias{Hasselfield_2013} (2.5$\arcmin$ matching
radius) with $M_{\rm 500c}$ values inferred from SZ decrement measurements made on D56 maps containing only ACTPol 
data, filtered at the $\theta_{\rm 500c} = 2.4\arcmin$ scale (this work). The data sets used for this test have independent detector noise. 
The red square marks the unweighted mean ratio ($\pm$ standard error) between the two sets of measurements. 
This test assumes that clusters are described by both the UPP and the \citetalias{Arnaud_2010} mass-scaling relation.}
\label{fig:massRecovery}
\end{figure}

\begin{figure*}
\includegraphics[width=\textwidth]{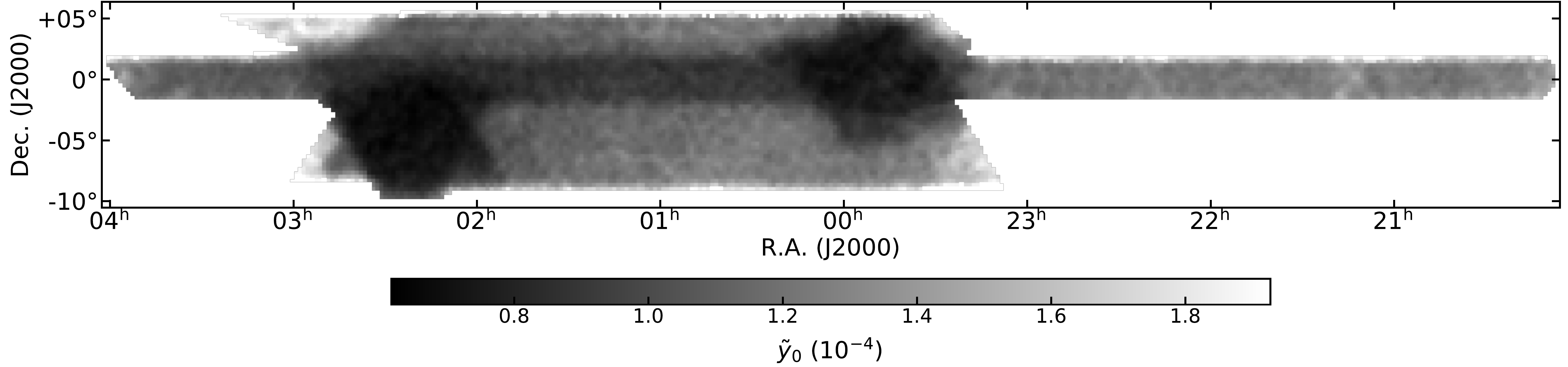}
\caption{Map of the $\tilde{y}_{0}$-limit corresponding to SNR$_{2.4} = 5$ across the ACTPol E-D56 field. In addition
to capturing the variation in the white noise level caused by the ACT scan strategy, noise on $20\arcmin$ scales from the CMB and 
Galactic dust emission is also visible.}
\label{fig:ycLimitMap}
\end{figure*}

Since we are using a different filtering and cluster finding scheme to that used in \citetalias{Hasselfield_2013}, and we have 296\,deg$^2$ of sky area in 
common between the \citetalias{Hasselfield_2013} ACT equatorial survey and the ACTPol observations, we performed an end-to-end check
of SZ signal measurement and mass recovery by using the ACT and ACTPol data independently. These are disjoint data sets with independent detector noise. 
For this test, we applied the $\theta_{\rm 500c} = 2.4\arcmin$ filtering scheme described in Section~\ref{sec:SZDetection} to ACTPol data 
alone, and cross-matched the detected cluster candidates with the \citetalias{Hasselfield_2013} cluster catalog using a 2.5$\arcmin$ matching radius, finding 
25 such clusters (the ACTPol observations only overlap with part of the \citetalias{Hasselfield_2013} map, and some low SNR objects reported in 
\citetalias{Hasselfield_2013} are not included in the ACTPol sample; see the discussion in 
Section~\ref{sec:Sample}). After estimating their masses using equations~(\ref{eq:y0}) and (\ref{eq:PM500}), we compare them with the UPP
masses listed in the \citetalias{Hasselfield_2013} cluster catalog (shown as $M^{\rm UPP}_{\rm 500c}\,[\rm H13]$ in this work). 
Fig.~\ref{fig:massRecovery} shows the result. Although the uncertainties on individual masses are large, the 
$M^{\rm UPP}_{\rm 500c}$ measurements inferred from the ACTPol data are unbiased with respect to the \citetalias{Hasselfield_2013} masses,
with an unweighted mean ratio of $\langle M^{\rm UPP}_{\rm 500c} / M^{\rm UPP}_{\rm 500c}\,[\rm H13] \rangle = 1.03 \pm 0.04$ 
(where the quoted uncertainty is the standard error on the mean, i.e., $\sigma/\sqrt{N}$, where $N = 25$). 
Moreover, the results of a two-sample Kolmogorov-Smirnov (K-S) test are consistent with the null hypothesis that both samples are drawn 
from the same mass distribution ($D = 0.12$, $p$-value $=0.99$).

Table~\ref{tab:masses} presents SZ mass estimates derived from $\tilde{y}_{0}$ measurements in the E-D56 map for all optically confirmed clusters
detected with ACTPol.

\subsection{Survey Completeness}
\label{sec:selFn}

We assess the completeness of the ACTPol cluster search by inserting UPP-model clusters into the real
ACTPol E-D56 map, after first inverting it to avoid any bias due to the presence of real clusters. 
Given the complications of inter-filter bias, we characterize the survey completeness using only the 
$\theta_{\rm 500c} = 2.4\arcmin$ filter. 

As can be seen from Fig~\ref{fig:noiseMap}, the white noise level in the map varies considerably, and so we break up the map 
into tiles that are $20\arcmin$ on a side and check the recovery of model clusters in each tile separately. 
We insert into each tile a UPP-model cluster with one of 20 linearly spaced $M_{\rm 500c}$ values between 
(0.5--10)$\times 10^{14}$\,M$_{\sun}$ in turn. We repeat this for each of a set of 15 different redshifts in the range
$0.05 < z < 2$, and for 80 randomly chosen positions within each tile, taking into account the survey and point source masks 
(Section~\ref{sec:ACTObs}). We then perform the same filtering operations on each tile that were applied
to the map in the cluster search (i.e., using the $\theta_{\rm 500c} = 2.4\arcmin$ real space matched filter kernel in combination
with the $\sigma = 2.5\arcmin$ high-pass filter),
and extract the SNR$_{2.4}$ and $\tilde{y}_{0}$ values at each of the 80 positions within each tile for each different
cluster model. We take the median SNR$_{2.4}$ and $\tilde{y}_{0}$ over the different positions within each tile,
and use these to perform a linear fit for $\tilde{y}_{0}$ as a function of SNR$_{2.4}$, in order
to determine the $\tilde{y}_{0}$ signal level corresponding to a chosen cut in SNR$_{2.4}$ in each tile. 
Fig.~\ref{fig:ycLimitMap} shows the resulting $\tilde{y}_{0}$-limit map corresponding to SNR$_{2.4} = 5$, which captures not
only the variation in the white noise level due to the ACT/ACTPol scan strategy, but also additional noise variation at the 
$20\arcmin$ scale, due to the CMB and galactic dust emission. 

\begin{figure}
\includegraphics[width=\columnwidth]{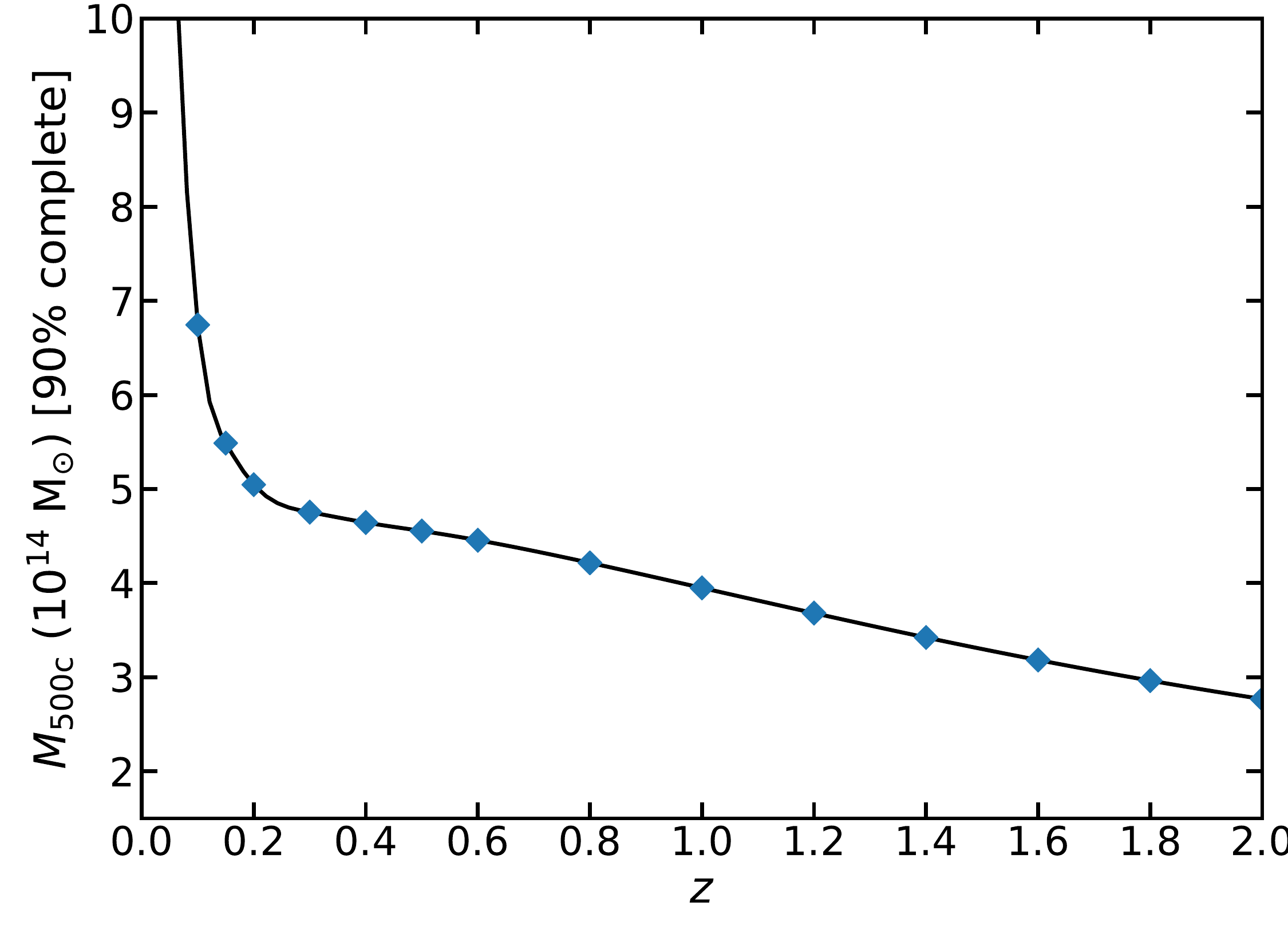}
\caption{Survey-averaged 90\% $M_{\rm 500c}$ completeness limit as a function of redshift, as assessed by inserting
UPP-model clusters into the map, filtering at the $\theta_{\rm 500c} = 2.4\arcmin$ scale, and assuming the \citetalias{Arnaud_2010} 
mass scaling relation holds. The blue diamonds mark the redshifts at which the limit was estimated, and the 
solid line is a spline fit. In the redshift range $0.2 < z < 1.0$, the average 90\% completeness limit is 
$M_{\rm 500c} > 4.5 \times 10^{14}$\,M$_\sun$ for SNR$_{2.4} > 5$.}
\label{fig:completenessByRedshift}
\end{figure}

\begin{figure}
\includegraphics[width=\columnwidth]{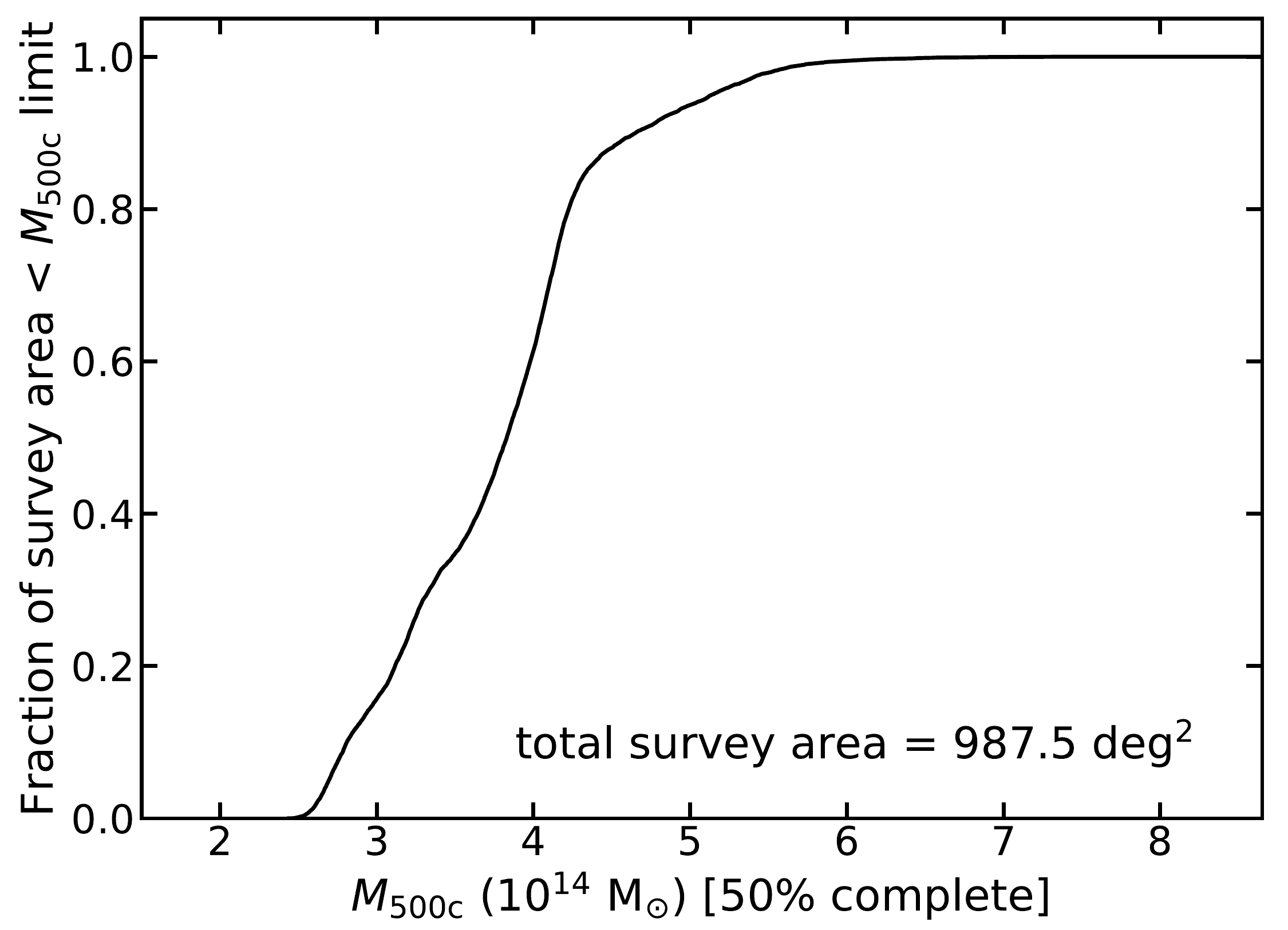}
\caption{Fraction of the survey area as a function of $M_{\rm 500c}$ 50\% completeness limit, averaged over the
redshift range $0.2 < z < 1$, as assessed from inserting UPP-model clusters into the E-D56 map, filtering
at the $\theta_{\rm 500c} = 2.4\arcmin$ scale, applying a cut of SNR$_{2.4} > 5$, and assuming 
the \citetalias{Arnaud_2010} mass scaling relation.}
\label{fig:massLimitArea}
\end{figure}

In order to express the survey-averaged completeness in terms of a mass limit, we apply equations~(\ref{eq:y0}) and 
(\ref{eq:PM500}) to the SNR$_{2.4}$ versus $\tilde{y}_{0}$ relation measured in each tile, over a grid of redshifts
spanning the range $0.05 < z < 2$, and weighting by fraction of the survey area. Fig.~\ref{fig:completenessByRedshift} 
shows the resulting survey-averaged 90\% completeness limit for a cut of SNR$_{2.4} > 5$. As seen in \citetalias{Hasselfield_2013}, the ACTPol 
cluster sample is expected to be incomplete for all but the most massive clusters 
at $z < 0.2$. This limitation is due to using only a spatial filter to remove the CMB, resulting in confusion 
when the angular size of low redshift clusters approaches that of CMB anisotropies. The SZ signal increases at fixed
$M_{\rm 500c}$ as redshift increases for our adopted scaling relation (equation~\ref{eq:y0}), and so lower mass clusters
are relatively easier to detect at higher redshift. Averaged over the redshift range
$0.2 < z < 1.0$, we estimate that the survey-averaged 90\% completeness limit is $M_{\rm 500c} > 4.5 \times 10^{14}$\,M$_\sun$
for SNR$_{2.4} > 5$. This mass limit is approximately 10\% lower than that found in \citetalias{Hasselfield_2013} in the S82 survey region, 
and reflects the lower average noise in the E-D56 map in comparison to the ACT maps used in that work. On this basis,
we expect the ACTPol sample to contain roughly 4.8 times as many SNR~$_{2.4} > 5$ clusters as the \citetalias{Hasselfield_2013} sample, 
after correcting for the differences in the depth and area between the two surveys (although the definitions of 
signal-to-noise are not exactly equivalent, as they are measured on different angular scales). A comparison of the two cluster catalogs
shows that this is the case.

We can similarly assess the variation in the mass limit across the survey area. Fig.~\ref{fig:massLimitArea} shows
the fraction of survey area as a function of the inferred 50\% completeness mass limit for a SNR$_{2.4} > 5$
cut, averaged over the redshift range $0.2 < z < 1$. Over 75\% of the map, the 50\% completeness limit is
$\approx 4.2 \times 10^{14}$\,M$_{\sun}$. In roughly 15\% of the map, corresponding to the ACTPol
D5 and D6 fields, the 50\% completeness limit is $M_{\rm 500c} \approx 3.0 \times 10^{14}$\,M$_{\sun}$ for SNR$_{2.4} > 5$.

\section{Confirmation and Redshifts from Large Public Surveys}
\label{sec:OpticalPublic}

As highlighted in Fig.~\ref{fig:E-D56}, one of the benefits of the location of the ACTPol E-D56 field is
its extensive overlap with public surveys. Almost the entire field is covered by the Sloan Digital Sky Survey
Data Release 13 \citep[SDSS DR13;][]{Albareti_2016}, which provides five-band ($ugriz$) photometry and 
spectroscopy. The deeper S82 region \citep{Annis_2014} also falls entirely within the survey area,
and there is partial overlap with the Canada-France-Hawaii Telescope Legacy Survey (CFHTLS) W1 field.
The ongoing Hyper Suprime-Cam Survey \citep[HSC;][]{Aihara_2017} has a few tens of square degrees 
of overlap with ACTPol observations at the time of writing, and this area will increase with time.
The entire field is covered by the first Pan-STARRS data release \citep[PS1;][]{Chambers_2016, Flewelling_2016},
although as this was made public recently, it is not used in this analysis, except for obtaining the redshift of one
cluster at low Galactic latitude, outside of SDSS (Section~\ref{sec:strayPS1}). In this Section, we describe how we use such 
surveys to provide confirmation and redshift measurements for the bulk of the ACTPol cluster candidates.

\subsection{Photometric Redshifts}
\label{sec:zClusterAlgorithm}
We now describe our algorithm, named zCluster,\footnote{\url{https://github.com/ACTCollaboration/zCluster}} 
for estimating cluster redshifts using multi-band optical/IR 
photometry. In this paper it has been applied to SDSS \citep{Albareti_2016}, S82 \citep{Annis_2014}, and 
CFHTLS survey data (we use the photometric catalogs of the CFHTLenS project; \citealt{Hildebrandt_2012, Erben_2013}), 
in addition to our own follow-up observations (Section~\ref{sec:APOSOAR}). The aim of 
zCluster is to use the full range of photometric information available, and to
make a minimal set of assumptions about the optical properties of clusters, since the algorithm is being
used to measure the redshifts of clusters selected by other methods (in this case via the SZ effect). 
This is a different approach to that used by redMaPPer \citep{Rykoff_2014}, for example, where the colors of cluster red-sequence
galaxies are used to find both the clusters themselves and to estimate the redshift. The approach we describe
here avoids modeling the evolution of the cluster red-sequence, but does require the choice of an 
appropriate set of spectral templates.

The first step in zCluster is to measure the redshift probability distribution $p(z)$ of each galaxy in the 
direction of each cluster candidate using a template-fitting method, as used in codes like \textsc{BPZ} 
\citep{Benitez_2000} and \textsc{EAZY} \citep{Brammer_2008}. In fact, we use the default set
of galaxy spectral energy distribution (SED) templates included with both of these 
codes.\footnote{These are the 6 empirical spectral templates of \citet{ColemanWuWeedman_1980} and \citet{Kinney_1996}, as included with BPZ, 
and the optimized set of 6 templates included with EAZY, which are derived from non-negative matrix 
factorization \citep{Blanton_2007} of stellar population synthesis models \citep{Fioc_1997}.}
For each template SED and filter transmission function 
($u$, $g$, $r$, $i$, $z$ in the case of SDSS, for which the filter curves are taken from \textsc{BPZ}), 
we calculate the AB magnitude that would be observed at each 
redshift $z_i$ over the range $0 < z < 3$, in steps of 0.01 in redshift. We then compare the observed 
broadband SED of each galaxy with each template SED at each $z_i$, and construct the
$p(z)$ distribution for each galaxy from the minimum $\chi^2$ value (over the template set) at each 
$z_i$. We apply a magnitude-based prior that sets $p(z) = 0$ at redshifts where the $r$-band absolute magnitude
is brighter than $-24$ (i.e., 2.5\,magnitudes brighter than the characteristic magnitude of the cluster galaxy
luminosity function, as measured by \citealt{Popesso_2005}), since the probability of observing such galaxies 
in reality is extremely small. Note that the peak of the $p(z)$ distribution gives the maximum likelihood galaxy redshift 
\citep[see, e.g.,][]{Benitez_2000}, although these are not what we use for estimating the cluster 
photometric redshift -- we make use of the full $p(z)$ distributions instead. 

\begin{figure*}
\includegraphics[width=\textwidth]{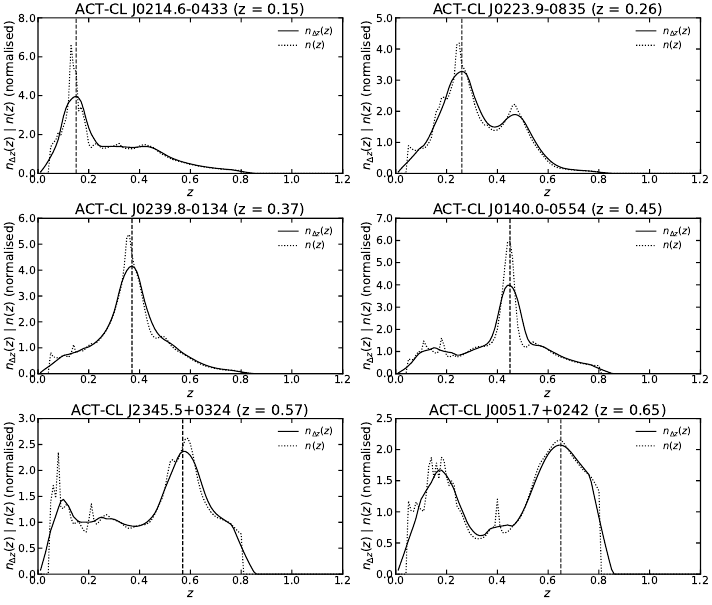}
\caption{Examples of normalized $n(z)$ and $n_{\Delta z}(z)$ distributions for several clusters at 
different redshifts (based on SDSS photometry), measured within 1\,Mpc projected radial distance. In
some cases, multiple peaks are seen; we adopt the maximum of $n_{\Delta z}(z)$ as the cluster photometric
redshift (shown as the vertical dashed line). Optical images corresponding to each
of the clusters shown here can be found in Fig.~\ref{fig:SDSSMontage}.}
\label{fig:nz}
\end{figure*}

We estimate the cluster photometric redshift from the weighted sum of the individual galaxy $p(z)$ 
distributions. For the case of SDSS DR13 data, we start with all galaxies within a 36$\arcmin$ radius of 
each cluster position. The reason for this large initial choice of aperture is for calculating the contrast
of each cluster above the local background (see Section~\ref{sec:delta} below). We define the weighted number
of galaxies $n(z)$ as
\begin{equation}
n(z) = P \sum_{k=0}^{N} p_k(z) w_k(z) s_k \, ,
\label{e_nz}
\end{equation}
where $z$ represents the array of $z_i$ values, $p_k(z)$ is the $p(z)$ distribution of the $k$th galaxy of $N$ galaxies
in the catalog;
$w_k(z)$ is a weight which depends on the projected radial distance $r$ of the $k$th galaxy from the 
cluster center, as determined by the SZ cluster detection algorithm, and calculated at $z_i$; $s_k$ is an overall `selection weight' (with value 1 or 0) for the $k$th 
galaxy; and $P$ is a prior distribution for the cluster redshift, which depends on the depth of the 
optical/IR survey.

\begin{figure*}
\begin{center}
\includegraphics[width=\textwidth]{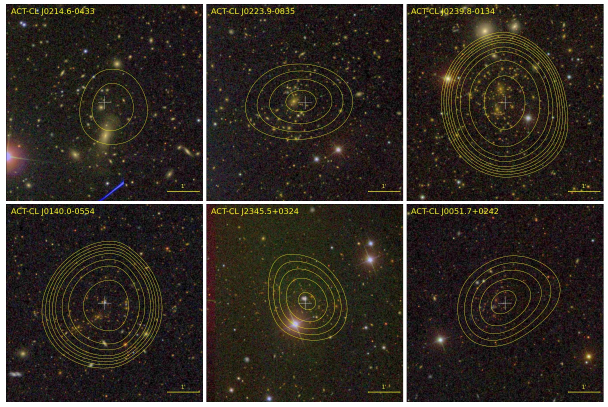}
\end{center}
\caption{Example optical $gri$ images of clusters confirmed in SDSS (these objects correspond to those shown in
Fig.~\ref{fig:nz}). Each image is $6\arcmin$ on a side, with North at the top and East at the left. The yellow 
contours (minimum 3$\sigma$, increasing in steps of 0.5$\sigma$ up to SNR~$=5$, and then by 1--2$\sigma$ thereafter) 
indicate the 
(smoothed) 148\,GHz decrement in the matched-filtered ACT map. The white cross indicates the ACT SZ cluster position. 
Note that ACT-CL\,J0051.7+0242 is a newly discovered cluster.}
\label{fig:SDSSMontage}
\end{figure*}

For the radial weights, $w_k(z)$, we assume that clusters follow a projected 2D Navarro-Frenk-White 
profile \citep[NFW;][]{NFW_1997}, as in \citet{Koester_2007} following \citet{Bartelmann_1996}. 
We adopt a scale radius of $r_s = R_{200}/c =150$\,kpc ($c$ is the 
concentration parameter). We define $w_k(z)$ such that $w_k(z) = 1$ for a galaxy located at the cluster
center ($r = 0$), and we set $w_k(z) = 0$ for galaxies with $r > 1$\,Mpc. Note that because of the 
way $w_k(z)$ is defined, different galaxies contribute to $n(z)$ at different redshifts.

For some galaxies, the $p(z)$ distribution can be relatively flat. In these cases, the photometric redshift 
of the galaxy itself is not well constrained, and including such objects only adds noise to $n(z)$. 
To mitigate this, we use an `odds' parameter $p_{\Delta z}$ (as introduced by \citealt{Benitez_2000} 
for \textsc{BPZ}, and also implemented in \textsc{EAZY}), where we define $p_{\Delta z}$ as the fraction 
of $p(z)$ found within $\Delta z = \pm 0.2$ of the maximum likelihood redshift of the galaxy. We set the 
selection weight $s_k = 1$ for galaxies with $p_{\Delta z} > 0.5$, and $s_k = 0$ otherwise to disregard
such galaxies.

The redshift distribution of clusters that we expect to find in a given survey depends upon its depth. For 
SDSS, for example, very few clusters can be detected in the optical data at $z > 0.5$ (as seen by the lack
of such objects in optical cluster catalogs based on these data; e.g., \citealt{Rykoff_2014}). We encode this 
information in the prior $P$, which for simplicity we take to have a uniform distribution. We adopt
(minimum $z$, maximum $z$) priors of (0.05. 0.8) in SDSS DR13; (0.2, 1.5) in S82; (0.05, 1.5) in CFHTLenS; and 
(0.5, 2.0) for our own APO/SOAR photometry (Section~\ref{sec:APOSOAR}). The maximum $z$-limits used for this
prior are quite generous, because in practice the magnitude-based prior prevents most contamination in the form
of spurious high-redshift estimates of individual galaxy photometric redshifts.

In principle, the cluster redshift can be estimated from the location of the peak of the $n(z)$ distribution. 
In practice, we have seen that, in a small number of cases, the maximum of $n(z)$ is identified with a sharp, 
thin peak that contains only a small fraction of the integrated $n(z)$ distribution. Hence, we define 
$n_{\Delta z}(z)$, which is the integral of $n(z)$ between $\Delta z = \pm 0.2$ calculated at each $z_i$
(this is similar to the definition of $p_{\Delta z}$, except $n_{\Delta z}(z)$ is evaluated over the whole 
redshift range). This procedure makes $n_{\Delta z}(z)$ a smoothed version of $n(z)$. Given the choice of 
$\Delta z$, this also changes the minimum and maximum possible cluster redshifts that can be obtained from a 
given survey by 0.1 compared to the redshift prior cuts. Fig.~\ref{fig:nz}
shows a comparison of $n_{\Delta z}(z)$ and $n(z)$ (normalized so that the integral of each is equal to 1) 
for a few example clusters to illustrate the difference.
However, for 6 clusters, we still found it necessary
to adjust the minimum redshift of the prior to avoid the algorithm selecting a spuriously low redshift.
We adopt the peak of $n_{\Delta z}(z)$ as the cluster redshift $z_{\rm c}$. We estimate the uncertainty
of $z_{\rm c}$ through comparison with the subset of clusters that also have spectroscopic redshift
measurements (see Section~\ref{sec:photoz-test} below).

\subsection{Cluster Confirmation and Archival Spectroscopic Redshifts}
\label{sec:delta}

To confirm the detected SZ candidates as bona fide clusters, and check the assignment of cluster redshifts, 
we used a combination of visual inspection of the available optical imaging, and more objective statistical
criteria. For the latter, we define an optical density contrast statistic $\delta$ 
\citep[e.g.,][]{Muldrew_2012}, which is evaluated for clusters with zCluster photometric redshifts,
\begin{equation}
\delta (z_{\rm c}) = \frac{n_{\rm 0.5\,Mpc}(z_{\rm c})}{A n_{\rm 3-4\,Mpc}(z_{\rm c})} - 1.
\label{eq:delta}
\end{equation}
Here, $n_{\rm 0.5\,Mpc}(z_{\rm c})$ is calculated using equation~(\ref{e_nz}) with uniform radial weights 
(i.e., $w_k(z_{\rm c}) = 1$ for galaxies within the specified projected distance of 0.5\,Mpc given in the 
subscript, and $w_k(z_{\rm c}) = 0$ otherwise). Similarly, $n_{\rm 3-4\,Mpc}(z_{\rm c})$ is the weighted 
number of galaxies at $z_{\rm c}$ in a circular annulus 3--4\,Mpc from the cluster position (taken to be 
the local background number of galaxies), and $A$ is a factor which accounts for the difference in area between these
two count measurements. The primary use of $\delta$ in this work is to flag unreliable photometric redshifts
(see Section~\ref{sec:photoz-test} below).

During the visual inspection stage, we checked that each SZ detection is associated with an optically identified
cluster. We inspected all SZ cluster candidates with SNR~$>5$. For candidates with $4 <$~SNR~$< 5$, we only inspected those
with $\delta > 2$ (as measured by zCluster), a spectroscopic redshift (see below), or with a possible match to a 
known cluster in another catalog.
We used a simple 2.5$\arcmin$ matching radius to search for possible cluster counterparts to ACTPol detections
in the NASA Extragalactic Database (NED\footnote{\url{http://ned.ipac.caltech.edu/}}), redMaPPer 
\citep[v5.10 in SDSS, and v6.3 in DES;][]{Rykoff_2014, Rykoff_2016}, CAMIRA \citep{Oguri_2017}, ACT \citep{Hasselfield_2013}, 
and various X-ray cluster surveys \citep{Piffaretti_2011, Mehrtens_2012, Liu_2015, Pacaud_2016}. 
The positions of SZ clusters detected by \textit{Planck} are more uncertain, and so we use a
10$\arcmin$ matching radius when matching to \textit{Planck} catalogs \citep{Planck_XXIX_2013, Planck2015_XXVII}.

For many objects, spectroscopic redshifts are available from large public surveys. We cross matched the
ACTPol cluster candidate list with SDSS DR13 and the VIMOS Public Extragalactic Redshift Survey 
\citep[VIPERS Public Data Release 2;][]{Scodeggio_2016}. We assign a redshift to each candidate using an 
iterative procedure. We first measure the cluster redshift, from all galaxy redshifts found within 1.5$\arcmin$
of the SZ candidate position, using the biweight location estimator \citep{Beers_1990}, 
which is robust to outliers. We then iterate, performing a cut of $\pm 3000$\,km\,s$^{-1}$
around the redshift estimate before re-measuring the cluster redshift using the biweight location estimate
of the remaining galaxies that are located within 1\,Mpc projected distance. For candidates with redshifts 
available from NED only, we checked the literature to ensure that the redshift was indeed spectroscopic 
before adopting it. We assigned spectroscopic redshifts to 142 clusters from publicly available
data or the literature (103 from SDSS DR13, 1 from VIPERS PDR2, 38 from other literature sources) by this process.
We obtained an additional 5 spectroscopic redshifts for clusters using our own SALT observations
(Section~\ref{sec:SALTSpec}).


At this stage, we also identified the brightest cluster galaxy (BCG) in each cluster, using a combination of visual inspection 
and the $i$, $r-i$ color--magnitude diagram, where available. This was done using the best data available
for each object (e.g., SDSS, S82, or our own follow-up observations; Section~\ref{sec:APOSOAR} 
below). For one cluster, ACT-CL~J0220.9-0333 ($z = 1.03$; first discovered as RCS~J0220.9-0333; see 
\citealt{Jee_2011}), we could not identify the BCG. \textit{Hubble Space Telescope} observations of this 
cluster suggest that the BCG may be hidden behind a foreground spiral galaxy \citep{Lidman_2013}.

Fig.~\ref{fig:SDSSMontage} presents some example optical images of ACTPol clusters confirmed in SDSS using
the process described above. Table~\ref{tab:redshifts} lists the cluster redshifts, $\delta$ measurements, 
and adopted BCG positions.

\subsection{Validation Checks}
\label{sec:validation}

\begin{figure}
\includegraphics[width=\columnwidth]{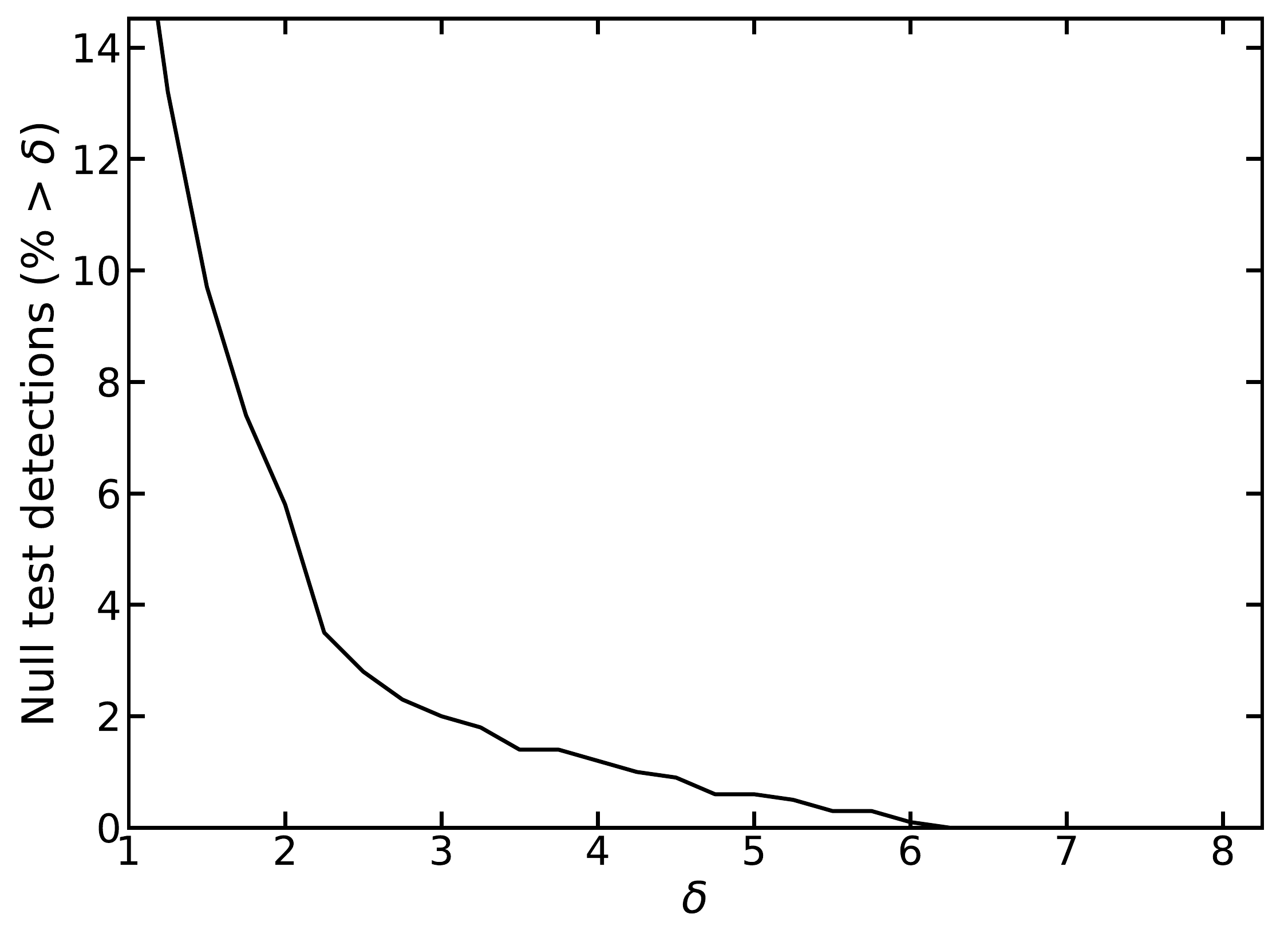}
\caption{The cumulative fraction of false detections (expressed as a percentage) at random positions in the 
SDSS zCluster null test (see Section~\ref{sec:validation}). For $\delta > 3$, this shows that the false 
detection rate is 2\%; this falls to 0.6\% per cent for $\delta > 5$.}
\label{fig:nullTest}
\end{figure}

We performed validation checks to test the performance of zCluster in both confirming clusters 
(using the $\delta$ statistic) and in photometric redshift accuracy.

\subsubsection{Null Test}

\begin{figure*}
\includegraphics[width=\textwidth]{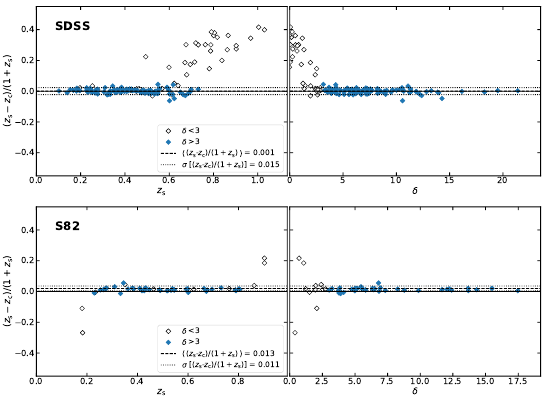}
\caption{Accuracy of photometric redshift recovery by zCluster, using SDSS (top) and S82 (bottom) data. 
Each data point represents a cluster in the E-D56 field with a spectroscopic redshift ($z_{\rm s}$).
The difference between the zCluster photometric redshift ($z_{\rm c}$) and the cluster spectroscopic redshift is plotted
on the vertical axis. Clusters with low density contrast ($\delta < 3$; equation~\ref{eq:delta}), as measured at the
photometric redshift, are shown as open diamonds. In the top panels, most of these objects are clusters with 
$z_{\rm s} > 0.5$, which is beyond the typical reach of SDSS photometry. As a result, their assigned photometric redshifts
are spurious, but are flagged by the $\delta < 3$ cut. For clusters with $\delta > 3$, $z_{\rm c}$ is unbiased when using
SDSS photometry, and has small scatter. However, as shown in the bottom panel,
the photometric redshifts are underestimated by $\Delta z / (1 + z) = 0.013$, when using S82 photometry.
}
\label{fig:zCluster_zpzs}
\end{figure*}

The $\delta$ statistic (Section~\ref{sec:zClusterAlgorithm}) measures the density contrast at a given 
(RA, Dec.) position, by comparison with a local background estimate. To be useful as an automated method of 
confirming SZ candidates as clusters, we would expect such a measurement to give a low value of $\delta$ at a 
position on the sky that is not associated with a galaxy cluster. Hence, we performed a null test, running the
zCluster algorithm on 1000 random positions in the SDSS DR12 survey region. Note that in building the catalog
of null test random positions, we rejected those that were located within 5$\arcmin$ of known clusters in NED
or the redMaPPer\ catalog. Fig.~\ref{fig:nullTest} shows the results. Interpreting the number of null test positions for which $\delta$
is greater than some chosen threshold as the false detection rate, 2\% of objects with
$\delta > 3$ are expected to be spurious. For $\delta > 5$, the false detection rate falls to 0.6\%,
and to zero for $\delta > 7$. Therefore, in the full list of 517 ACTPol cluster candidates with SNR~$ > 4$,
we would expect 11 of the objects with $\delta > 3$ to be spurious. Based on visual inspection, 
we find only 5 candidates that are not clusters, but have $\delta > 3$ as measured in SDSS photometry, 
in agreement with the null test.


\subsubsection{Photometric Redshift Accuracy}
\label{sec:photoz-test}
We used the 147 ACTPol clusters with spectroscopic redshifts to characterize the photometric redshift accuracy
of the zCluster algorithm. Fig.~\ref{fig:zCluster_zpzs} shows the comparison between $z_{\rm c}$, as measured using 
SDSS or S82 data, and spectroscopic redshift $z_{\rm s}$. Clusters with $\delta > 3$ are highlighted. 

Using SDSS photometry, we found that the zCluster redshift estimates are unbiased, with small scatter. The typical scatter 
$\sigma_z$ in the photometric redshift residuals $(z_{\rm s} - z_{\rm_c}) / (1 + z_{\rm s})$ is $\sigma_z = 0.015$, for
objects with $\delta > 3$. We adopt this $\sigma_z$ as the measurement of the redshift uncertainty for the 11 clusters
in the final catalog that are assigned zCluster SDSS redshifts, as no spectroscopic redshift is available for them (Section~\ref{sec:Sample}). 
As can be seen in Fig.~\ref{fig:zCluster_zpzs}, some clusters with $z_{\rm s} > 0.5$ (beyond the reach of SDSS) are assigned
erroneous redshifts by zCluster, but these are easily identified and rejected because they have low $\delta$ values.

We see similarly small scatter in the comparison of zCluster redshifts measured in S82 with the spectroscopic redshifts,
with $\sigma_z = 0.011$ for objects with $\delta > 3$ over the full redshift range. We adopt this as the redshift 
uncertainty for the 9 clusters assigned zCluster S82 redshifts in the final cluster catalog. However, as 
Fig.~\ref{fig:zCluster_zpzs} shows, on average the zCluster S82 photometric redshifts are underestimated 
by $\Delta z / (1 + z) = 0.013$. We therefore correct the redshifts recorded for these 9 clusters in the final catalog to account for this bias.

Using CFHTLenS photometry, we see no evidence that the zCluster redshifts are biased, although the comparison sample is small, with
only 5 objects with spectroscopic redshifts having $\delta > 3$. We adopt the measured scatter of $\sigma_z = 0.07$ as the photometric 
redshift error. Only one object in the final catalog is assigned a zCluster CFHTLenS redshift.

\section{Confirmation and Redshifts From Follow-up Observations}
\label{sec:OpticalPrivate}

Using large optical surveys, we obtained confirmation and redshifts for 170 clusters with SNR~$> 4$,
with the vast majority of these coming from SDSS. However, SDSS is only deep enough to confirm clusters up
to $z \approx 0.5$, and in principle the SZ selection of the ACTPol sample can detect clusters at any 
redshift. In this section we describe follow-up observations that we performed to confirm clusters at higher
redshift. These included optical/IR imaging with the Southern Astrophysical Research Telescope (SOAR) and
the Astrophysical Research Consortium 3.5\,m telescope at Apache Point Observatory (APO), and optical
spectroscopy using the Southern African Large Telescope (SALT).

\subsection{APO/SOAR Imaging and Photometric Redshifts}
\label{sec:APOSOAR}

\subsubsection{SOAR Observations}

We obtained $riz$ imaging of 24 cluster candidates located within the ACTPol E-D56 survey area using the SOAR telescope. 
The targets were selected from preliminary versions of the candidate list, and only 12 candidates remain in the final
list that we report in this paper, with 10/12 of these being confirmed as clusters (see below). 
The candidates have $4.3 <$~SNR~$< 7.3$ in the final list. Of the 12 targets from the preliminary lists
that were not subsequently detected with SNR~$>4$, three appear to be genuine high-redshift ($z \sim 1$) clusters on
the basis of their optical/IR imaging. We will report on these objects in a future publication, if they are detected
with higher SNR in Advanced ACTPol observations \citep{DeBernardis_2016}.

We used the SOAR Optical Imager \citep[SOI;][]{Walker_2003} for the 
first observing run, during 2015 October 31 -- 2015 November 2. Half of the time was lost due to bad weather, and
the seeing was poor on average (typically $>1.5\arcsec$), being at its best $1.0-1.3 \arcsec$ during 
2015 October 31. For the second run, which took place during 2017 January 5--9, we used the Goodman 
Spectrograph \citep{Clemens_2004} in imaging mode, using a new, red-sensitive detector with negligible fringing at
red wavelengths. During this second run the seeing was between $0.7-1.4 \arcsec$, with median $1.0\arcsec$,
and only the first night was adversely affected by non-photometric conditions. We spent roughly half of the time 
during the second observing run observing an additional 19 cluster candidates located in the ACTPol BOSS-N field; 
we will present the clusters discovered in these data in a future publication. 

We obtained images with total integration times of 750\,s, 1200\,s, 1800\,s in the $r$, 
$i$, and $z$ bands respectively for each candidate during both runs. These integration times were chosen to allow us to reach sufficient depth to 
detect clusters at $z = 1$ using the SOAR data alone. Each observation was broken down into a number of 
exposures, typically 6--12, the exact number depending upon the presence of any bright stars in a given 
field. We used a 3-step dither pattern that offset the telescope by 15$\arcsec$ during each observation, in 
order to cover the gap between the two CCDs in the SOI camera, and allow us to later construct fringe frames from
the $i$ and $z$-band data.

The data were reduced using \textsc{PyRAF/IRAF} routines,\footnote{IRAF is distributed by the National Optical Astronomy Observatories,which are operated 
by the Association of Universities for Research in Astronomy, Inc., under cooperative agreement with the 
National Science Foundation.} in particular making use of the
\textsc{mscred} package \citep{Valdes_1998}. The data were bias subtracted and initially flat-fielded 
using dome flats. After this initial processing, we constructed object masks for every image. These were used 
in the creation of fringe frames for the $i$ and $z$-band science observations, which were applied to the
$i$ and $z$-band science frames taken with the SOI instrument. We found that no fringing correction was necessary
for the images taken with the Goodman Spectrograph. The object masks were then used in the creation of sky flats in each band, which 
were applied to the appropriate science frames. We performed astrometric calibration with the 
\textsc{SCAMP} software \citep{Bertin_2006}, using SDSS DR9 as the astrometric reference catalog, and stacked the 
images for each candidate in each band using \textsc{SWARP} \citep{Bertin_2002}.

\begin{figure*}
\begin{center}
\includegraphics[width=\textwidth]{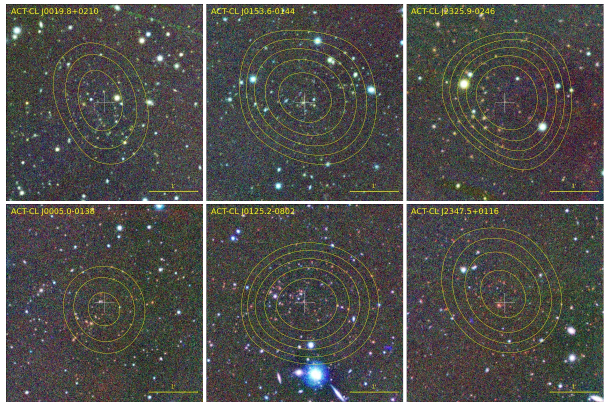}
\end{center}
\caption{Images of newly discovered $z > 0.7$ clusters, confirmed with imaging from the ARC\,3.5m and SOAR telescopes.
Each image is 4$\arcmin$ on a side, with North at the top and East at the left. The top row shows SOAR $riz$ images,
while the bottom row shows SOAR+ARC 3.5\,m $riK_s$ images, with the $K_s$-band channel coming from the latter.
The yellow contours (minimum 3$\sigma$, increasing in steps of 0.5$\sigma$) indicate the (smoothed) 148\,GHz decrement in the matched-filtered
ACT map. The white cross indicates the ACT SZ cluster position.}
\label{fig:APOSOARMontage}
\end{figure*}

The photometric zero point for each stacked image was bootstrapped from the magnitudes of SDSS stars detected
with SNR~$> 5$ in SDSS. There were 2--63 such stars in each field, with a median number of 
26 stars per field. The uncertainties in the zero points across all bands cover the range 0.001--0.017\,mag, 
with median uncertainty 0.004, 0.003, 0.004\,mag in the $r$, $i$, $z$ zero points respectively. 
The final depths of the stacked images were estimated in each band by placing 1000 3$\arcsec$ diameter 
apertures in each image at random positions where objects were not detected. We found that the images reach 
median $5\sigma$ depths of 23.0, 22.9, and 22.3 mag in the $r$, $i$, and $z$-bands respectively.

\subsubsection{ARC 3.5\,m Observations}

\begin{figure*}
\includegraphics[width=\textwidth]{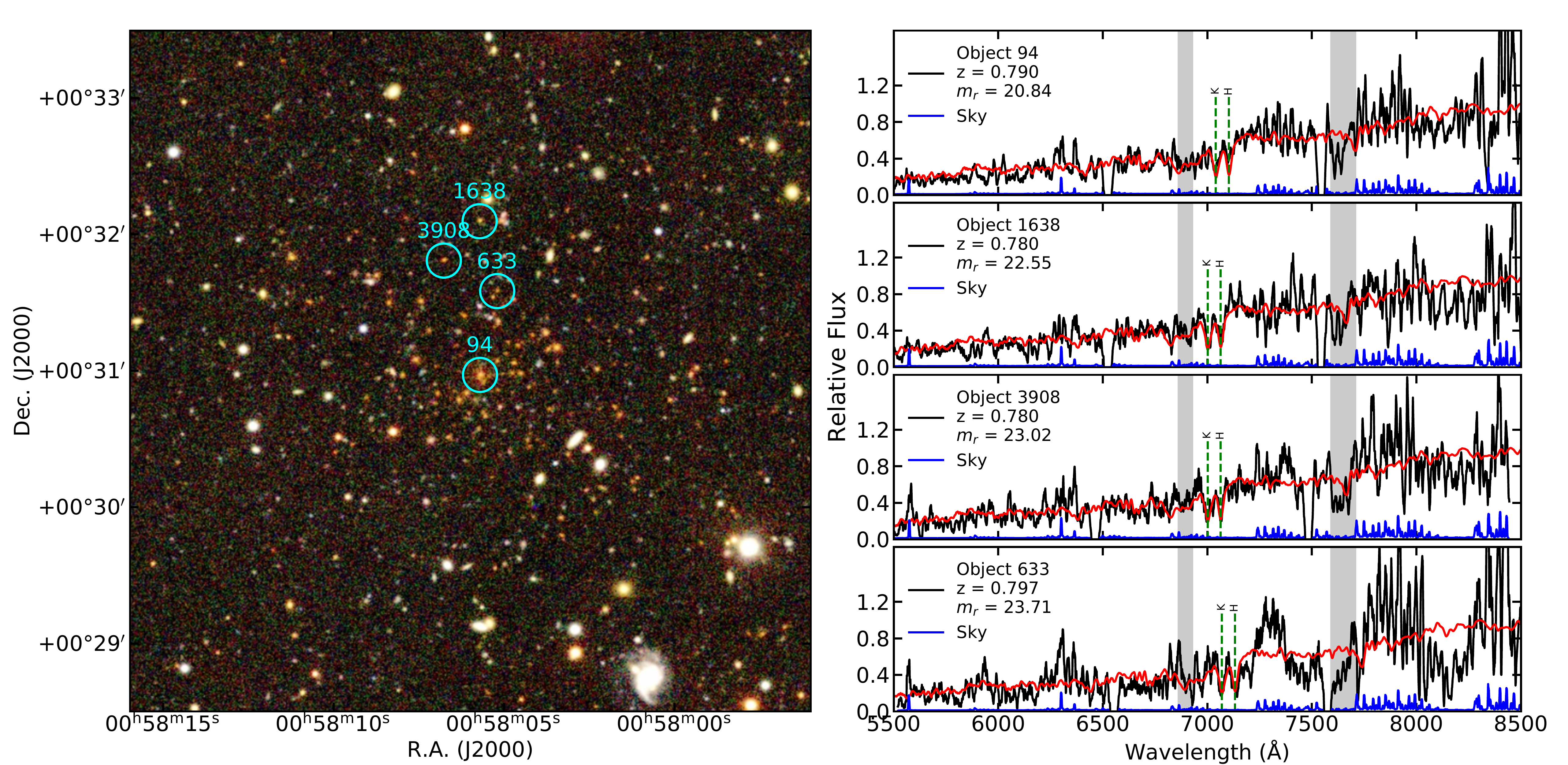}
\caption{The $z = 0.79$ cluster ACT-CL J$0058.1+0031$.  Secure spectroscopic redshifts have been obtained for 7 member 
galaxies in this cluster. The left hand panel shows a $5 \arcmin \times 5 \arcmin$
false color S82 optical image ($g$, $r$, $i$). SALT spectra for the four galaxies highlighted by the cyan circles are shown in the right hand panel.
Here, the black lines are the SALT RSS spectra (smoothed with a 15 pixel boxcar), while red lines show the best 
match redshifted SDSS spectral template in each case. The blue line is the sky spectrum, and the gray bands
indicate regions strongly affected by absorption features in the atmosphere.}
\label{fig:specImagePlot}
\end{figure*}

We observed 7 candidates in the $K_s$-band with the Near-Infrared Camera and Fabry-Perot 
Spectrometer (NICFPS) at the ARC 3.5\,m telescope on 2015 October 2 (0.8$\arcsec$ seeing) and 2015 November 23
(1.3$\arcsec$ seeing). To enable good sky subtraction, we used a cycling 5-point dither pattern, offsetting 
the telescope by 20$\arcsec$ after every 1-2 exposures. Each exposure was 20\,s in length, with eight Fowler 
samples per exposure. We obtained total integration times of 1760--2120\,s on each candidate. 

The data were reduced as described in \citet{Menanteau_2013}. Each science frame was dark subtracted, 
distortion corrected, flat fielded (using a sky flat constructed from the science frames after masking out 
detected objects), and then sky subtracted (using a running median method). Each individual frame was
astrometrically calibrated using \textsc{SCAMP}, using 2MASS \citep{Skrutskie_2006} as the reference 
catalog, before stacking using \textsc{SWARP}. 

Photometric calibration for all but one field was performed 
by bootstrapping the zero point from comparison with stars identified in Data Release 3 of the VISTA 
Hemisphere Survey \citep[VHS;][]{McMahon_2013}. In the case of ACT-CL~J0125.3-0802, we used 2MASS instead.
The zero points were converted to AB magnitudes using $K_s {\rm(AB)} = K_s {\rm(Vega)} + 1.86$ 
\citep{TokunagaVacca_2005}. The median zero point uncertainty is 0.008\,mag, and the range of zero point
uncertainties is 0.004--0.014\,mag. Each field contained 6--24 stars (median 14) that were used for the zero 
point determination. The final depths of the stacked images were estimated to be 21--21.5\,mag (5$\sigma$, AB),
by placing 1000 3$\arcsec$ diameter apertures in each image at random positions where objects were not detected.


\subsubsection{Photometric Redshifts from APO/SOAR Observations}

We performed matched aperture photometry on all available $rizK_s$ imaging using \textsc{SExtractor v2.19.5} 
\citep{BertinArnouts_1996}. We used \textsc{SWARP} to first rebin all images for a given field onto a common 
coordinate grid, so that the images are aligned at the pixel level. We used \textsc{SExtractor} in dual-image
mode, using the reddest available band ($z$ or $K_s$) as the detection image. We adopt \textsc{MAG\_AUTO}
as the magnitude measurement that we use in computing photometric redshifts, after first
correcting for Galactic extinction using the maps and software of \citet{Schlegel_1998}. 

We estimated photometric redshifts by applying the zCluster algorithm described in 
Section~\ref{sec:zClusterAlgorithm}. Given the small field of view for both the APO and SOAR imaging,
we were not able to define a background galaxy sample within an annulus for the measurement of
$\delta$ (equation~\ref{eq:delta}). Instead, we created a separate background galaxy sample from observations of 8 candidates
that were found not to contain clusters. The total area covered by this background galaxy
sample is $0.238$\,deg$^{2}$. We visually inspected the APO/SOAR images, and confirmed the presence of high-redshift clusters
for 10/12 candidates, with 9/10 of these having $\delta > 2.5$, and the remaining cluster being spectroscopically
confirmed with SALT (Section~\ref{sec:SALTSpec}). Fig.~\ref{fig:APOSOARMontage} shows some examples. 
These objects have photometric redshifts in the range 
0.70--1.12 (median $z_{\rm c} = 0.94$). We have obtained spectroscopic redshifts for only three of these clusters,
and find that they are all within $|z_{\rm s}-z_{\rm_c}| < 0.05$ of the photometric
redshift estimates. We adopt this as the photometric redshift uncertainty.

\subsection{SALT Spectroscopic Redshifts}
\label{sec:SALTSpec}

We obtained spectroscopic redshifts for five clusters with the Southern African Large Telescope (SALT), using the 
Robert Stobie Spectrograph (RSS) in its multi-object spectroscopy (MOS) mode. The observations were obtained
in programmes 2015-2-MLT-003 and 2016-1-MLT-008. The design of 
SALT limits the maximum observing time for our targets to blocks of less than one hour duration, and so targets were 
visited several times during each observing semester to build up the integration time, taking advantage of queue 
scheduling. The total integration times varied between 1950--5850\,s, depending on the number of blocks observed.  
The observations were conducted in dark time, with a maximum seeing constraint of 2$\arcsec$. For all observations, 
we used the PG0900 grating with the PC04600 order blocking filter, and $2\times2$ binning of the RSS detectors, 
giving a dispersion of 0.96\,\AA{} per binned pixel. 

The MOS mode of SALT uses custom-designed slit masks. Target galaxies were selected using color--magnitude cuts
applied to photometric catalogs, either from public surveys (S82, CFHTLenS), or from our own APO/SOAR observations
(Section~\ref{sec:APOSOAR}). In every cluster, the BCG was selected, with
remaining slits being placed on galaxies fainter than the BCG and with $r-i > 1.0$, using the same automated
algorithm for target selection as in \citet{Kirk_2015}. Each slit was 1.5$\arcsec$ wide and $10\arcsec$ long.
We observed 17--26 target galaxies per slit mask, observing one slit mask per target.

The data were reduced using a pipeline that operates on the basic data products delivered from 
SALT. The initial processing is carried out using the \textsc{PySALT} package \citep{Crawford_2010}, which 
prepares the image headers, applies CCD amplifier gain and crosstalk corrections, and performs bias subtraction.
The \textsc{PySALT} data products are then passed into a fully automated pipeline\footnote{\url{https://github.com/mattyowl/RSSMOSPipeline}} 
that performs flat field corrections, wavelength calibration, and extraction and stacking of one dimensional spectra.

Redshifts were measured using the \textsc{XCSAO} task of the \textsc{RVSAO} \textsc{IRAF} package \citep{KurtzMink_1998}, and verified 
by visual inspection. We consider redshifts measured from spectra in which two or more 
strongly detected features were identified (for example, the H and K lines due to Ca\textsc{ii}) to be secure.
We successfully measured secure redshifts for 2--7 member galaxies, including the BCG, in each cluster. We 
adopt the biweight location of the member redshifts as the final spectroscopic redshift for each cluster 
(listed in Table~\ref{tab:redshifts} in the appendix). Fig.~\ref{fig:specImagePlot} shows some examples of SALT spectra for 
members identified in one of the observed clusters.

\section{The E-D56 Field Cluster Catalog}
\label{sec:Sample}

Tables~\ref{tab:detections}--\ref{tab:masses} in the Appendix present the ACTPol two-season cluster catalog in the 
987.5\,deg$^{2}$ E-D56 field. The catalog consists of the 182 clusters detected with SNR~$> 4$ that have been
optically confirmed and have a redshift measurement at the time of writing. A cluster is considered to be confirmed based
on visual inspection of all available optical/IR imaging, the availability of a spectroscopic redshift measurement, 
and/or a match to another cluster catalog, as described in Sections~\ref{sec:OpticalPublic}
and \ref{sec:OpticalPrivate}. Table~\ref{tab:zSources} 
provides a breakdown of the redshift sources used and the number of clusters with redshifts drawn from each source. Where possible, 
spectroscopic redshifts are preferred, followed by zCluster photometric redshifts as measured in this work, and then other literature sources
of photometric redshifts.


\begin{deluxetable}{rcl}
\small
\tablecaption{Number of clusters by redshift source in the E-D56 cluster catalog.
\label{tab:zSources}}
\tablewidth{0pt}
\decimals
\tablehead{
\colhead{Source}       &
\colhead{Number} &
\colhead{Reference} \\
}
\startdata
Lit. (spec)     & 11  & See Table~\ref{tab:redshifts}\\
SALT (spec)     & 5   & This work\\
SDSS (spec)     & 103 & This work$^*$\\
S16 (spec)      & 27  & \citet{Sifon_2016}\\
VIPERS (spec)   & 1   & \citet{Scodeggio_2016}\\
CAMIRA (phot)   & 2   & \citet{Oguri_2017}\\
M13 (phot)      & 6   & \citet{Menanteau_2013}\\
zCluster (phot) & 27  & This work\\
\enddata
\tablecomments{\,$^*$Based on DR13 \citep{Albareti_2016}.}
\end{deluxetable}

Table~\ref{tab:detections} lists the positions of the detected clusters, their SNR values, and our chosen SZ
observable, the central Compton parameter $\tilde{y}_0$ extracted at the $2.4\arcmin$ filter scale. 
We also note ACTPol clusters that are cross-matched against clusters detected in other catalogs, specifically
highlighting those reported previously by ACT (in \citetalias{Hasselfield_2013}), \textit{Planck} \citep[PSZ2;][]{Planck2015_XXVII} and redMaPPer
\citep[v5.10;][]{Rykoff_2014}, as well as listing the nearest cluster counterpart found in NED.

The E-D56 sample contains
53/68 clusters reported by ACT in \citetalias{Hasselfield_2013}. We list the 15 \citetalias{Hasselfield_2013} clusters that are 
not detected with SNR~$> 4$ in this work in Table~\ref{tab:H13Missed}. We note
that all of these clusters are optically confirmed and are thus `real'. However, the SZ cluster detection pipeline used in this study 
differs enough from that used in \citetalias{Hasselfield_2013} that they do not all appear with SNR~$> 4$. 
Of the missing 15 
\citetalias{Hasselfield_2013} clusters, 4 (ACT-CL\,J0308.1+0103, 
ACT-CL\,J2025.2+0030, ACT-CL\,J2051.1+0215, and ACT-CL\,J2135.1$-$0102) are not in the E-D56 survey footprint, with
3/4 of these being masked due to nearby point sources. With the exception of these 4 objects, all \citetalias{Hasselfield_2013}
clusters with SNR~$>5$ are recovered. We recover 9/11 of the missing 
\citetalias{Hasselfield_2013} clusters by decreasing the SNR threshold used for candidate selection in the 
E-D56 field from SNR~$>4$ to SNR~$> 3$. Most of these objects (7/11) are located in regions covered only by ACT observations, and therefore the 
reason they are not detected with SNR~$> 4$ in the E-D56 map is ascribed to differences between the cluster detection pipelines used in 
\citetalias{Hasselfield_2013} and this work (see Section~\ref{sec:SZDetection}). We checked for pipeline-versus-pipeline
differences by considering the regions of the E-D56 map that contain only ACT data, and comparing the SNR values reported in H13 
with those measured using the method described in this work. From the 24 clusters that fall in such regions, 
the median SNR measured by the pipeline used in this work is 5\% lower than \citetalias{Hasselfield_2013}, 
with $\approx 10$\% scatter around this value. The lower SNR measured in the E-D56 map may be a result of the different noise estimation method, 
or indicates that the filtering scheme used here is slightly less effective than the Fourier-space matched filter used in \citetalias{Hasselfield_2013}. 
We verified that the SZ masses of the clusters listed in Table~\ref{tab:H13Missed} measured by the pipeline used in this work are consistent
(well within $< 1\sigma$) with the UPP masses reported in \citetalias{Hasselfield_2013} for these objects.

\begin{figure}
\includegraphics[width=\columnwidth]{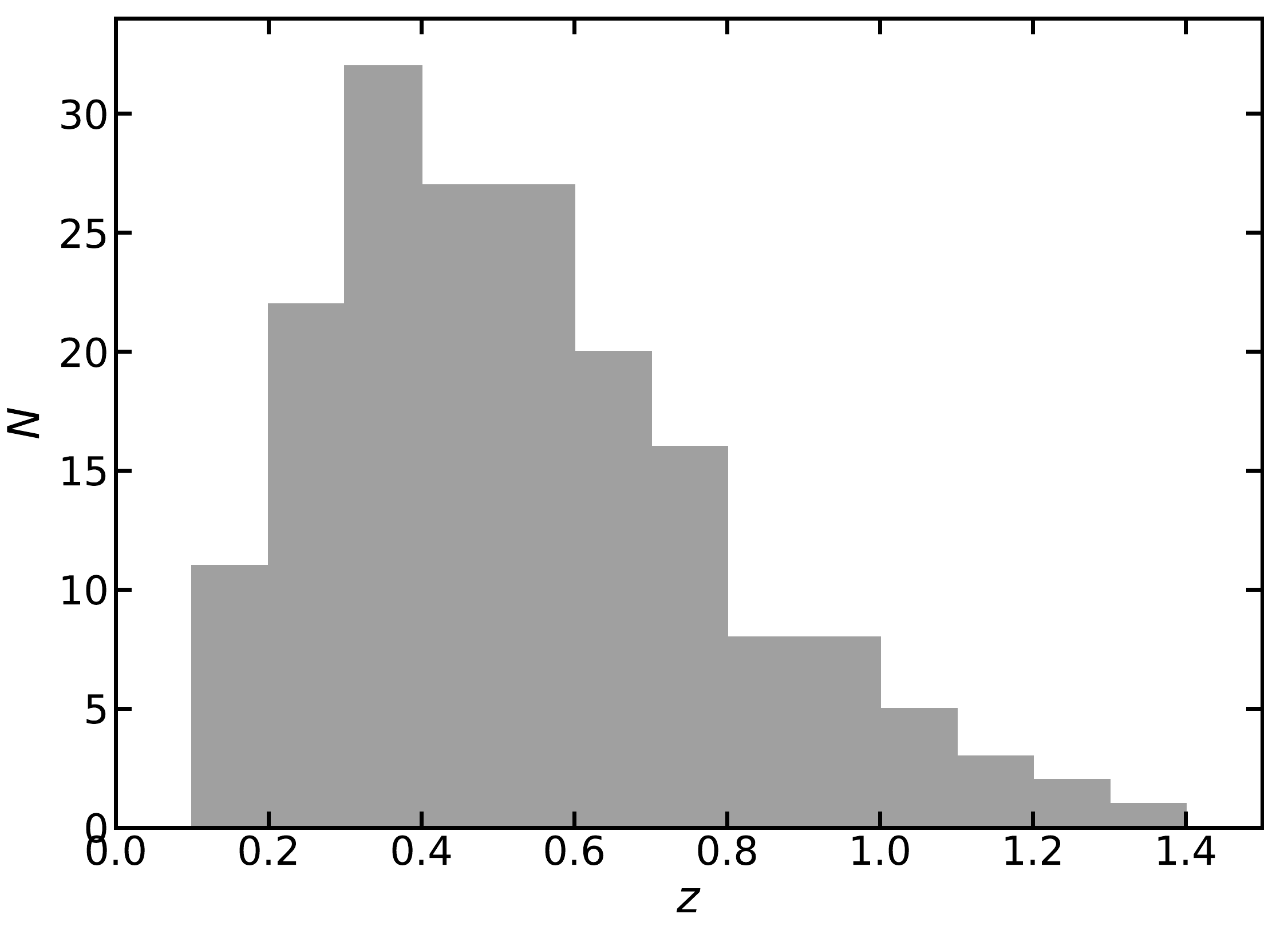}
\caption{The redshift distribution of the 182 clusters in E-D56 cluster catalog. The median redshift is 0.49. The lack of
clusters at low redshift ($z < 0.2$) is largely a selection effect, due to the angular size of such clusters being similar to 
CMB anisotropies (see Section~\ref{sec:selFn}).}
\label{fig:zHist}
\end{figure}

\begin{figure}
\includegraphics[width=\columnwidth]{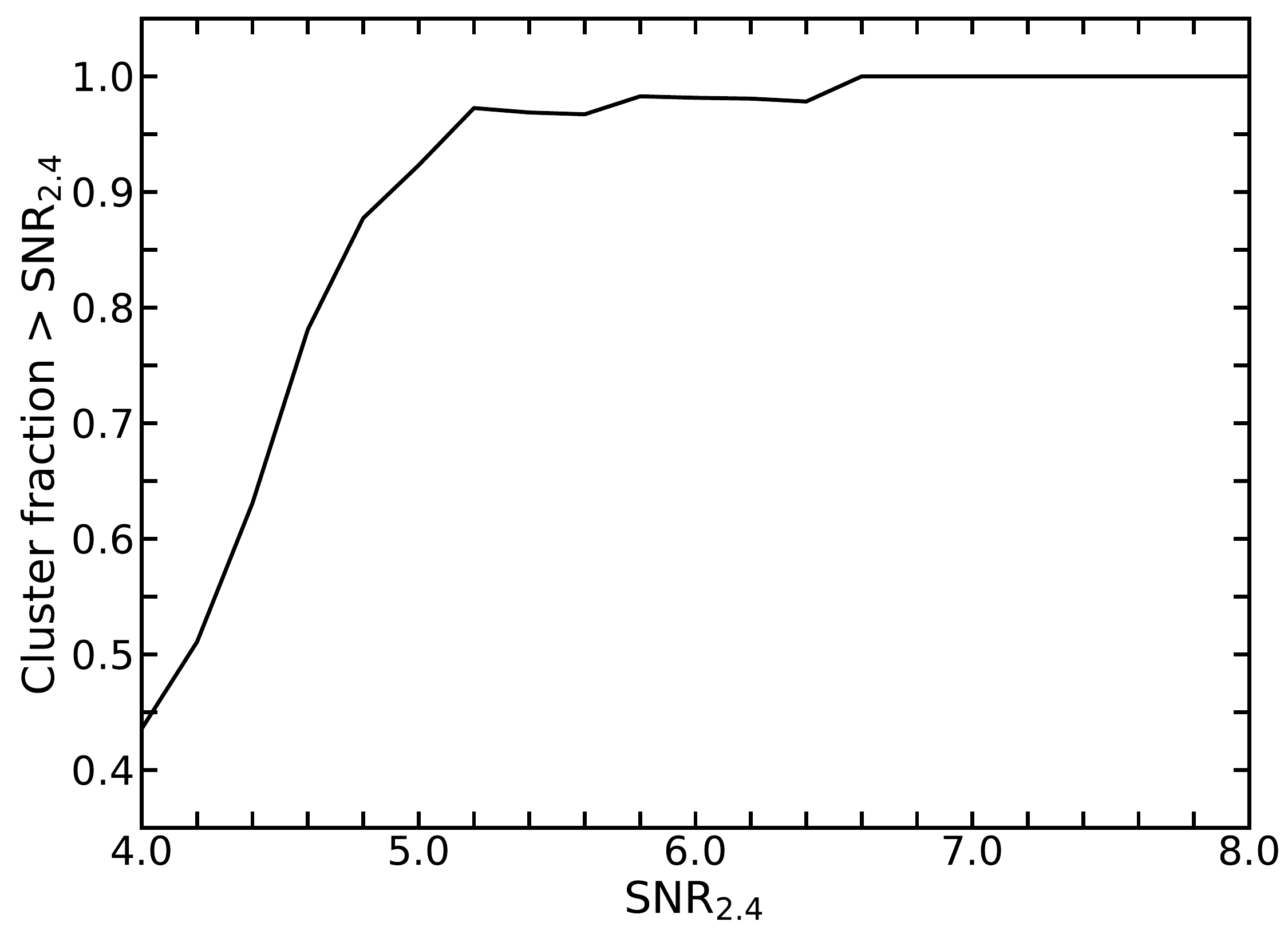}
\caption{The cumulative fraction of candidates that are confirmed clusters as a function of SNR$_{2.4}$. 
For SNR$_{2.4} > 5$, the fraction is less than 1 because of incomplete redshift follow-up; there is evidence 
from e.g., WISE imaging that these candidates are likely to be high-redshift ($z > 1$) clusters. At SNR$_{2.4} < 5$, 
the dominant effect is sample impurity (see Fig.~\ref{fig:contamination}).}
\label{fig:purity}
\end{figure}

We detect 30/45 of the subset of PSZ2 candidates that fall within the E-D56 survey footprint. Of the 15 missed PSZ2 
candidates, 6/15 have not been optically confirmed, and so may be spurious. These are listed in Table~\ref{tab:PSZ2NotConfirmed}. 
The other 9 objects are confirmed clusters, with median $z = 0.09$, and 7/9 of these objects are located at $z < 0.2$.
It is not surprising that these larger angular size clusters are not detected by ACTPol, due to the lack of multi-frequency data 
and the resulting confusion with CMB anisotropies (Section~\ref{sec:selFn} and Fig.~\ref{fig:completenessByRedshift}).
However, two clusters with $0.2 < z < 0.3$ (PSZ2\,G083.85-55.43 and PSZ2\,G052.35-31.98) are also not detected by ACTPol.
We discuss the comparison with PSZ2 further in Section~\ref{sec:SZComparison} below.

Objects that were not detected in PSZ2 or previously with ACT in \citetalias{Hasselfield_2013}, but were detected in previous optical or X-ray surveys,
are new SZ detections. These make up 113/182 clusters in the E-D56 sample. 

\begin{figure}
\begin{centering}
\includegraphics[width=\columnwidth]{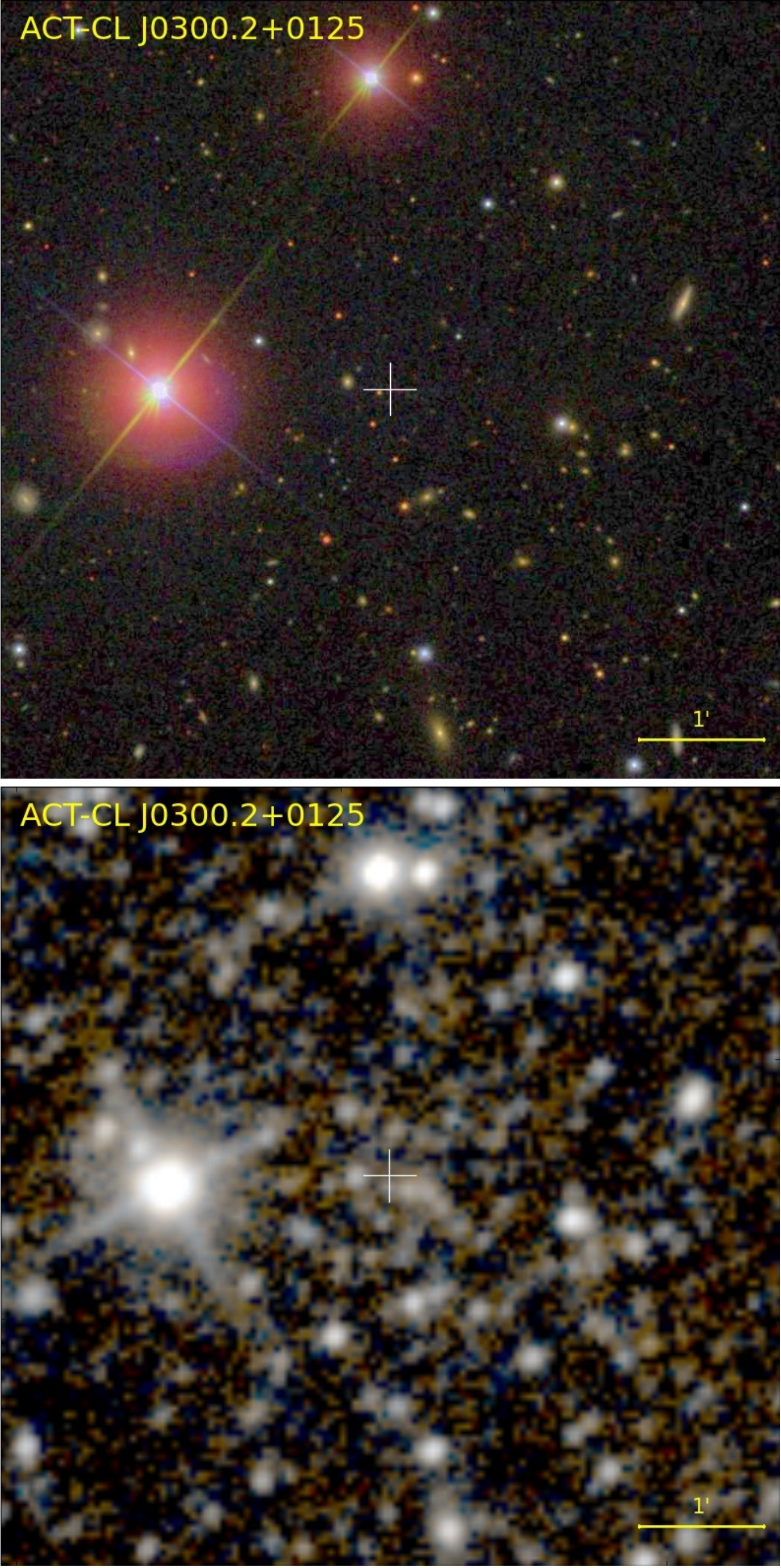}
\caption{SDSS ($gri$; top) and WISE (W1/W2; bottom) imaging of ACT-CL\,J0300.2+0125, the candidate detected with the
highest SNR (6.6) that does not yet have a redshift measurement. Each image is 6$\arcmin$ on a side, with North at
the top and East at the left. IR-bright but optically faint galaxies, with IR-colors consistent with those expected 
for early-type galaxies at $z > 1$, are clearly visible close to the position of the SZ detection, which is marked with
the white cross. The false color WISE image is taken from the unWISE project \citep{Lang_2014}.}
\label{fig:J0300}
\end{centering}
\end{figure}

Newly discovered clusters make up roughly 15\% of the catalog (28/182 clusters). These
are mostly at high redshift, with median $z = 0.80$, since the vast majority of clusters at $z < 0.5$ have previously been discovered
in optical surveys based on SDSS \citep{Goto_2002, Koester_2007, WHL_2009, Hao_2010, Geach_2011, Szabo_2011, WHL_2012, Oguri_2014, WH_2015}.
For example, 99/182 of the ACTPol clusters in the E-D56 field are also found in the redMaPPer catalog \citep{Rykoff_2014}, which
is based on SDSS legacy survey data.
Nevertheless, we do find 10 ACTPol clusters (median $z = 0.80$) using only SDSS/S82 data that have not been found in these previous surveys.

We find that 16/182 clusters have matches with the CAMIRA catalog \citep{Oguri_2017}, although the overlap of the E-D56 map with
the HSC survey is currently only a few tens of degrees. The detected CAMIRA clusters cover a wide redshift range ($0.14 < z < 1.04$), and
the HSC observations of these objects will be used for future studies of the weak-lensing mass calibration.

Table~\ref{tab:redshifts} lists the redshifts and the BCG coordinates for each cluster in the E-D56 catalog. As noted earlier, 
80\% of the clusters in the sample have spectroscopic redshifts (147/182), largely due to the overlap with SDSS DR13. 
Fig.~\ref{fig:zHist} presents the redshift distribution of the sample, which covers the range $0.1 < z < 1.4$ 
(median $z = 0.49$). 

Fig.~\ref{fig:purity} shows the fraction of confirmed clusters as a function of SNR$_{2.4}$. This plot reflects the combined
effects of the purity of the sample, and the completeness of the redshift follow-up. The redshift follow-up is complete for all 
candidates detected with SNR$_{2.4} > 6.6$, with all 41 objects above this cut being confirmed as clusters. 
For SNR$_{2.4} > 5.7$, only one candidate is detected that currently
does not have a redshift: ACT-CL\,J0300.2+0125, which is shown in Fig.~\ref{fig:J0300}. This object appears to be a 
$z \approx 1$ cluster, based on WISE imaging and the infrared colors of galaxies near the SZ candidate position. There are
only 7/91 candidates in total with SNR$_{2.4} > 5$ that currently lack a redshift. We are in the process of following-up
a few other similar cases to ACT-CL\,J0300.2+0125, but we note that we expect roughly this number of candidates to be false
positives, based on running the cluster detection algorithm over simulated signal-free maps (Section~\ref{sec:SZDetection} and 
Fig.~\ref{fig:contamination}). At SNR$_{2.4} < 5$, the dominant effect contributing to the decreasing cluster fraction 
is contamination. The cluster fraction here is just under half that implied by Fig.~\ref{fig:contamination},
but we expect a number of these candidates will also be high-redshift clusters, so Fig.~\ref{fig:purity} represents a 
lower limit on the purity of the sample.

\begin{figure}
\includegraphics[width=\columnwidth]{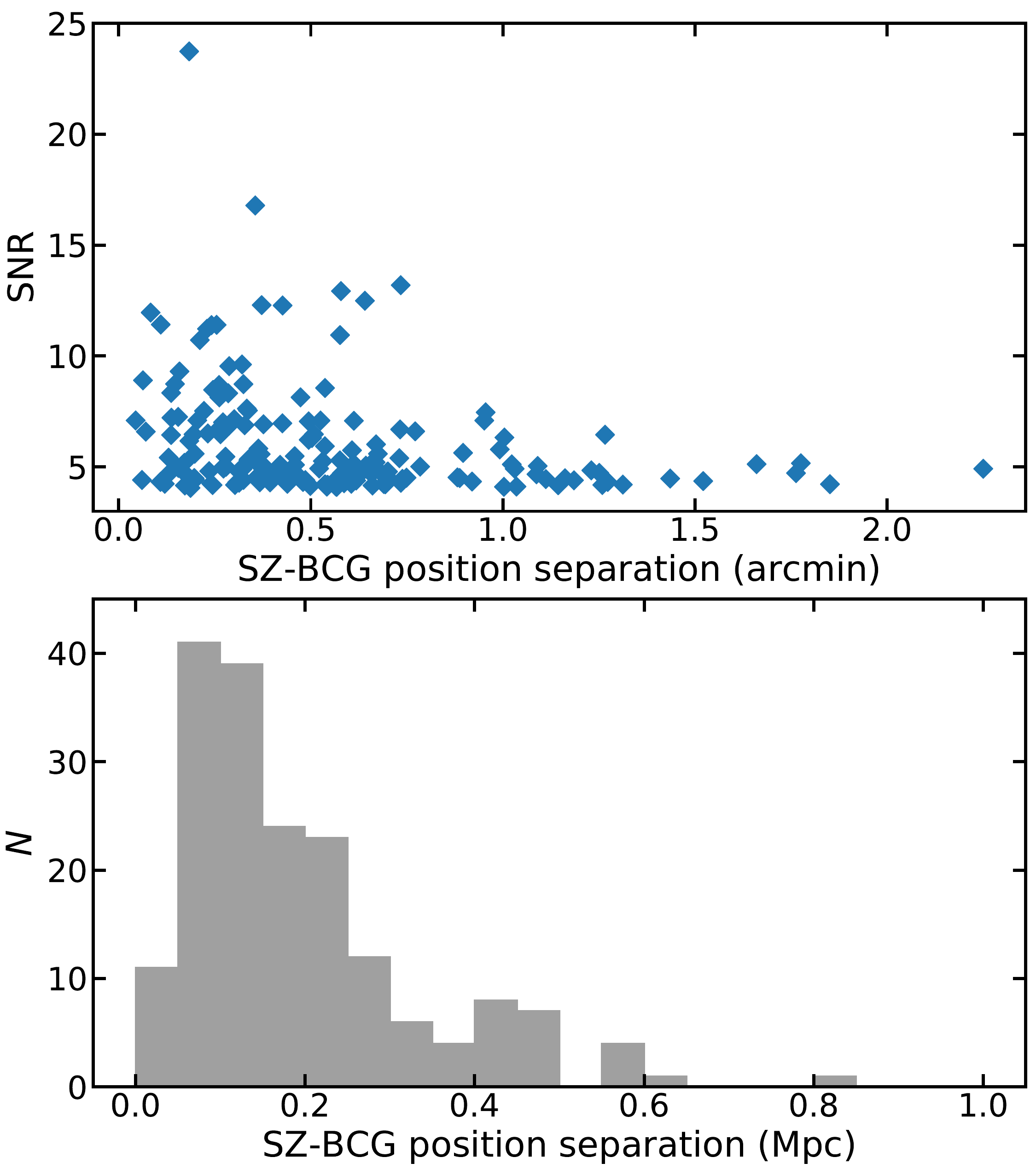}
\caption{The separation between BCG positions and the position at which each cluster was detected via the SZ. The top panel
shows this in terms of arcminutes as a function of SNR, while the bottom panel shows the distribution in terms
of projected radial distance. The typical offset is $<150$\,kpc.}
\label{fig:BCGOffsets}
\end{figure}

Fig.~\ref{fig:BCGOffsets} presents a comparison of the offset between the SZ cluster candidate position and the BCG.
The median offset for the whole sample is $0.46\arcmin$, which is equivalent to 
$\approx 1$\,pixel in the 148\,GHz maps. The top panel of Fig.~\ref{fig:BCGOffsets} shows that the typical size of the 
offset varies with SNR, with the highest SNR detections having smaller offsets. In terms of projected radial distance
from the SZ cluster position, the median offset is 148\,kpc.

\begin{figure}
\includegraphics[width=\columnwidth]{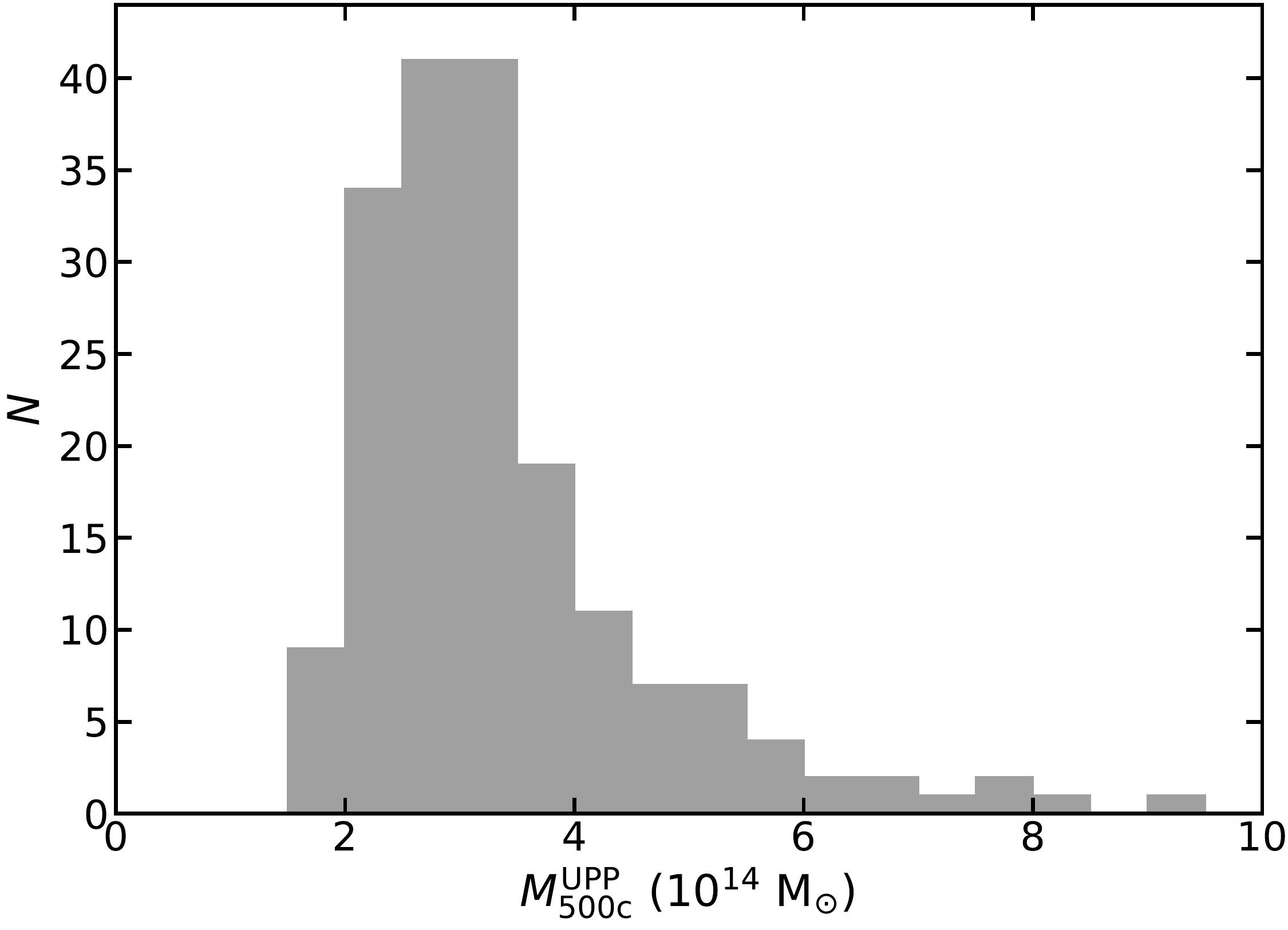}
\caption{The mass distribution of the 182 clusters in E-D56 cluster catalog (median $M^{\rm UPP}_{\rm 500c} = 3.1 \times 10^{14}$\,M$_{\sun}$), 
estimated from the central Compton parameter $\tilde{y}_0$ measured at the 2.4$\arcmin$ filter scale, assuming the \citetalias{Arnaud_2010} scaling 
relation.}
\label{fig:M500Hist}
\end{figure}

Table~\ref{tab:masses} lists the SZ-derived masses for clusters in the E-D56 sample, following the methods described in 
Section~\ref{sec:SZMass}. Fig.~\ref{fig:M500Hist} shows
the mass distribution, which spans the range $1.7 < M^{\rm UPP}_{\rm 500c} / (10^{14}$\,M$_{\sun}) < 9$, with 
median $M^{\rm UPP}_{\rm 500c} = 3.1 \times 10^{14}$\,M$_{\sun}$. We discuss the ACTPol mass distribution in the context of other SZ surveys
in Section~\ref{sec:SZComparison} below. For comparison with other studies (e.g., Section~\ref{sec:WLComparison}), 
in Table~\ref{tab:masses} we also list masses measured within a radius that encloses 200 times the mean density at each cluster
redshift ($M_{\rm 200m}$). These are converted from the $M_{\rm 500c}$ values by assuming
the concentration--mass relation of \citet{Bhattacharya_2013} and following the methodology of \citet{HuKravtsov_2003}.

\section{Discussion}
\label{sec:Discussion}

\subsection{Mass Calibration and Comparison with Weak-lensing Studies}
\label{sec:WLComparison}

Throughout this work we have modeled the SZ signal using the UPP, and have related this to mass using the \citetalias{Arnaud_2010} scaling
relation (Section~\ref{sec:SZMass}).
However, several works have noted that this mass--scaling relation typically underestimates cluster masses inferred from weak-lensing measurements by 
$\approx 30$\% \citep[e.g.,][]{vonDerLinden_2014, Hoekstra_2015, Planck2015_XXIV, Penna-Lima_2017}, while other studies, based on either weak-lensing measurements
or dynamical mass estimates, have not found evidence of a significant bias, although the uncertainties are quite
large \citep[$\approx 10-30$\%;][]{Battaglia_2016, Sifon_2016, Rines_2016}. It is possible that the bias depends on the dynamical states of clusters 
(e.g., the fraction of cool-core versus non-cool-core clusters in a sample; \citealt{Andrade-Santos_2017}) or is redshift dependent; for an analysis
restricted to $z < 0.3$, \citet{Smith_2016} found no evidence for a bias, at the 5\% level, between weak-lensing masses and 
\textit{Planck} SZ masses.

The ratio of SZ mass to weak-lensing mass, i.e., the mass bias $\langle M^{\rm SZ}_{\rm 500c} \rangle / \langle M^{\rm WL}_{\rm 500c} \rangle$, 
is often parametrized as (1-$b$), where $b$ is the fraction by which the `true' mass (typically taken as corresponding with the weak-lensing mass) is underestimated \citep{Planck_XX_2013}. 
Hydrodynamical simulations have shown that X-ray analyses, which assume hydrostatic equilibrium and on which the \citetalias{Arnaud_2010} scaling relation is based, 
underestimate the `true' mass in the simulations by $\approx 10-20$\% \citep[e.g,][]{Biffi_2016, Henson_2017}, and so if this were the only source of bias,
$b = 0.1-0.2$ would be expected. Instrument calibration issues affecting X-ray telescopes \citep[e.g.,][]{Mahdavi_2013, Israel_2015, Madsen_2017} are another potential source of bias. Given the location of the E-D56 field on the sky and its
large size, there are a number of published weak-lensing masses and weak-lensing calibrated cluster mass measurements with which we can compare our 
UPP/\citetalias{Arnaud_2010}-scaling-relation-based SZ masses. Here, we compare against the CoMaLit \citep{Sereno_2015} public compilation of weak-lensing mass measurements, and the
\citet{Simet_2017} optical richness ($\lambda$)--mass relation, which was measured via a stacked weak-lensing analysis of redMaPPer \citep{Rykoff_2014}
clusters detected in the SDSS.

\begin{figure}
\includegraphics[width=\columnwidth]{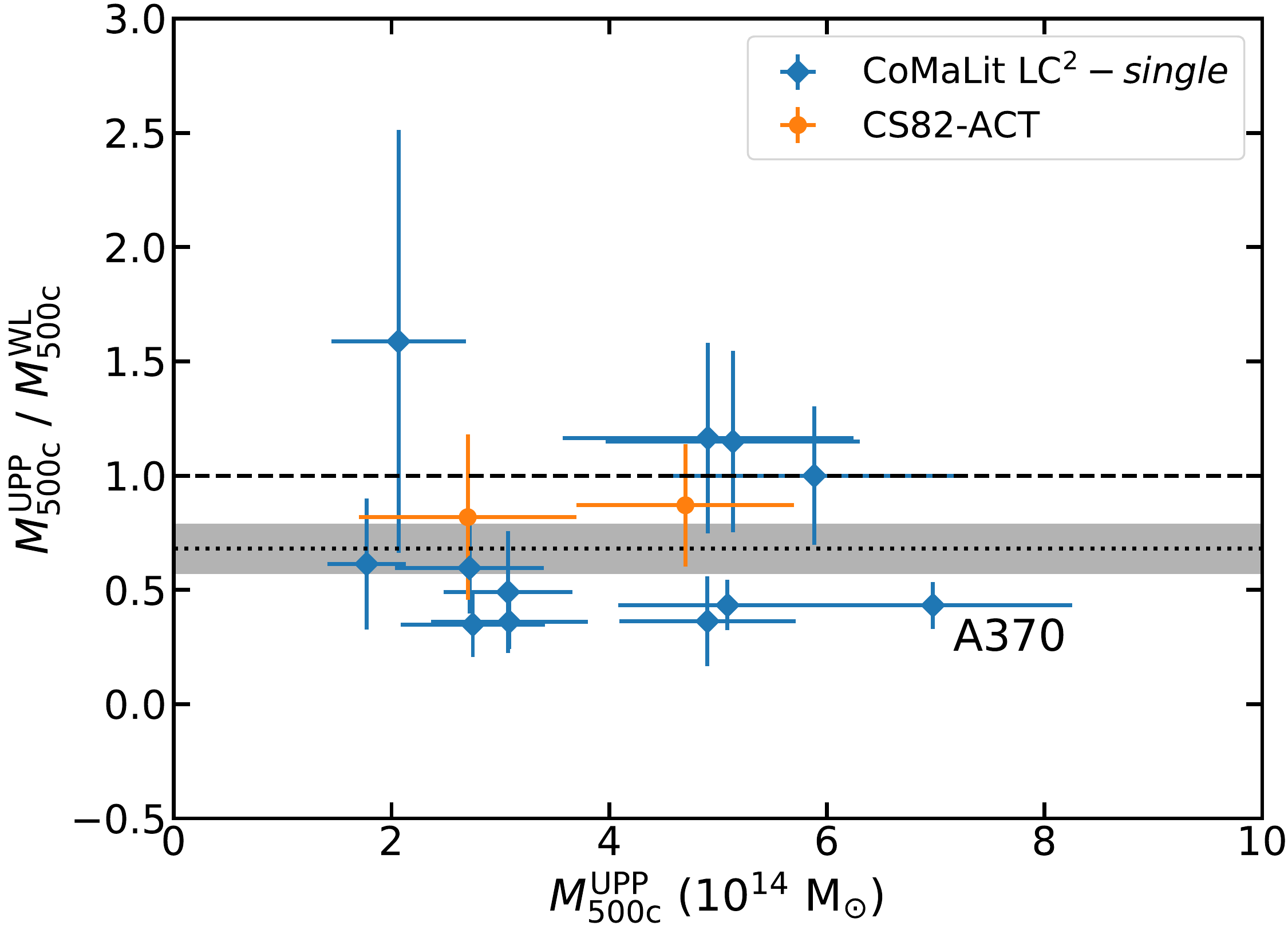}
\caption{Comparison of weak-lensing masses from the CoMaLit database \citep[blue]{Sereno_2015} and the stacked weak-lensing analysis
of \citet[CS82-ACT;][orange]{Battaglia_2016} with ACTPol SZ masses based on the UPP and \citetalias{Arnaud_2010} mass--scaling relation. The CS82-ACT masses 
plotted here are from NFW profile fits to the stacked weak-lensing signal. Here we used the LC$^2$--\textit{single} catalog from CoMaLit, which consists of 
objects modeled using a single halo. The dotted line and shaded area indicates the richness-based weak-lensing mass calibration factor 
and its uncertainty ($0.68 \pm 0.11$), obtained independently from these data by applying the \citet{Simet_2017} scaling relation to ACTPol clusters cross matched with the redMaPPer catalog 
(see Section~\ref{sec:WLComparison}).}
\label{fig:comalit}
\end{figure}

Fig.~\ref{fig:comalit} shows the ACTPol--CoMaLit comparison, including previous stacked weak-lensing masses of ACT clusters reported in 
\citet{Battaglia_2016}, labeled as CS82-ACT. Here we used the LC$^2$--\textit{single} catalog from CoMaLit, which consists of objects modeled using 
a single halo. Inspection of Fig.~\ref{fig:comalit} shows that the majority of the weak-lensing masses are larger than the SZ masses. One of the most 
significant outliers, with a very high weak-lensing mass, is Abell 370 (ACT-CL\,J0239.8$-$0134). We note that this cluster has been observed with the
\textit{Hubble Space Telescope} as part of the Frontier Fields initiative, and initial results show that a complicated, multi-component lensing model is required to describe the mass distribution
in this cluster \citep{Lagattuta_2017}. Given the heterogeneous nature of the CoMaLit catalog, we limit this comparison to a qualitative one, since modeling the 
selection function between ACTPol clusters and pointed weak-lensing observations of individual clusters analysed by several groups is non-trivial. 

\begin{figure}
\includegraphics[width=\columnwidth]{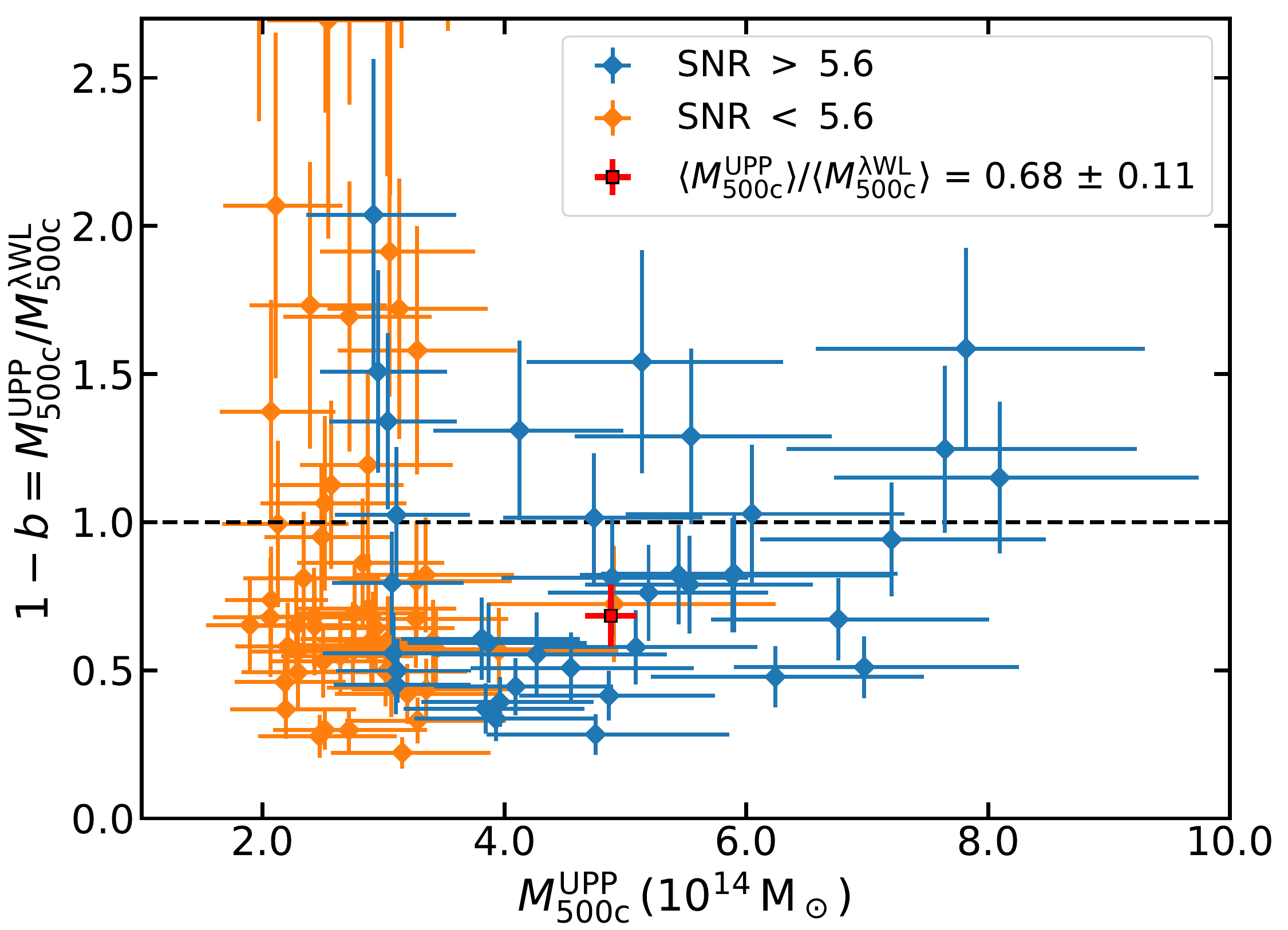}
\caption{Comparison of richness-based weak-lensing masses ($M^{\rm \lambda WL}_{\rm 500c}$), derived from applying the \citet{Simet_2017} scaling
relation to ACTPol clusters in common with redMaPPer, with ACTPol UPP/\citetalias{Arnaud_2010} SZ masses. The red square marks the ratio
$\langle M^{\rm UPP}_{\rm 500c} \rangle / \langle M^{\rm \lambda WL}_{\rm 500c} \rangle = 0.68 \pm 0.11$ for the SNR~$ > 5.6$ subsample, 
which is complete at $z < 0.6$. The effect of a Malmquist-type bias in the SZ selection can be seen on the clusters with SNR~$ < 5.6$,
many of which have SNR close to the detection threshold.
}
\label{fig:redmapper}
\end{figure}

In Fig.~\ref{fig:redmapper}, we compare our SZ-based masses to the redMaPPer richness-based masses that were calibrated with stacked weak-lensing 
measurements by \citet{Simet_2017}. Although the analysis of \citet{Simet_2017} is restricted to $z < 0.3$, we applied this relation to the full
subsample of ACTPol clusters with redMaPPer richness measurements (using an extended version of the \citet{Rykoff_2014} redMaPPer v5.10 catalog, 
which contains objects down to $\lambda = 5$), since a similar study using deeper DES data found no evidence that the $\lambda$--mass relation evolves with 
redshift \citep{Melchior_2017}. Note that as masses from the \citet{Simet_2017} scaling relation are defined within a radius $R_{\rm 200m}$ 
(within which the average density is 200 times the mean density of the Universe at the cluster redshift), we apply the concentration--mass relation 
of \citet{Bhattacharya_2013} to scale them to measurements within $R_{\rm 500c}$. We label these richness-based weak-lensing masses as 
$M^{\rm \lambda WL}_{\rm 500c}$. Within $z < 0.6$, there are 101 ACTPol clusters that have 
redMaPPer counterparts with $\lambda > 5$ and 4 that do not. Out of the 4 ACTPol clusters in the common ACTPol/redMaPPer survey area without a redMaPPer match, 
2 of them were probably masked in the redMaPPer optical cluster search, as they are within a few arcminutes of a bright star and a low redshift dwarf galaxy, and
another object (ACT-CL\,J2342.4+0406 at $z = 0.57$) does have a match in v6.3 of the redMaPPer catalog, but not in v5.10.
We discard these objects. 




To quantify the mass bias, we compute the ratio of the average SZ-mass to the average richness-based, weak-lensing calibrated mass 
$\langle M^{\rm UPP}_{\rm 500c} \rangle / \langle M^{\rm \lambda WL}_{\rm 500c} \rangle$, following the methodology and reasoning 
presented in \citet{Sifon_2016}. Computing the ratio of the averages, with uniform weighting of each measurement, has the advantage that 
many of the uncertainties related to the selection of these clusters and the underlying mass function are removed 
\citep[see the Appendix in][]{Sifon_2016}. This ratio is then used to calibrate the
normalization of the \citet{Arnaud_2010} relation we use to infer SZ masses. 

Using the subsample of SNR~$ > 5.6$ ACTPol clusters that is both 100\% pure and complete for $z < 0.6$, we find 
$\langle M^{\rm UPP}_{\rm 500c} \rangle = (4.88 \pm 0.21) \times 10^{14}$\,M$_{\sun}$ and 
$\langle M^{\rm \lambda WL}_{\rm 500c} \rangle = (7.13 \pm 1.05) \times 10^{14}$\,M$_{\sun}$, and their ratio is
$\langle M^{\rm UPP}_{\rm 500c} \rangle / \langle M^{\rm \lambda WL}_{\rm 500c} \rangle = 0.68 \pm 0.11$. The uncertainty quoted on each
average mass is the standard error on the mean, to which we have added the 7\% systematic uncertainty in the richness-based weak-lensing
masses \citep{Simet_2017}. As Fig.~\ref{fig:redmapper} shows, there is clearly intrinsic scatter between 
$M^{\rm UPP}_{\rm 500c}$ and $M^{\rm \lambda WL}_{\rm 500c}$, in addition to the scatter caused by the measurement uncertainties.
We stress that the purpose of this exercise is to obtain an overall re-scaling factor for application to the cluster population as a whole, and 
not to examine the scatter between the different mass estimates for any individual cluster. The intrinsic scatter should 
not in principle affect our measurement of the ratio of the average masses. We obtain consistent results 
(well within the uncertainties) if we repeat this analysis using either the entire sample, a higher cut in SNR ($> 8$), or 
split into two $M^{\rm UPP}_{\rm 500c}$ bins. If we split the sample at $z = 0.3$ into two redshift bins, we again 
find consistent results within 1$\sigma$, although we note that the lower redshift bin favors a mass ratio that is closer to unity
($\langle M^{\rm UPP}_{\rm 500c} \rangle / \langle M^{\rm \lambda WL}_{\rm 500c} \rangle = 0.88 \pm 0.18$ using 28 $z < 0.3$ clusters, and
$\langle M^{\rm UPP}_{\rm 500c} \rangle / \langle M^{\rm \lambda WL}_{\rm 500c} \rangle = 0.66 \pm 0.10$ using 73 $z > 0.3$ clusters).
A larger sample is needed to test for significant redshift evolution.

The mass bias that we measure is consistent 
with the value of $\langle M^{\rm UPP}_{\rm 500c} \rangle / \langle M^{\rm WL}_{\rm 500c} \rangle = 0.97 \pm 0.26$ measured by \citet{Battaglia_2016} 
using a stacked weak-lensing analysis of ACT clusters in the CS82 survey region, although the uncertainties are large. 
We also plot the measured mass bias in Fig.~\ref{fig:comalit}, for comparison with the CoMaLit sample, which is an independent dataset.

We use our $\langle M^{\rm UPP}_{\rm 500c} \rangle / \langle M^{\rm \lambda WL}_{\rm 500c} \rangle$ measurement to re-scale the 
ACTPol UPP/\citetalias{Arnaud_2010} scaling relation based SZ-derived masses and record these as $M_{\rm 500c}^{\rm Cal}$ in 
Table~\ref{tab:masses} in the Appendix.

\subsection{Comparison with SPT and \textit{Planck}}
\label{sec:SZComparison}

\begin{figure}
\includegraphics[width=\columnwidth]{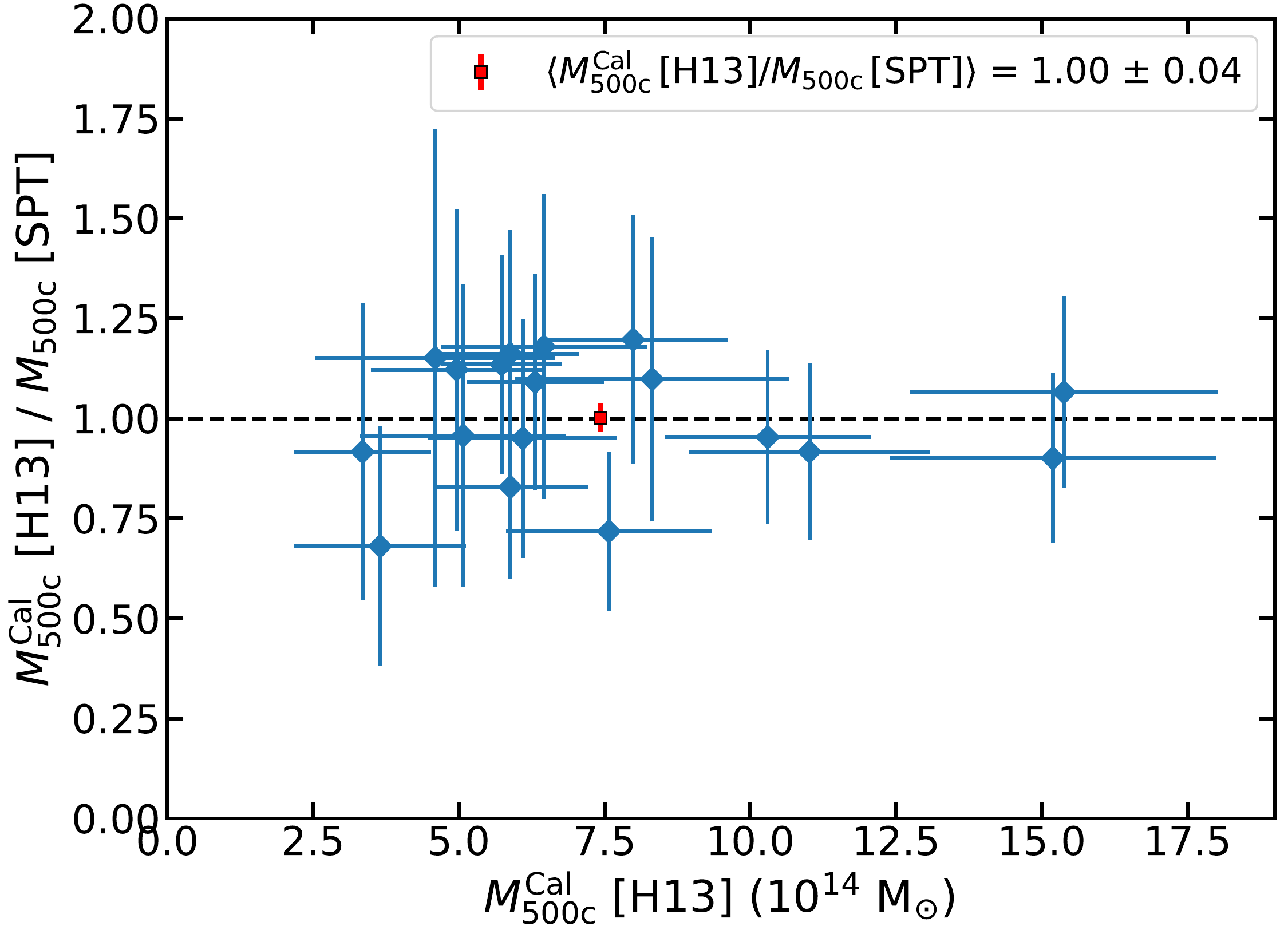}
\caption{Comparison of the ratio of SPT masses reported in \citet{Bleem_2015} to the ACTPol UPP-based masses, re-scaled using the
richness-based weak-lensing mass calibration ($M^{\rm Cal}_{\rm 500c}$; Section~\ref{sec:WLComparison}), for southern ACT clusters
in \citetalias{Hasselfield_2013}, for 18 objects cross matched between the samples. The red square marks the unweighted mean ratio ($\pm$ 
standard error) between the two sets of measurements.}
\label{fig:SPTRatio}
\end{figure}

\begin{figure*}
\begin{center}
\includegraphics[width=\textwidth]{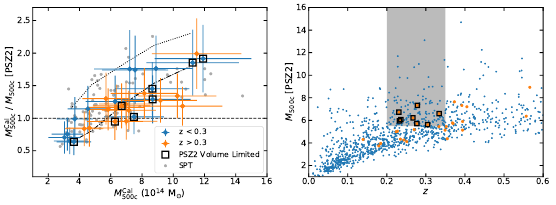}
\end{center}
\caption{Left panel: Comparison of the ratio of PSZ2 masses \citep{Planck2015_XXVII} to the ACTPol UPP-based masses, re-scaled using the
richness-based weak-lensing mass calibration ($M^{\rm Cal}_{\rm 500c}$; Section~\ref{sec:WLComparison}). 
Clearly there is a mass-dependent trend, with ACTPol mass estimates being progressively larger than PSZ2 with mass, which persists when
the sample is split by redshift. The \citet{Bleem_2015} SPT catalog, cross-matched with PSZ2 using a 10$\arcmin$ matching radius, follows
a similar trend (gray points). The dotted (dot-dashed) line shows the limit obtained by assigning masses at the 2$\sigma$ (5$\sigma$) PSZ2 detection
threshold to clusters that were detected by ACTPol but not PSZ2, averaged in $M^{\rm Cal}_{\rm 500c}$ bins (see the text). Right panel: 
The distribution of the whole PSZ2 catalog in the mass, redshift plane (small blue points). 
Clusters that are detected by both ACTPol and \textit{Planck} are shown as the larger yellow points. The shaded area shows a volume-limited
sample defined by $0.2 < z < 0.35$ and $M_{\rm 500c}\,{\rm[PSZ2]} > 5.5 \times 10^{14}$\,M$_{\sun}$. The 8 clusters in this region,
detected by both ACT and \textit{Planck}, are highlighted in both panels by black squares. The lower redshift limit accounts for the
fact that $z < 0.2$ clusters are underrepresented in the ACTPol sample (see Fig.~\ref{fig:completenessByRedshift}).}
\label{fig:PSZ2Ratio}
\end{figure*}

We now compare the ACTPol E-D56 cluster sample against the most recent cluster catalogs from other blind SZ surveys: 
the \citet{Bleem_2015} SPT catalog, and the PSZ2 catalog \citep{Planck2015_XXVII}. Ideally, one would compare the distributions
of the SZ cluster signals measured by the surveys; however, each project quantifies the SZ signal differently, and in
a model-dependent way, and so it is just as straightforward to compare the mass distributions (in any case the quantity of interest for cosmological studies) 
derived from the SZ measurements. In order to do this, a scaling relation between the chosen SZ observable and mass must be assumed,
and each survey has made different assumptions. Therefore we first make a comparison of the SZ masses measured by each survey,
to test if any correction is necessary to place them on an equivalent mass scale to this work.

In the case of SPT, there is no overlap between the \citet{Bleem_2015} catalog and the ACTPol E-D56 field. However, there is
an overlapping sample of 18 clusters in common with the southern ACT survey \citep{Marriage_2011}, for which \citetalias{Hasselfield_2013} provided
revised $M_{\rm 500c}$ measurements using the same PBAA method we have used to estimate $M^{\rm UPP}_{\rm 500c}$ in this work (Section~\ref{sec:SZMass}).
Moreover, we have
shown (Fig.~\ref{fig:massRecovery}) that the E-D56 $M^{\rm UPP}_{\rm 500c}$ mass measurements are on the same mass scale as the UPP masses
tabulated in \citetalias{Hasselfield_2013}. We therefore re-scale the \citetalias{Hasselfield_2013} UPP masses by 
the factor of 1/0.68 determined from comparing the ACTPol UPP masses with the richness-based weak-lensing masses 
(Section~\ref{sec:WLComparison}). Fig.~\ref{fig:SPTRatio} plots the 
ratio $M^{\rm Cal}_{\rm 500c}\,[{\rm H13}] / M_{\rm 500c}\,[{\rm SPT}]$ versus 
$M^{\rm Cal}_{\rm 500c}\,[{\rm H13}]$. We see that the mass ratio 
is constant over the mass range, and the unweighted mean ratio 
$\langle M^{\rm Cal}_{\rm 500c}\,[{\rm H13}] / M_{\rm 500c}\,[{\rm SPT}] \rangle = 1.00 \pm 0.04$
(where the quoted uncertainty is the standard error on the mean). Therefore, the SPT masses listed in the \citet{Bleem_2015}
catalog are consistent with the $M^{\rm Cal}_{\rm 500c}$ mass scale, and the two samples can be directly compared. This
agreement is remarkable, given that the mass calibration in each case has been arrived at from two very different directions.
The scaling relation used to calculate the SPT masses as listed in \citet{Bleem_2015} is derived from a Monte Carlo Markov
Chain analysis of the \citet{Reichardt_2013} cluster counts, with the cosmological parameters fixed to 
$\sigma_8 = 0.80$, $\Omega_{\rm m} = 0.3$, $\Omega_{\rm \Lambda} = 0.7$ and $H_0 = 70$\,km\,s$^{-1}$\,Mpc$^{-1}$. This
contrasts with the richness-based weak-lensing mass calibration, using an independent external dataset, that we have applied 
to the ACTPol sample. \citet{Bleem_2015} also used the projected isothermal $\beta$-model \citep{Cavaliere_1976}, 
rather than the UPP, to describe the expected cluster signal.

We perform a similar exercise with the PSZ2 Union catalog, this time using the 30 clusters in common with the ACTPol E-D56
catalog (Section~\ref{sec:Sample}). 
We compare the ACTPol SZ masses, after re-scaling by the richness-based weak-lensing mass calibration factor ($M^{\rm Cal}_{\rm 500c}$),
with the PSZ2 SZ masses as listed in \citet{Planck2015_XXVII}. The left panel of Fig.~\ref{fig:PSZ2Ratio} shows the result. The most striking feature
of this plot is the mass-dependent trend, with the ACTPol masses becoming larger in comparison to PSZ2 with mass 
(although the uncertainties are large). Although we have plotted the comparison with $M^{\rm Cal}_{\rm 500c}$ in Fig.~\ref{fig:PSZ2Ratio},
the systematic trend is still present if comparing to the ACTPol $M^{\rm UPP}_{\rm 500c}$ measurements, as the former results from
changing only the normalization of the scaling relation, and not its slope. The mass-dependent bias is surprising, given
that the UPP and the associated \citet{Arnaud_2010} scaling relation are used in both the ACTPol and \textit{Planck} analyses. This
bias does not seem to depend on redshift, angular size (as inferred from the recorded PSZ2 mass), or the detection significance in the PSZ2 catalog. 

A mass-dependent trend is also seen in the comparison of the \citet{Bleem_2015} SPT sample with PSZ2 (shown as the gray points in the left panel of
Fig.~\ref{fig:PSZ2Ratio}, where we plot $M_{\rm 500c}\, {\rm [SPT]} \, / \, M_{\rm 500c}\, {\rm [PSZ2]}$ versus $M_{\rm 500c}\, {\rm [SPT]}$).
Despite the differences between the SPT and ACT analyses, including in the modeling of the SZ signal itself, we do not see a similar
mass-dependent trend when comparing to SPT (Fig.~\ref{fig:SPTRatio}), nor do we see a mass-dependent trend when comparing ACTPol masses
to weak-lensing measurements, although the cross-matched sample is small (Fig.~\ref{fig:comalit}). 



A mass-dependent trend between weak-lensing mass and \textit{Planck} SZ-based masses has previously been noted in other studies 
\citep{vonDerLinden_2014, Hoekstra_2015, Mantz_2016}, with \citet{Mantz_2016} finding $M_{\rm 500c}\, [{\rm PSZ2}]  \propto M_{\rm WL}\,^{0.73 \pm 0.02}$ 
(a similar mass-dependent trend is also seen by \citet{Schellenberger_2017} when comparing the \textit{Planck} SZ-based masses with
hydrostatic mass estimates derived from \textit{Chandra} X-ray data).
Using the \citet{Kelly_2007} regression method, we find a non-linear slope, 
$M_{\rm 500c}\, [{\rm PSZ2}] \propto M^{\rm Cal}_{\rm 500c}\,^{0.55 \pm 0.18}$. We caution that this result, 
which is significant at the 2.5$\sigma$ level, does not account for selection effects. This is a concern because Fig.~\ref{fig:PSZ2Ratio}
shows the intersection of the PSZ2 and ACTPol cluster samples, and therefore clusters that were detected in one survey, but not the other,
could potentially drive the mass-dependent trend that we see. 

To mitigate selection effects, we define a volume-limited sample of PSZ2
clusters, adopting limits of $M_{\rm 500c} \, {\rm [PSZ2]} > 5.5 \times 10^{14}$\,M$_{\sun}$ and $0.2 < z < 0.35$, where the low redshift limit is
set to avoid the underrepresentation of such clusters in the ACTPol sample (see Fig.~\ref{fig:completenessByRedshift}). The chosen
mass limit is well above the apparent mass limit of the PSZ2 sample, as shown in the right panel of Fig.~\ref{fig:PSZ2Ratio}, and 
all of the PSZ2 clusters within this volume-limited sample are detected by ACTPol. These objects are highlighted using black boxes in
Fig.~\ref{fig:PSZ2Ratio}, and again, follow the same mass-dependent trend. We also considered the effect of clusters that were detected by ACT, but which are below the PSZ2 mass threshold. 
For the purposes of calculating the average ratio $M^{\rm Cal}_{\rm 500c} / M_{\rm 500c} \, {\rm [PSZ2]}$ in bins of $M^{\rm Cal}_{\rm 500c}$, 
we assigned PSZ2 masses at the approximate 2$\sigma$ detection threshold for the PSZ2 sample (estimated from the PSZ2 mass, redshift distribution shown in the right panel of 
Fig.~\ref{fig:PSZ2Ratio}) to those clusters that were detected by ACT, but not PSZ2. The corresponding upper limit is shown as the dotted
line in the left panel of Fig.~\ref{fig:PSZ2Ratio}. Similarly, we show the result of assigning PSZ2 masses at the estimated 
5$\sigma$ detection threshold for the PSZ2 sample as the dot-dashed line in the left panel of Fig.~\ref{fig:PSZ2Ratio}. 
Again, these follow the mass-dependent trend seen for the clusters that were detected in both catalogs.
Nevertheless, given the relatively simple nature of these tests, and the relative complexity of the PSZ2 cluster selection compared to the method
used in this work, we cannot completely rule out selection effects as the cause of the effect seen in Fig.~\ref{fig:PSZ2Ratio}.

\begin{figure}
\includegraphics[width=\columnwidth]{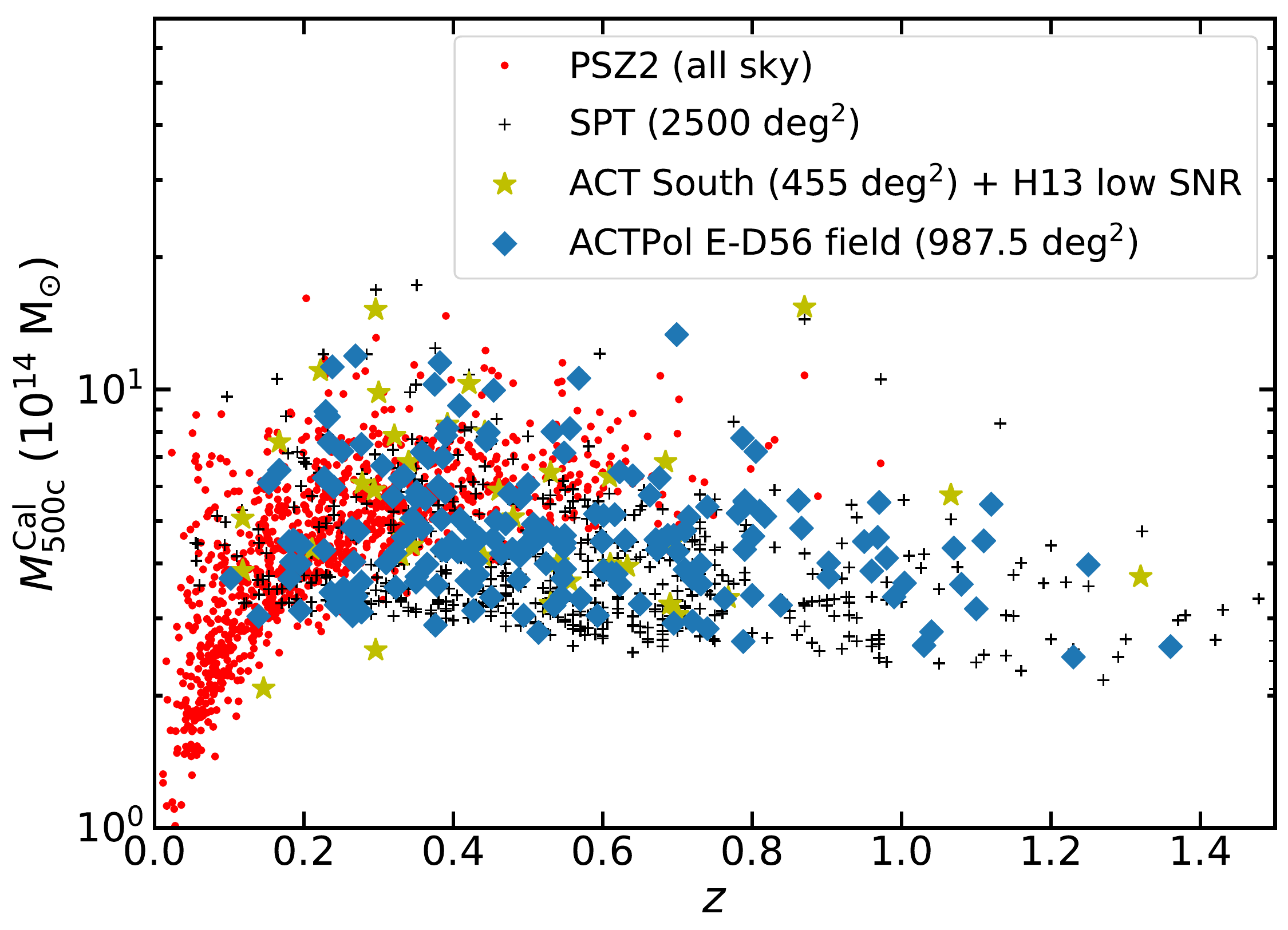}
\caption{Comparison of the ACTPol E-D56 cluster sample in the (mass, redshift) plane with other blind SZ surveys: SPT \citep{Bleem_2015},
and PSZ2 \citep{Planck2015_XXVII}. Additional clusters from the southern ACT field (\citealt{Marriage_2011}; 23 objects) 
and equatorial clusters that were masked/not detected in the E-D56 field with SNR~$> 4$ (Table~\ref{tab:H13Missed}; 15 objects) are shown as 
yellow stars, using the masses and redshifts as listed in \citetalias{Hasselfield_2013}. 
Here, all the ACT SZ masses have been re-scaled according to a richness-based weak-lensing 
mass calibration (Section~\ref{sec:WLComparison}). The SPT and PSZ2 mass measurements are as reported in \citet{Bleem_2015} and
\citet{Planck2015_XXVII} respectively (see Section~\ref{sec:SZComparison}). }
\label{fig:massVsRedshift}
\end{figure}

\begin{figure*}
\includegraphics[width=\textwidth]{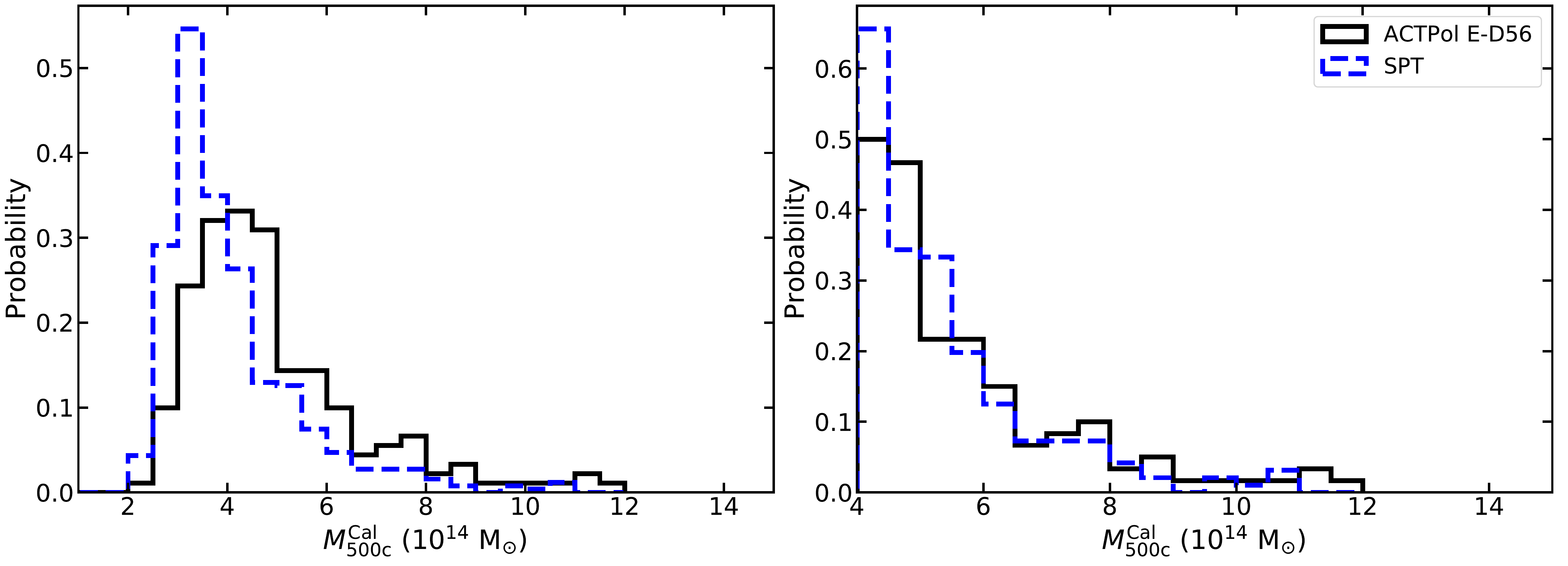}
\caption{Comparison of the ACTPol E-D56 mass distribution after applying the richness-based weak-lensing mass calibration (black) 
with SPT (blue; \citealt{Bleem_2015}). The left panel shows the whole distribution; here it is clear that the SPT sample
contains the larger fraction of lower mass clusters, with the ACTPol mass distribution becoming incomplete for 
$M^{\rm Cal}_{\rm 500c} < 4 \times 10^{14}$\,M$_{\sun}$. The right panel shows both distributions after applying a 
$M^{\rm Cal}_{\rm 500c} > 4 \times 10^{14}$\,M$_{\sun}$ cut. A two-sample KS test shows that in this
case, both samples are consistent with being drawn from the same mass distribution.}
\label{fig:SPT-ACT-MDist}
\end{figure*}

One possible explanation of the mass-dependent bias seen in the comparison between \textit{Planck} and weak-lensing mass measurements
\citep[e.g.,][]{Mantz_2016}
is unknown systematics in the weak-lensing analyses. However, this cannot explain Fig.~\ref{fig:PSZ2Ratio}, where we are comparing
SZ-based masses from two experiments that have made similar assumptions in modeling the SZ signal and mass-scaling relation. The most
obvious difference between the two experiments is angular resolution, with ACT having 1.4$\arcmin$ resolution compared to
$\approx 7 \arcmin$ for \textit{Planck}. Perhaps the key difference in terms of the analysis is the handling of the SZ-signal--size
degeneracy. Following \citetalias{Hasselfield_2013}, we do not attempt to measure $R_{\rm 500c}$ from the ACTPol data, and 
assume the combination of the UPP and the \citetalias{Arnaud_2010} scaling relation to model how the cluster signal changes 
with mass (and size), for a map filtered at a single reference angular scale. In contrast, in the \textit{Planck} analysis,
$R_{\rm 500c}$ and in turn the integrated SZ signal $Y_{\rm 500c}$ are inferred from the filtered map that optimizes the 
detection SNR. If the underlying average cluster profile is the UPP, as assumed in both analyses, then this should yield
consistent results. However, the difference in angular resolution between the experiments means that \textit{Planck} is
more sensitive to emission at the outskirts of clusters, while the SZ signal measured by ACT is dominated by emission from
within $R_{\rm 500c}$. In fact, for the ACTPol clusters that are cross matched with PSZ2, their PSZ2 masses imply 
$2.7 < \theta_{\rm 500c}$ (arcmin)~$< 7.4$, and so they are not resolved by \textit{Planck}. Therefore, one possible 
explanation of the trend seen in Fig.~\ref{fig:PSZ2Ratio} is that the true SZ signal in the outskirts of clusters differs 
from that implied by the UPP, and varies with mass. Simulations have shown that this could result from the effects of 
non-gravitational physics on the intracluster medium, such as the level of AGN feedback \citep[e.g.,][]{LeBrun_2015}.
Alternatively, it could be the case that the signal from within $R_{\rm 500c}$ is on average higher than expected compared to the UPP,
perhaps as a result of shocks from cluster mergers. This could bias the SZ masses measured by ACT high in comparison to
the PSZ2 masses, although it is not obvious why such a scenario would depend on cluster mass, and the lifetimes of such merger boosts
to the SZ signal are short \citep[e.g.,][]{Poole_2007, Wik_2008, Yang_2010, Nelson_2012}.
We are investigating this by measuring the stacked profiles of ACT clusters
beyond $R_{\rm 500c}$, and the results of this work will appear in a future publication. 
Alternatively, high resolution measurements of the SZ pressure profile, as will be provided by MUSTANG-2 \citep{Mason_2016}
and NIKA2 \citep{Mayet_2017}, could resolve this issue.

Fig.~\ref{fig:massVsRedshift} shows a comparison of the ACTPol E-D56, SPT, and PSZ2 cluster samples in the (mass, redshift) plane.
For ACTPol, we plot the masses after re-scaling by the richness-based weak-lensing mass calibration ($M^{\rm Cal}_{\rm 500c}$). 
We do not apply any re-scaling to the \citet{Bleem_2015} SPT masses or the PSZ2 masses.
Fig.~\ref{fig:massVsRedshift} shows the 
complementary nature of the ACT and SPT samples to PSZ2, with the former detecting clusters
at lower mass and at higher redshift, with only a weak dependence of the mass threshold with redshift. PSZ2, on the other hand, is not
biased against the detection of larger angular size, lower redshift clusters, owing to its extensive multi-frequency coverage and the absence
of atmospheric noise in the \textit{Planck} sky maps.

Fig.~\ref{fig:massVsRedshift} also suggests that SPT detects a greater number of lower mass clusters than ACTPol, while having an otherwise
similar selection function. We investigate this by directly comparing the mass distributions of the two samples. This is shown in the 
left panel of Fig.~\ref{fig:SPT-ACT-MDist}. We see that the number of
clusters in the ACTPol sample begins to fall for $M^{\rm Cal}_{\rm 500c} < 4 \times 10^{14}$\,M$_{\sun}$, indicating that below this mass limit the sample
is largely incomplete. In contrast, the SPT sample contains a larger fraction of clusters below this mass limit. This is expected, as the average
white-noise level of the E-D56 field is 18\,$\mu$K.arcmin \citep{Louis_2016}, compared to 15.5\,$\mu$K.arcmin for SPT \citep{Bleem_2015} at the
same frequency. In addition, the SPT cluster search benefits from the use of multi-frequency (95, 220\,GHz) data, and SPT's smaller beam size
(1.1$\arcmin$ at 150\,GHz). However, we do expect both ACTPol
and SPT to detect similar numbers of clusters above a mass threshold where neither survey is incomplete. We tested this by applying a mass cut
of $M^{\rm Cal}_{\rm 500c} > 4 \times 10^{14}$\,M$_{\sun}$ to both samples; the right panel of Fig.~\ref{fig:SPT-ACT-MDist} shows the result. 
Both cluster samples are consistent with being drawn from the same population after applying this cut. This is confirmed by a two-sample KS test,
which is not able to reject the null hypothesis that both samples are drawn from the same parent distribution ($D = 0.10$, $p$-value~=~0.49).

\subsection{Notable Clusters}
\label{sec:Notable}

In this Section we comment on a few notable clusters in the E-D56 field, including pairs of clusters, and very high-redshift ($z > 1.5$) clusters that were detected at
other wavelengths, but are not currently detected via the SZ by ACTPol.

\begin{figure}
\begin{center}
\includegraphics[width=\columnwidth]{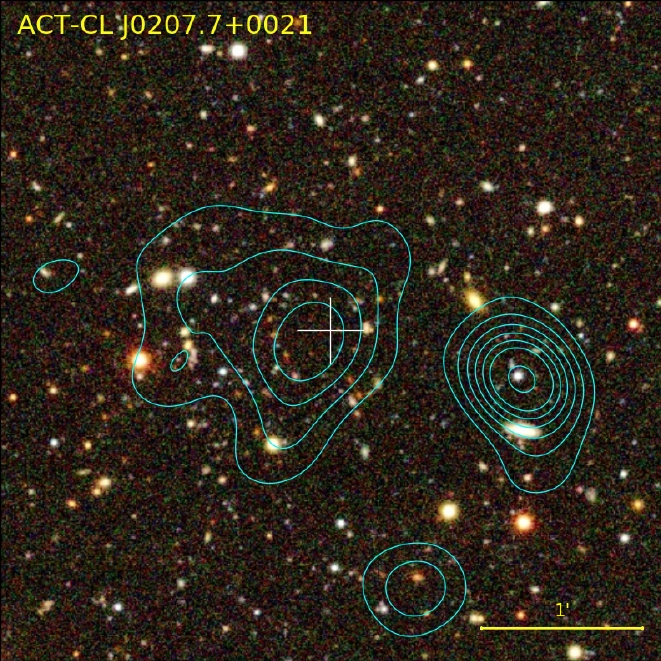}
\end{center}
\caption{S82 $gri$ image of ACT-CL\,J0207.7+0024 ($z = 1.10$), with blue contours (arbitrary levels) showing the extended X-ray emission (smoothed at 12$\arcsec$ scale)
detected by SWIFT. The image is $4\arcmin$ on a side, with North at the top and East at the left. The white cross marks the SZ cluster position. 
An unassociated X-ray point source, centered on a blue star-like object, is seen to the West. While J0207.7+0024 was previously 
reported as an X-ray cluster candidate by \citet{Liu_2015}, we present the first optical confirmation and redshift estimate for 
this cluster.}
\label{fig:J0207}
\end{figure}

\begin{figure}
\begin{center}
\includegraphics[width=\columnwidth]{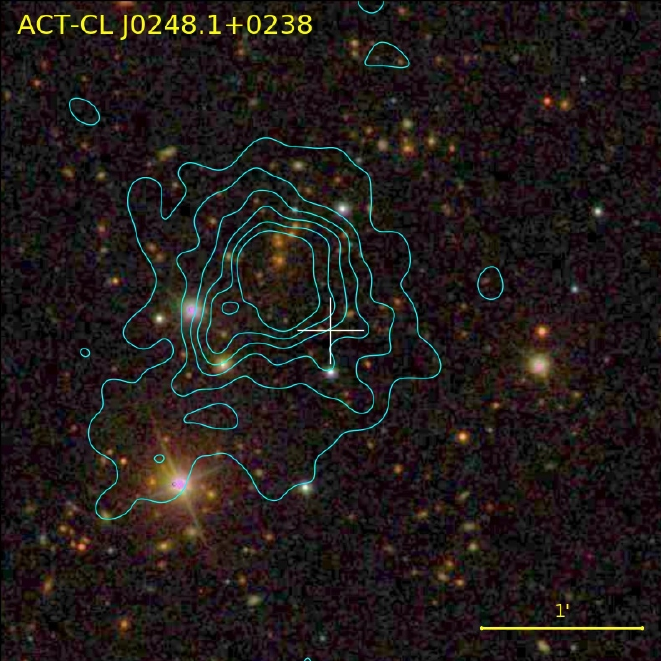}
\end{center}
\caption{SDSS $gri$ image of the massive cluster ACT-CL\,J0248.1+0238 ($z = 0.556$), with contours showing the extended X-ray emission 
detected by \textit{Chandra} (arbitrary levels; smoothed at 5$\arcsec$ scale).
The cluster is morphologically disturbed, and has a high X-ray temperature ($T = 8.4^{+1.4}_{-1.0}$\,keV). The image is $4\arcmin$ on a side, with North at the 
top and East at the left. The white cross marks the SZ cluster position.}
\label{fig:J0248}
\end{figure}

\subsubsection{ACT-CL\,J0012.1$-$0046}
This is the highest redshift cluster reported in the sample (photometric $z = 1.36 \pm 0.06$), and was first reported in \citet{Menanteau_2013} and \citetalias{Hasselfield_2013},
where it was detected with SNR~$=5.3$. In this work, using deeper data, it is detected with SNR~$=4.2$, which implies 
$M^{\rm UPP}_{\rm 500c} = (1.8^{+0.4}_{-0.3}) \times 10^{14}$\,M$_{\sun}$. 
This is roughly 70\% lower than the UPP-based mass estimate reported in \citetalias{Hasselfield_2013}, but differs at $<2 \sigma$ significance. Inspection of the deeper ACTPol data reveals that this cluster 
sits close to the center of a CMB cold spot, and is detected at SNR~$> 4$ using larger scale filters only. This perhaps caused the 
previously reported SNR to be `boosted' above the value we find here.

\subsubsection{ACT-CL\,J0207.7+0024}
This cluster, detected at SNR~$=5.3$ by ACTPol, was previously identified as an extended X-ray source, detected at SNR~$=9.7$, in the Swift X-ray Clusters Survey 
\citep[SWXCS;][]{Tundo_2012, Liu_2015}. However, no optical confirmation or redshift has previously been reported for this object.
\citet{Liu_2015} measured the (0.5--2.0\,keV) X-ray flux of J0207.7+0024 to be 
$F_{\rm X} = (4.5 \pm 0.5) \times 10^{-14}$\,erg\,cm$^{-2}$\,s$^{-1}$ within an effective radius of 76.6$\arcsec$, using data with an
effective exposure time of 84\,ks. For our photometric redshift estimate of $z = 1.10$, this implies the cluster has (0.5--2.0\,keV) luminosity
$L_{\rm X} = (2.3 \pm 0.3) \times 10^{44}$\,erg\,s$^{-1}$ (assuming temperature $T = 5$\,keV for the purpose of calculating the $k$-correction,
and neglecting the uncertainty on the photometric redshift). Based on the cluster's SZ signal, we estimate 
$M^{\rm UPP}_{\rm 500c} = (2.1^{+0.4}_{-0.3}) \times 10^{14}$\,M$_{\sun}$ for this object. 
Fig.~\ref{fig:J0207} shows the S82 optical image of the cluster, with the SWIFT 
X-ray contours overlaid.

\subsubsection{ACT-CL\,J0248.1+0238}
This $z = 0.556$ cluster has previously been identified in optical surveys by \citet{Lopes_2004} and \citet{Rykoff_2014}. Our SZ observations indicate this
is a massive object ($M^{\rm UPP}_{\rm 500c} = (5.5^{+1.0}_{-0.9}) \times 10^{14}$\,M$_{\sun}$), although it is not found in the PSZ2 sample or
ROSAT X-ray selected cluster
catalogs. We have obtained \textit{Chandra} observations of this object, and an X-ray spectral analysis confirms that this is a massive object, 
particularly given its redshift, with X-ray temperature $T = 8.4^{+1.4}_{-1.0}$\,keV (more details will be presented in a future publication). 
Fig.~\ref{fig:J0248} shows an optical image with overlaid X-ray contours; clearly, the cluster is somewhat morphologically disturbed.

\subsubsection{ACT-CL\,J2015.3$-$0126}
\label{sec:strayPS1}
\begin{figure}
\begin{center}
\includegraphics[width=\columnwidth]{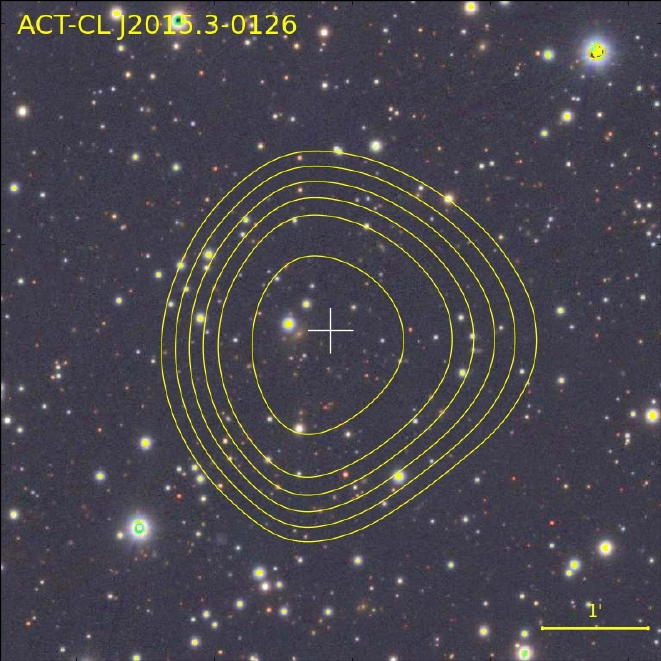}
\end{center}
\caption{PS1 $gri$ image of the newly discovered, massive, low Galactic latitude cluster ACT-CL\,J2015.3$-$0126. 
The image is $6\arcmin$ on a side, with North at the top and East at the left. The markings and contours are as 
indicated in Fig.~\ref{fig:SDSSMontage}.}
\label{fig:PS1}
\end{figure}

This is a newly discovered, massive ($M^{\rm UPP}_{\rm 500c} \approx 5 \times 10^{14}$\,M$_{\sun}$) cluster at low Galactic latitude 
($b = -19.3\deg$), detected at SNR~$=7.4$. 
Since it lies outside of the SDSS footprint, 
we visually confirmed this object through Pan-STARRS imaging (Fig.~\ref{fig:PS1}) and photometry \citep[PS1;][]{Chambers_2016, Flewelling_2016}.
We estimated the redshift ($z = 0.39$) of this cluster using the zCluster algorithm (Section~\ref{sec:zClusterAlgorithm}), but since we have not yet fully
tested zCluster using the PS1 photometry, which was released only recently, we adopt a conservative error of $\pm 0.1$ on the cluster 
redshift for now.

\begin{deluxetable}{p{4cm}cc}
\small
\tablecaption{Cluster pairs in the ACTPol E-D56 field\label{tab:pairs}}
\tablewidth{0pt}
\decimals
\tablehead{
\colhead{Cluster Pair}        &
\colhead{$z$}                 & 
\colhead{Projected Separation}\\
\colhead{}                    &  
\colhead{}                    & 
\colhead{(Mpc)} 
}
\startdata
ACT-CL\,J0034.4+0225/ ACT-CL\,J0034.9+0233 & 0.38 & 3.7\\
ACT-CL\,J0247.4$-$0156/ ACT-CL\,J0248.1$-$0216 & 0.24 & 5.2\\
ACT-CL\,J0301.6+0155/ ACT-CL\,J0303.3+0155 & 0.15 & 4.0\\
ACT-CL\,J2050.7+0122/ ACT-CL\,J2051.1+0057 & 0.33 & 7.5\\
ACT-CL\,J2319.7+0030/ ACT-CL\,J2320.0+0033 & 0.90 & 2.1\\
\enddata
\tablecomments{Only clusters with spectroscopic redshifts were considered. Each pair of clusters is within 
$\pm 3000$\,km$^{-1}$ of each other in terms of peculiar velocity.}
\end{deluxetable}

\subsubsection{Cluster Pairs}
\label{sec:pairs}
Since the E-D56 cluster search region covers a large, contiguous area, we conducted a search for pairs of clusters
that could be either physically associated or part of a supercluster. These objects may be of interest for future searches for
the warm-hot intergalactic medium (WHIM) associated with filaments between clusters \citep[e.g.,][]{Jauzac_2012, Eckert_2015}, or 
targeted kinetic-SZ studies \citep[e.g.,][]{Sayers_2016, Adam_2017}. Using only the subset of clusters with spectroscopic redshifts,
we matched pairs of clusters located within a  10\,Mpc projected radius \citep[cf.,][]{Eckert_2015}, and within $\pm 3000$\,km\,s$^{-1}$ of each other. 
We find 5 pairs of clusters matching these criteria, listed in Table~\ref{tab:pairs}. Of these, only 
ACT-CL\,J2319.7+0030/ACT-CL\,J2320.0+0033 at $z = 0.9$ is associated with a known supercluster \citep{Gilbank_2008}.
  
\subsubsection{Non-detected $z > 1.5$ Clusters}
Since the SZ effect is redshift independent, we checked the SZ signal measured by ACTPol at the locations of 
three relatively well known, very high redshift ($z > 1.5$) clusters that fall within the E-D56 footprint,
which are not detected with SNR~$> 4$ in our current data.

ClG\,J0218.3-0510 at $z = 1.63$ \citep{Papovich_2010, Tanaka_2010} and JKCS\,041 at $z = 1.80$ 
\citep{Andreon_2008, Newman_2014} are spectroscopically confirmed, IR-selected clusters. The $\tilde{y}_{0}$
signals that we measure at the reported positions of these clusters are consistent with zero, indicating they are
likely to be well below our mass threshold. This is as expected, given that X-ray analyses indicate that these clusters
have $M_{\rm 500c} \lesssim 10^{14}$\,M$_{\odot}$ \citep{Pierre_2012, Andreon_2014}. 

XLSSU\,J021744.1-034536 at $z = 1.9$ (photometric redshift) is an X-ray selected cluster detected in the XMM Large
Scale Structure survey \citep{Willis_2013}. At the reported position of this object, we measure 
$\tilde{y}_{0} = (0.47 \pm 0.13) \times 10^{-4}$, which implies 
$M^{\rm UPP}_{\rm 500c} \approx 1.5 \times 10^{14}$\,M$_{\sun}$. \citet{Mantz_2014} report an SZ detection of this cluster at 
30\,GHz using the Combined Array for Research in Millimeter-wave Astronomy (CARMA). Their mass estimate of 
$(1-2) \times 10^{14}$\,M$_{\sun}$, based on both SZ and X-ray data, is consistent with our measurement. Given
that this object is currently detected at SNR$_{2.4} = 3.5$, there is a good chance that this object will be included
in a future ACTPol cluster catalog, as the observations in this region become deeper.


\section{Summary}
\label{sec:Summary}

This work presents a catalog of 182 optically confirmed clusters, selected using the SZ effect with SNR~$> 4$, 
from the combination of the first two seasons of ACTPol observations with the original ACT equatorial survey at 148\,GHz. 
The cluster candidates were selected by applying a spatial matched filter to the maps in real space, using the UPP \citep{Arnaud_2010} to model
the cluster signal. Optical confirmation and redshifts were obtained largely from public surveys, with only a small number of 
clusters being followed-up using 4\,m-class telescopes for imaging and SALT for spectroscopy. The final sample 
spans the redshift range $0.1 < z < 1.4$, with median $z = 0.49$. Largely due to the overlap with SDSS, 80\%
of the clusters in the final sample have spectroscopic redshifts. We report the new discovery of 28 clusters (median $z = 0.80$), roughly one third
of which are confirmed through public SDSS/S82 data.

We characterized the relation between cluster mass and our chosen SZ observable, the central Compton
parameter measured in maps filtered at a scale of 2.4$\arcmin$, through the PBAA approach introduced by \citetalias{Hasselfield_2013} and
the application of the \citetalias{Arnaud_2010} scaling relation. The resulting mass distribution covers the 
range $1.6 < M^{\rm UPP}_{\rm 500c} / 10^{14} {\rm M}_{\sun} < 9.1$, with median $M^{\rm UPP}_{\rm 500c} = 3.1 \times 10^{14}$\,M$_{\sun}$.
We assessed the completeness of the cluster catalog as a function of mass and redshift by inserting UPP-model clusters
into the real data, and taking into account the variation in the noise level across the map. We estimate that the 
survey-averaged 90\% completeness limit of the survey is $M^{\rm UPP}_{\rm 500c} > 4.5 \times 10^{14}$\,M$_\sun$ 
for SNR$_{2.4} > 5$.

Comparing our UPP/\citetalias{Arnaud_2010} scaling relation based SZ masses with a richness-based, weak-lensing 
mass calibration, we found $\langle M^{\rm UPP}_{\rm 500c} \rangle / \langle M^{\rm \lambda WL}_{\rm 500c} \rangle = 0.68 \pm 0.11$. 
This is in line with the findings of some previous weak-lensing studies, although note that here we do not make a direct
comparison with weak-lensing mass measurements. We used this result to re-scale our UPP-based SZ mass estimates and report 
a set of richness-based, weak-lensing mass calibrated measurements, labeled as $M^{\rm Cal}_{\rm 500c}$ in the cluster
catalog.

We compared the ACTPol E-D56 cluster sample with the SPT and \textit{Planck} SZ-selected cluster catalogs. We found that
the ACTPol $M^{\rm Cal}_{\rm 500c}$ masses are on the same average mass scale as the \citet{Bleem_2015} SPT catalog, which is remarkable
given that the mass calibration of the \citet{Bleem_2015} sample was chosen to match the \citet{Reichardt_2013} cluster counts
for a fixed $\Lambda$CDM cosmology, whereas the richness-based, weak-lensing mass calibration used here relies on an independent dataset.
The mass distribution of our sample is consistent with the results of the SPT SZ cluster search for 
$M^{\rm Cal}_{\rm 500c} > 4 \times 10^{14}$\,M$_{\sun}$, a mass limit above which both surveys have a large degree of 
completeness. In the comparison with PSZ2 SZ masses, we find there is a mass-dependent trend, despite the fact that the UPP has been used to model the
cluster signal in both the ACTPol and \textit{Planck} analyses. The cause of this is being
investigated, but can perhaps be explained by a higher than average SZ signal in the cluster outskirts than is expected from the UPP model.

One of the principal aims of the ACTPol SZ cluster survey is to use clusters to constrain cosmological parameters;
such an analysis will be presented in future work. The sample presented here, with its clean, 
well-characterized SZ selection, can also be used for a number of other studies of the evolution of clusters
over most of cosmic time, and benefits from its overlap with a number of large, public surveys at many wavelengths
(Fig.~\ref{fig:E-D56}). While this catalog represents a significant step forward in terms of the cluster yield
in comparison to the previous \citetalias{Hasselfield_2013} cluster catalog, much more ACTPol data remains to be analyzed.
In addition, Advanced ACTPol \citep{DeBernardis_2016} has
already begun its survey of 15,000\,deg$^{2}$ of the Southern sky, and will produce an SZ cluster 
sample that is much larger than the catalog presented in this work.

\acknowledgments

We thank the referee for several comments that helped to improve this paper. 
We thank E. Rozo for providing the redMaPPer catalog down to $\lambda = 5$, and S. Bhargava for providing the
DES footprint as plotted in Fig.~\ref{fig:E-D56}. 
We thank the staff at APO, SALT, and SOAR for their help in conducting the optical/IR observations.
%
This work was supported by the U.S. National Science Foundation through awards AST-1440226, AST-0965625 and AST-0408698 for the 
ACT project, as well as awards PHY-1214379 and PHY-0855887. Funding was also provided by Princeton University, the University of 
Pennsylvania, and a Canada Foundation for Innovation (CFI) award to UBC. ACT operates in the Parque Astronómico Atacama in northern Chile under
the auspices of the Comisión Nacional de Investigación Cientı́fica y Tecnológica de Chile (CONICYT). Computations were performed on the
GPC supercomputer at the SciNet HPC Consortium and on the hippo cluster at the University of KwaZulu-Natal. SciNet is funded by the
CFI under the auspices of Compute Canada, the Government of Ontario, the Ontario Research Fund -- Research Excellence; and the 
University of Toronto. The development of multichroic detectors and lenses was supported by NASA grants NNX13AE56G and NNX14AB58G. 
MHi, DC, KM, and JLS acknowledge financial support from the National Research Foundation, the South African Square Kilometre Array project, and the University of KwaZulu-Natal. 
NB acknowledges the support from the Lyman Spitzer Jr. Fellowship.
JPH is supported by NASA grant NNX14AF73G and NSF grant AST-1615657.
JD and SN are supported by ERC grant 259505. 
TL is supported by ERC grant 267117, by ERC grant 259505, and by the Labex ILP (reference ANR-10-LABX-63) part of the Idex SUPER, and received financial state aid from ANR-11-IDEX-0004-02. 
HT is supported by grants NASA ATP NNX14AB57G, DOE DE-SC0011114, and NSF AST-1312991. 
AK has been supported by grant NSF AST-1312380. 
BS and BK are funded by NASA Space Technology Research Fellowships. 
RD acknowledges CONICYT for grants FONDECYT 1141113, Anillo ACT-1417, QUIMAL 160009 and  BASAL PFB-06 CATA. 
LM is funded by CONICYT FONDECYT grant 3170846.
HM is supported by the Jet Propulsion Laboratory, California Institute of Technology, under a contract with the National Aeronautics and Space Administration.
We thank our many colleagues from ABS, ALMA, APEX, and Polarbear who have helped us at critical junctures. Colleagues at AstroNorte and RadioSky provide logistical support
and keep operations in Chile running smoothly. We also thank the Mishrahi Fund and the Wilkinson Fund for their generous support of the project.
This work is based in part on observations obtained with the Southern African Large Telescope (SALT). 
Funding for SALT is provided in part by Rutgers University, a founding member of the SALT consortium, and the National Research Foundation.
Based in part on observations obtained at the Southern Astrophysical Research (SOAR) telescope, which is a 
joint project of the Minist\'{e}rio da Ci\^{e}ncia, Tecnologia, e Inova\c{c}\~{a}o (MCTI) da 
Rep\'{u}blica Federativa do Brasil, the U.S. National Optical Astronomy Observatory (NOAO), the University of
North Carolina at Chapel Hill (UNC), and Michigan State University (MSU). 
Based on observations obtained with the Apache Point Observatory 3.5-meter telescope, which is owned and operated by the Astrophysical 
Research Consortium. 
This research has made use of the NASA/IPAC Extragalactic Database (NED) which is 
operated by the Jet Propulsion Laboratory, California Institute of Technology, under contract with the 
National Aeronautics and Space Administration. 
Funding for SDSS-III has been provided by the Alfred P. 
Sloan Foundation, the Participating Institutions, the National Science Foundation, and the U.S. Department 
of Energy Office of Science. The SDSS-III web site is \url{http://www.sdss3.org/}. SDSS-III is managed by 
the Astrophysical Research Consortium for the Participating Institutions of the SDSS-III Collaboration 
(see the SDSS-III web site for details).  
Based in part on observations obtained with MegaPrime/MegaCam, 
a joint project of CFHT and CEA/IRFU, at the Canada-France-Hawaii Telescope (CFHT) which is operated by the 
National Research Council (NRC) of Canada, the Institut National des Science de l'Univers of the Centre National
de la Recherche Scientifique (CNRS) of France, and the University of Hawaii. This work is based in part on 
data products produced at Terapix available at the Canadian Astronomy Data Centre as part of the 
Canada-France-Hawaii Telescope Legacy Survey, a collaborative project of NRC and CNRS.
This paper uses data from the VIMOS Public Extragalactic Redshift Survey (VIPERS). VIPERS has been 
performed using the ESO Very Large Telescope, under the "Large Programme" 182.A-0886. The participating 
institutions and funding agencies are listed at \url{http://vipers.inaf.it}.
Based on data collected at the Subaru Telescope and retrieved from the HSC data archive system, 
which is operated by Subaru Telescope and Astronomy Data Center, National Astronomical Observatory of Japan.
The Hyper Suprime-Cam (HSC) collaboration includes the astronomical communities of Japan and Taiwan,
and Princeton University. The HSC instrumentation and software were developed by the National
Astronomical Observatory of Japan (NAOJ), the Kavli Institute for the Physics and Mathematics of the
Universe (Kavli IPMU), the University of Tokyo, the High Energy Accelerator Research Organization
(KEK), the Academia Sinica Institute for Astronomy and Astrophysics in Taiwan (ASIAA), and Princeton
University. Funding was contributed by the FIRST program from Japanese Cabinet Office, the Ministry
of Education, Culture, Sports, Science and Technology (MEXT), the Japan Society for the Promotion of
Science (JSPS), Japan Science and Technology Agency (JST), the Toray Science Foundation, NAOJ, Kavli
IPMU, KEK, ASIAA, and Princeton University.
This paper makes use of software developed for the Large Synoptic Survey Telescope. We thank the
LSST Project for making their code available as free software at \texttt{http://dm.lsst.org}.
The Pan-STARRS1 Surveys (PS1) and the PS1 public science archive have been made possible through contributions 
by the Institute for Astronomy, the University of Hawaii, the Pan-STARRS Project Office, the Max-Planck Society 
and its participating institutes, the Max Planck Institute for Astronomy, Heidelberg and the Max Planck Institute 
for Extraterrestrial Physics, Garching, The Johns Hopkins University, Durham University, the University of Edinburgh,
the Queen's University Belfast, the Harvard-Smithsonian Center for Astrophysics, the Las Cumbres Observatory Global 
Telescope Network Incorporated, the National Central University of Taiwan, the Space Telescope Science Institute, the 
National Aeronautics and Space Administration under Grant No. NNX08AR22G issued through the Planetary Science Division of the 
NASA Science Mission Directorate, the National Science Foundation Grant No. AST-1238877, the University of Maryland, 
Eotvos Lorand University (ELTE), the Los Alamos National Laboratory, and the Gordon and Betty Moore Foundation.

\software{
SExtractor \citep{BertinArnouts_1996},
RVSAO IRAF \citep{KurtzMink_1998},
SWARP \citep{Bertin_2002},
SCAMP \citep{Bertin_2006}, 
PySALT \citep{Crawford_2010},
astropy \citep{Astropy_2013},
hmf \citep{Murray_2013}
}





\appendix

\section{The ACTPol Cluster Catalog in the E-D56 Field}

\setcounter{table}{0}
\renewcommand{\thetable}{A\arabic{table}}



\begin{deluxetable*}{cccccccccp{4cm}}
\small
\tablecaption{Clusters detected with SNR~$>4$ in the ACTPol E-D56 field\label{tab:detections}}
\tablewidth{0pt}
\decimals
\tablehead{
\colhead{ACT-CL}      & 
\colhead{RA}          & 
\colhead{Dec}         & 
\colhead{SNR}             & 
\colhead{SNR$_{2.4}$}     &
\colhead{$\tilde{y}_{0}$} & 
\colhead{ACT?}        & 
\colhead{PSZ2?}       & 
\colhead{RM?}         & 
\colhead{Alt ID}      \\
\colhead{}            &
\colhead{(deg)}       & 
\colhead{(deg)}       & 
\colhead{}            & 
\colhead{}            & 
\colhead{($10^{-4}$)} & 
\colhead{}            &
\colhead{}            & 
\colhead{}            & 
\colhead{}            
}
\decimalcolnumbers
\startdata
$^*$J$0001.4-0306$ & $0.3633$ & $-3.1016$ & 4.3 & 4.1 & 0.68 $\pm$ 0.17 & \nodata & \nodata & \nodata & \nodata \\
\phantom{0}J$0003.1-0605$ & $0.7993$ & $-6.0877$ & 8.5 & 8.1 & 2.03 $\pm$ 0.25 & \nodata & $\checkmark$ & $\checkmark$ & {\scriptsize Abell 2697} \\
$^*$J$0005.0-0138$ & $1.2690$ & $-1.6379$ & 7.1 & 6.3 & 0.99 $\pm$ 0.16 & \nodata & \nodata & \nodata & \nodata \\
\phantom{0}J$0006.0-0231$ & $1.5190$ & $-2.5285$ & 4.8 & 4.5 & 0.79 $\pm$ 0.18 & \nodata & \nodata & $\checkmark$ & \nodata \\
\phantom{0}J$0006.9-0041$ & $1.7269$ & $-0.6864$ & 5.3 & 5.3 & 0.73 $\pm$ 0.14 & \nodata & \nodata & $\checkmark$ & {\scriptsize GMBCG\,J001.72541-00.68874} \\
\enddata
\tablecomments{The right ascension and declination coordinates in this table are for 
the ACT SZ detection position, given for the J2000 equinox; SNR is the SZ detection signal-to-noise optimized over all filter scales; 
SNR$_{2.4}$ is the SZ detection signal-to-noise ratio at the 2.4$\arcmin$ filter scale; 
$\tilde{y}_0$ is the cluster central Compton parameter measured at the 
2.4$\arcmin$ filter scale. Cross matches to other cluster catalogues are flagged in the ACT? \citep{Hasselfield_2013}, 
PSZ2? \citep{Planck2015_XXVII}, and RM? (redMaPPer v5.10; \citealt{Rykoff_2014}) columns. The Alt ID column gives the closest match
listed in the Nasa Extragalactic Database. Newly discovered clusters are indicated with the prefix $^*$ in column (1). Table~\ref{tab:detections} is published in its entirety in machine-readable format. A portion is shown here for guidance regarding its form and content.}
\end{deluxetable*}

\begin{deluxetable*}{cccccccccc}
\small
\tablecaption{Redshifts for clusters detected with SNR~$>4$ in the ACTPol E-D56 field\label{tab:redshifts}}
\tablewidth{0pt}
\decimals
\tablehead{
\colhead{ACT-CL}              & 
\colhead{BCG RA}              & 
\colhead{BCG Dec}             & 
\colhead{$z$}                 & 
\colhead{$z$ Type}            &
\colhead{$z$ Source}          & 
\colhead{$\delta_{\rm SDSS}$} & 
\colhead{$\delta_{\rm S82}$}  & 
\colhead{$\delta_{\rm CFHT}$} & 
\colhead{$\delta_{\rm SOAR}$} \\
\colhead{}                    & 
\colhead{(deg)}               & 
\colhead{(deg)}               & 
\colhead{}                    & 
\colhead{}                    &
\colhead{}                    & 
\colhead{}                    & 
\colhead{}                    & 
\colhead{}                    & 
\colhead{} 
}
\decimalcolnumbers
\startdata
J$0001.4-0306$ & $0.36493$ & $-3.08636$ & 0.102 & spec & SDSS & \phantom{0}$3.9 \pm 0.2$ & \nodata & \nodata & \nodata \\
J$0003.1-0605$ & $0.79826$ & $-6.09170$ & 0.233 & spec & SDSS & $13.9 \pm 0.9$ & \nodata & \nodata & \nodata \\
J$0005.0-0138$ & $1.27419$ & $-1.64499$ & 0.98 $\pm$ 0.05 & phot & zC$_{\rm SOAR}$ & \nodata & \nodata & \nodata & $18.0 \pm 2.4$ \\
J$0006.0-0231$ & $1.53010$ & $-2.52497$ & 0.618 & spec & SDSS & \nodata & \nodata & \nodata & \nodata \\
J$0006.9-0041$ & $1.73389$ & $-0.68106$ & 0.546 & spec & SDSS & \phantom{0}$6.2 \pm 0.9$ & \phantom{0}$5.8 \pm 0.4$ & \nodata & \nodata \\
\enddata
\tablecomments{The right ascension and declination
coordinates in this table are for the BCG position, given for the J2000 equinox. The $z$ column contains the
adopted `best' redshift, and $z$ Type indicates whether the redshift is spectroscopic (`spec') or photometric (`phot'). Uncertainties are only
quoted for photometric redshifts. The $z$ Source column indicates the source of the redshift: 
SDSS = spectroscopic redshift from SDSS (see Section~\ref{sec:delta}); 
VIPERS = spectroscopic redshift from VIPERS (Section~\ref{sec:delta}); 
CAMIRA = photometric redshift from \citet{Oguri_2017}; 
SALT = SALT spectroscopic redshift (Section~\ref{sec:SALTSpec});
S16 = spectroscopic redshift from \citet{Sifon_2016};
M13 = photometric redshift from \citet{Menanteau_2013};
zC = zCluster photometric redshift, from SDSS, S82, CFHTLenS, PS1, APO/SOAR data as indicated (Sections~\ref{sec:zClusterAlgorithm} and~\ref{sec:APOSOAR});
Lit = redshift from the literature, drawn from the following sources: 
    (1) \citet{Bohringer_2000}, 
    (2) \citet{Piffaretti_2011}, 
    (3) \citet{Muzzin_2012}, 
    (4) \citet{Dawson_2009}, \citet{Gilbank_2011},
    (5) \citet{Rykoff_2016},
    (6) \citet{Valtchanov_2004},
    (7) \citet{Crawford_1995},
    (8) \citet{Struble_1999},
    (9) \citet{Gilbank_2008},
    (10) \citet{Hoag_2015}.
Columns (7--10) list the density contrast statistic (equation~\ref{eq:delta}), measured at the zCluster redshift using the photometric data indicated in the subscript, and is shown where the zCluster photometric redshift is within $| \Delta z | < 0.05$ of the redshift listed in column (4). Table~\ref{tab:redshifts} is published in its entirety in machine-readable format. A portion is shown here for guidance regarding its form and content.
}
\end{deluxetable*}

\begin{deluxetable*}{cccccc}
\small
\tablecaption{Masses of clusters detected with SNR~$>4$ in the ACTPol E-D56 field\label{tab:masses}}
\tablewidth{0pt}
\decimals
\tablehead{
\colhead{ACT-CL}                & 
\colhead{$M^{\rm UPP}_{\rm 500c}$}             &
\colhead{$M_{\rm 500c}^{\rm Unc}$}   &
\colhead{$M^{\rm UPP}_{\rm 200m}$}            &
\colhead{$M_{\rm 200m}^{\rm Unc}$}  &
\colhead{$M_{\rm 500c}^{\rm Cal}$}   \\
\colhead{}                      & 
\colhead{($10^{14}$\,M$_{\sun}$)} &
\colhead{($10^{14}$\,M$_{\sun}$)} &
\colhead{($10^{14}$\,M$_{\sun}$)} &
\colhead{($10^{14}$\,M$_{\sun}$)} &
\colhead{($10^{14}$\,M$_{\sun}$)}         
}
\decimalcolnumbers
\startdata
J$0001.4-0306$ & $2.5^{+0.8}_{-0.6}$ & $3.1^{+1.1}_{-0.8}$ & $5.0^{+1.6}_{-1.2}$ & $6.1^{+2.2}_{-1.6}$ & $3.7^{+1.4}_{-1.1}$\\
J$0003.1-0605$ & $5.9^{+1.3}_{-1.1}$ & $6.8^{+1.6}_{-1.3}$ & $11.3^{+2.5}_{-2.1}$ & $13.2^{+3.1}_{-2.5}$ & $8.7^{+2.4}_{-2.1}$\\
J$0005.0-0138$ & $2.8^{+0.5}_{-0.4}$ & $3.1^{+0.6}_{-0.5}$ & $4.8^{+0.9}_{-0.7}$ & $5.4^{+1.0}_{-0.8}$ & $4.1^{+1.0}_{-0.9}$\\
J$0006.0-0231$ & $2.6^{+0.6}_{-0.5}$ & $2.9^{+0.7}_{-0.5}$ & $4.5^{+1.0}_{-0.8}$ & $5.1^{+1.1}_{-0.9}$ & $3.8^{+1.0}_{-0.9}$\\
J$0006.9-0041$ & $2.5^{+0.5}_{-0.4}$ & $2.8^{+0.6}_{-0.5}$ & $4.4^{+0.9}_{-0.7}$ & $4.9^{+1.0}_{-0.9}$ & $3.7^{+1.0}_{-0.9}$\\
\enddata
\tablecomments{Masses reported here assume the SZ signal is described by the UPP and the \citet{Arnaud_2010} scaling relation - 
refer to Section~\ref{sec:SZMass} for details. $M^{\rm UPP}_{\rm 500c}$ is measured with respect to the critical density at the cluster redshift; 
$M^{\rm UPP}_{\rm 200m}$ is measured with respect to the mean density at the cluster redshift, and is obtained from $M^{\rm UPP}_{\rm 500c}$ through the
concentration--mass relation of \citet{Bhattacharya_2013}, following \citet{HuKravtsov_2003}. Columns (2) and (4) report values
that have been corrected for the bias due to the steepness of the halo mass function, using the results of \citet{Tinker_2008}.
Columns (3) and (5) have not had this correction applied. Column (6) gives
$M^{\rm UPP}_{\rm 500c}$ re-scaled by the richness-based weak-lensing mass calibration factor of 1/0.68 
(see Section~\ref{sec:WLComparison}). Table~\ref{tab:masses} is published in its entirety in machine-readable format. A portion is shown here for guidance regarding its form and content.}
\end{deluxetable*}

\begin{deluxetable*}{cccccl}
\small
\tablecaption{Clusters in the \citetalias{Hasselfield_2013} catalog that are not included in the cluster catalog presented in this work.
\label{tab:H13Missed}}
\tablewidth{0pt}
\decimals
\tablehead{
\colhead{\citetalias{Hasselfield_2013} ID}        &
\colhead{SNR} &
\colhead{SNR} &
\colhead{$M^{\rm UPP}_{\rm 500c}$ [\citetalias{Hasselfield_2013}]} &
\colhead{$M^{\rm UPP}_{\rm 500c}$ [this work]}                     &
\colhead{Reason for Exclusion} \\
\colhead{} &
\colhead{(\citetalias{Hasselfield_2013})} &
\colhead{(this work)} &
\colhead{($10^{14}\,$M$_{\sun}$)} &
\colhead{($10^{14}\,$M$_{\sun}$)} &
\colhead{} 
}
\startdata
ACT-CL\,J$0017.6-0051$ & 4.2 & 3.8     & $2.9 \pm 1.0$ & $1.9^{+0.6}_{-0.4}$ & Low SNR\\
ACT-CL\,J$0051.1+0055$ & 4.2 & $<3$    & $2.2 \pm 0.8$ & \nodata & Low SNR\\
ACT-CL\,J$0139.3-0128$ & 4.3 & 3.2     & $2.1 \pm 0.9$ & $1.9^{+0.5}_{-0.4}$ & Low SNR\\ 
ACT-CL\,J$0230.9-0024$ & 4.2 & 3.3     & $2.8 \pm 0.9$ & $1.8^{+0.5}_{-0.4}$ & Low SNR\\
ACT-CL\,J$0301.1-0110$ & 4.2 & $<3$    & $2.2 \pm 0.8$ & \nodata  & Low SNR\\
ACT-CL\,J$0308.1+0103$ & 4.8 & \nodata & $2.7 \pm 0.8$ & \nodata  & Point source mask\\
ACT-CL\,J$0336.9-0110$ & 4.8 & 3.9     & $2.5 \pm 0.7$ & $2.4^{+0.5}_{-0.4}$  & Low SNR\\
ACT-CL\,J$0348.6-0028$ & 4.7 & 3.9     & $3.1 \pm 0.9$ & $3.5^{+0.9}_{-0.7}$ & Low SNR\\
ACT-CL\,J$2025.2+0030$ & 6.4 & \nodata & $4.6 \pm 1.0$  & \nodata & Point source mask\\  
ACT-CL\,J$2051.1+0215$ & 5.2 & \nodata & $5.3 \pm 1.4$ & \nodata & Outside E-D56 sky area\\
ACT-CL\,J$2135.1-0102$ & 4.1 & \nodata & $2.8 \pm 1.0$ & \nodata & Point source mask\\
ACT-CL\,J$2135.7+0009$ & 4.0 & 3.2     & $6.3 \pm 1.2$ & $5.6^{+1.3}_{-1.1}$ & Low SNR\\
ACT-CL\,J$2152.9-0114$ & 4.4 & 3.9     & $3.0 \pm 0.9$ & $2.9^{+0.7}_{-0.5}$ & Low SNR\\
ACT-CL\,J$2229.2-0004$ & 4.0 & 3.7     & $2.7 \pm 1.0$ & $2.2^{+0.6}_{-0.5}$ & Low SNR\\
ACT-CL\,J$2253.3-0031$ & 4.0 & 3.4     & $2.7 \pm 0.9$ & $2.5^{+0.6}_{-0.5}$ & Low SNR\\
\enddata
\end{deluxetable*}

\begin{deluxetable*}{cc}
\small
\tablecaption{PSZ2 candidates in the ACTPol survey area that were not optically confirmed in the PSZ2 catalog, 
and are not detected/confirmed by ACTPol.\label{tab:PSZ2NotConfirmed}}
\tablewidth{0pt}
\decimals
\tablehead{
\colhead{Name}        &
\colhead{PSZ2 SNR}    
}
\startdata
PSZ2\,G045.96$-$26.94 & 5.1  \\ 
PSZ2\,G051.48$-$30.87 & 5.0  \\ 
PSZ2\,G084.69$-$58.60 & 4.7  \\ 
PSZ2\,G135.94$-$68.22 & 6.9  \\ 
PSZ2\,G146.10$-$55.55 & 4.7  \\ 
PSZ2\,G167.43$-$53.67 & 4.6  \\ 
\enddata
\end{deluxetable*}


\clearpage

\begin{thebibliography}{142}
\expandafter\ifx\csname natexlab\endcsname\relax\def\natexlab#1{#1}\fi

\bibitem[{{Adam} et~al.(2017)}]{Adam_2017}
{Adam}, R., et~al. 2017, \texttt{arXiv:1606.07721}, \aap, 598, A115

\bibitem[{{Aihara} et~al.(2017{\natexlab{a}})}]{HSC_2017}
{Aihara}, H., et~al. 2017{\natexlab{a}}, \texttt{arXiv:1702.08449}, ArXiv
  e-prints

\bibitem[{{Aihara} et~al.(2017{\natexlab{b}})}]{Aihara_2017}
---. 2017{\natexlab{b}}, \texttt{arXiv:1704.05858}, PASJ submitted
  (arXiv:1704.05858)

\bibitem[{{Andrade-Santos} et~al.(2017)}]{Andrade-Santos_2017}
{Andrade-Santos}, F., et~al. 2017, \texttt{arXiv:1703.08690}, \apj, 843, 76

\bibitem[{{Andreon}(2008)}]{Andreon_2008}
{Andreon}, S. 2008, \texttt{arXiv:0710.2737}, MNRAS, 386, 1045

\bibitem[{{Andreon} et~al.(2014){Andreon}, {Newman}, {Trinchieri}, {Raichoor},
  {Ellis}, \& {Treu}}]{Andreon_2014}
{Andreon}, S., {Newman}, A.~B., {Trinchieri}, G., {Raichoor}, A., {Ellis},
  R.~S., \& {Treu}, T. 2014, \texttt{arXiv:1311.4361}, \aap, 565, A120

\bibitem[{{Annis} et~al.(2014)}]{Annis_2014}
{Annis}, J., et~al. 2014, \texttt{arXiv:1111.6619}, ApJ, 794, 120

\bibitem[{{Arnaud} et~al.(2005){Arnaud}, {Pointecouteau}, \&
  {Pratt}}]{Arnaud_2005}
{Arnaud}, M., {Pointecouteau}, E., \& {Pratt}, G.~W. 2005,
  \texttt{arXiv:astro-ph/0502210}, A\&A, 441, 893

\bibitem[{{Arnaud} et~al.(2010){Arnaud}, {Pratt}, {Piffaretti},
  {B{\"o}hringer}, {Croston}, \& {Pointecouteau}}]{Arnaud_2010}
{Arnaud}, M., {Pratt}, G.~W., {Piffaretti}, R., {B{\"o}hringer}, H., {Croston},
  J.~H., \& {Pointecouteau}, E. 2010, \texttt{arXiv:0910.1234}, A\&A, 517, A92

\bibitem[{{Astropy Collaboration} et~al.(2013)}]{Astropy_2013}
{Astropy Collaboration}, et~al. 2013, \texttt{arXiv:1307.6212}, \aap, 558, A33

\bibitem[{{Bartelmann}(1996)}]{Bartelmann_1996}
{Bartelmann}, M. 1996, \texttt{astro-ph/9602053}, \aap, 313, 697

\bibitem[{{Battaglia} et~al.(2016)}]{Battaglia_2016}
{Battaglia}, N., et~al. 2016, \texttt{arXiv:1509.08930}, \jcap, 8, 013

\bibitem[{{Beers} et~al.(1990){Beers}, {Flynn}, \& {Gebhardt}}]{Beers_1990}
{Beers}, T.~C., {Flynn}, K., \& {Gebhardt}, K. 1990, AJ, 100, 32

\bibitem[{{Ben{\'{\i}}tez}(2000)}]{Benitez_2000}
{Ben{\'{\i}}tez}, N. 2000, \apj, 536, 571

\bibitem[{{Bertin}(2006)}]{Bertin_2006}
{Bertin}, E. 2006, in Astronomical Society of the Pacific Conference Series,
  Vol. 351, Astronomical Data Analysis Software and Systems XV, ed.
  C.~{Gabriel}, C.~{Arviset}, D.~{Ponz}, \& S.~{Enrique},  112

\bibitem[{{Bertin} \& {Arnouts}(1996)}]{BertinArnouts_1996}
{Bertin}, E. \& {Arnouts}, S. 1996, A\&AS, 117, 393

\bibitem[{{Bertin} et~al.(2002){Bertin}, {Mellier}, {Radovich}, {Missonnier},
  {Didelon}, \& {Morin}}]{Bertin_2002}
{Bertin}, E., {Mellier}, Y., {Radovich}, M., {Missonnier}, G., {Didelon}, P.,
  \& {Morin}, B. 2002, in Astronomical Society of the Pacific Conference
  Series, Vol. 281, Astronomical Data Analysis Software and Systems XI, ed.
  D.~A. {Bohlender}, D.~{Durand}, \& T.~H. {Handley},  228

\bibitem[{{Bhattacharya} et~al.(2013){Bhattacharya}, {Habib}, {Heitmann}, \&
  {Vikhlinin}}]{Bhattacharya_2013}
{Bhattacharya}, S., {Habib}, S., {Heitmann}, K., \& {Vikhlinin}, A. 2013,
  \texttt{arXiv:1112.5479}, \apj, 766, 32

\bibitem[{{Biffi} et~al.(2016)}]{Biffi_2016}
{Biffi}, V., et~al. 2016, \texttt{arXiv:1606.02293}, \apj, 827, 112

\bibitem[{{Birkinshaw}(1999)}]{Birkinshaw_1999}
{Birkinshaw}, M. 1999, \texttt{arXiv:astro-ph/9808050}, Physics Reports, 310,
  97

\bibitem[{{Blanton} \& {Roweis}(2007)}]{Blanton_2007}
{Blanton}, M.~R. \& {Roweis}, S. 2007, \texttt{arXiv:astro-ph/0606170}, AJ,
  133, 734

\bibitem[{{Bleem} et~al.(2015)}]{Bleem_2015}
{Bleem}, L.~E., et~al. 2015, \texttt{arXiv:1409.0850}, \apjs, 216, 27

\bibitem[{{Bode} et~al.(2012){Bode}, {Ostriker}, {Cen}, \& {Trac}}]{Bode_2012}
{Bode}, P., {Ostriker}, J.~P., {Cen}, R., \& {Trac}, H. 2012,
  \texttt{arXiv:1204.1762}, ArXiv e-prints

\bibitem[{{B{\"o}hringer} et~al.(2000)}]{Bohringer_2000}
{B{\"o}hringer}, H., et~al. 2000, \texttt{astro-ph/0003219}, \apjs, 129, 435

\bibitem[{{Brammer} et~al.(2008){Brammer}, {van Dokkum}, \&
  {Coppi}}]{Brammer_2008}
{Brammer}, G.~B., {van Dokkum}, P.~G., \& {Coppi}, P. 2008,
  \texttt{arXiv:0807.1533}, \apj, 686, 1503

\bibitem[{{Cavaliere} \& {Fusco-Femiano}(1976)}]{Cavaliere_1976}
{Cavaliere}, A. \& {Fusco-Femiano}, R. 1976, \aap, 49, 137

\bibitem[{{Chambers} et~al.(2016)}]{Chambers_2016}
{Chambers}, K.~C., et~al. 2016, \texttt{arXiv:1612.05560}, ArXiv e-prints

\bibitem[{{Clemens} et~al.(2004){Clemens}, {Crain}, \&
  {Anderson}}]{Clemens_2004}
{Clemens}, J.~C., {Crain}, J.~A., \& {Anderson}, R. 2004, in \procspie, Vol.
  5492, Ground-based Instrumentation for Astronomy, ed. A.~F.~M. {Moorwood} \&
  M.~{Iye},  331--340

\bibitem[{{Coleman} et~al.(1980){Coleman}, {Wu}, \&
  {Weedman}}]{ColemanWuWeedman_1980}
{Coleman}, G.~D., {Wu}, C.-C., \& {Weedman}, D.~W. 1980, \apjs, 43, 393

\bibitem[{{Crawford} et~al.(1995){Crawford}, {Edge}, {Fabian}, {Allen},
  {Bohringer}, {Ebeling}, {McMahon}, \& {Voges}}]{Crawford_1995}
{Crawford}, C.~S., {Edge}, A.~C., {Fabian}, A.~C., {Allen}, S.~W., {Bohringer},
  H., {Ebeling}, H., {McMahon}, R.~G., \& {Voges}, W. 1995, \mnras, 274, 75

\bibitem[{{Crawford} et~al.(2010)}]{Crawford_2010}
{Crawford}, S.~M., et~al. 2010, in Society of Photo-Optical Instrumentation
  Engineers (SPIE) Conference Series, Vol. 7737, Society of Photo-Optical
  Instrumentation Engineers (SPIE) Conference Series

\bibitem[{{Dawson} et~al.(2009)}]{Dawson_2009}
{Dawson}, K.~S., et~al. 2009, \texttt{arXiv:0908.3928}, \aj, 138, 1271

\bibitem[{{De Bernardis} et~al.(2016)}]{DeBernardis_2016}
{De Bernardis}, F., et~al. 2016, \texttt{arXiv:1607.02120}, in \procspie, Vol.
  9910, Observatory Operations: Strategies, Processes, and Systems VI,  991014

\bibitem[{{de Haan} et~al.(2016)}]{deHaan_2016}
{de Haan}, T., et~al. 2016, \texttt{arXiv:1603.06522}, ApJ submitted
  (arXiv:1603.06522)

\bibitem[{{Diehl} et~al.(2016)}]{Diehl_2016}
{Diehl}, H.~T., et~al. 2016, in \procspie, Vol. 9910, Observatory Operations:
  Strategies, Processes, and Systems VI,  99101D

\bibitem[{{D{\"u}nner} et~al.(2013)}]{Dunner_2013}
{D{\"u}nner}, R. et~al. 2013, \texttt{arXiv:1208.0050}, ApJ, 762, 10

\bibitem[{{Eckert} et~al.(2015)}]{Eckert_2015}
{Eckert}, D., et~al. 2015, \texttt{arXiv:1512.00454}, \nat, 528, 105

\bibitem[{{Erben} et~al.(2013)}]{Erben_2013}
{Erben}, T., et~al. 2013, \texttt{arXiv:1210.8156}, MNRAS, 433, 2545

\bibitem[{{Fioc} \& {Rocca-Volmerange}(1997)}]{Fioc_1997}
{Fioc}, M. \& {Rocca-Volmerange}, B. 1997, \texttt{arXiv:astro-ph/9707017},
  A\&A, 326, 950

\bibitem[{{Flewelling} et~al.(2016)}]{Flewelling_2016}
{Flewelling}, H.~A., et~al. 2016, \texttt{arXiv:1612.05243}, ArXiv e-prints

\bibitem[{{Geach} et~al.(2011){Geach}, {Murphy}, \& {Bower}}]{Geach_2011}
{Geach}, J.~E., {Murphy}, D.~N.~A., \& {Bower}, R.~G. 2011,
  \texttt{arXiv:1101.4585}, \mnras, 413, 3059

\bibitem[{{Gilbank} et~al.(2011){Gilbank}, {Gladders}, {Yee}, \&
  {Hsieh}}]{Gilbank_2011}
{Gilbank}, D.~G., {Gladders}, M.~D., {Yee}, H.~K.~C., \& {Hsieh}, B.~C. 2011,
  \texttt{arXiv:1012.3470}, \aj, 141, 94

\bibitem[{{Gilbank} et~al.(2008){Gilbank}, {Yee}, {Ellingson}, {Hicks},
  {Gladders}, {Barrientos}, \& {Keeney}}]{Gilbank_2008}
{Gilbank}, D.~G., {Yee}, H.~K.~C., {Ellingson}, E., {Hicks}, A.~K., {Gladders},
  M.~D., {Barrientos}, L.~F., \& {Keeney}, B. 2008, \texttt{arXiv:0803.1675},
  \apjl, 677, L89

\bibitem[{{Goto} et~al.(2002)}]{Goto_2002}
{Goto}, T., et~al. 2002, \texttt{astro-ph/0112482}, \aj, 123, 1807

\bibitem[{{Hao} et~al.(2010)}]{Hao_2010}
{Hao}, J., et~al. 2010, \texttt{arXiv:1010.5503}, \apjs, 191, 254

\bibitem[{{Hasselfield} et~al.(2013)}]{Hasselfield_2013}
{Hasselfield}, M. et~al. 2013, \texttt{arXiv:1301.0816}, JCAP, 7, 8

\bibitem[{{Henson} et~al.(2017){Henson}, {Barnes}, {Kay}, {McCarthy}, \&
  {Schaye}}]{Henson_2017}
{Henson}, M.~A., {Barnes}, D.~J., {Kay}, S.~T., {McCarthy}, I.~G., \& {Schaye},
  J. 2017, \texttt{arXiv:1607.08550}, \mnras, 465, 3361

\bibitem[{{Hildebrandt} et~al.(2012)}]{Hildebrandt_2012}
{Hildebrandt}, H., et~al. 2012, \texttt{arXiv:1111.4434}, MNRAS, 421, 2355

\bibitem[{{Hoag} et~al.(2015)}]{Hoag_2015}
{Hoag}, A., et~al. 2015, \texttt{arXiv:1503.02670}, \apj, 813, 37

\bibitem[{{Hoekstra} et~al.(2015){Hoekstra}, {Herbonnet}, {Muzzin}, {Babul},
  {Mahdavi}, {Viola}, \& {Cacciato}}]{Hoekstra_2015}
{Hoekstra}, H., {Herbonnet}, R., {Muzzin}, A., {Babul}, A., {Mahdavi}, A.,
  {Viola}, M., \& {Cacciato}, M. 2015, \texttt{arXiv:1502.01883}, \mnras, 449,
  685

\bibitem[{{Hu} \& {Kravtsov}(2003)}]{HuKravtsov_2003}
{Hu}, W. \& {Kravtsov}, A.~V. 2003, \texttt{astro-ph/0203169}, \apj, 584, 702

\bibitem[{{Israel} et~al.(2015){Israel}, {Schellenberger}, {Nevalainen},
  {Massey}, \& {Reiprich}}]{Israel_2015}
{Israel}, H., {Schellenberger}, G., {Nevalainen}, J., {Massey}, R., \&
  {Reiprich}, T.~H. 2015, \texttt{arXiv:1408.4758}, \mnras, 448, 814

\bibitem[{{Itoh} et~al.(1998){Itoh}, {Kohyama}, \& {Nozawa}}]{Itoh_1998}
{Itoh}, N., {Kohyama}, Y., \& {Nozawa}, S. 1998, \texttt{astro-ph/9712289},
  \apj, 502, 7

\bibitem[{{Jauzac} et~al.(2012)}]{Jauzac_2012}
{Jauzac}, M., et~al. 2012, \texttt{arXiv:1208.4323}, \mnras, 426, 3369

\bibitem[{{Jee} et~al.(2011)}]{Jee_2011}
{Jee}, M.~J., et~al. 2011, \texttt{arXiv:1105.3186}, \apj, 737, 59

\bibitem[{{Kelly}(2007)}]{Kelly_2007}
{Kelly}, B.~C. 2007, \texttt{arXiv:0705.2774}, ApJ, 665, 1489

\bibitem[{{Kinney} et~al.(1996){Kinney}, {Calzetti}, {Bohlin}
  et~al.}]{Kinney_1996}
{Kinney}, A.~L., {Calzetti}, D., {Bohlin}, R.~C., et~al. 1996, ApJ, 467, 38

\bibitem[{{Kirk} et~al.(2015)}]{Kirk_2015}
{Kirk}, B., et~al. 2015, \texttt{arXiv:1410.7887}, \mnras, 449, 4010

\bibitem[{{Koester} et~al.(2007)}]{Koester_2007}
{Koester}, B.~P., et~al. 2007, \texttt{astro-ph/0701265}, \apj, 660, 239

\bibitem[{{Kurtz} \& {Mink}(1998)}]{KurtzMink_1998}
{Kurtz}, M.~J. \& {Mink}, D.~J. 1998, \texttt{arXiv:astro-ph/9803252}, PASP,
  110, 934

\bibitem[{{Lagattuta} et~al.(2017)}]{Lagattuta_2017}
{Lagattuta}, D.~J., et~al. 2017, \texttt{arXiv:1611.01513}, \mnras, 469, 3946

\bibitem[{{Lang}(2014)}]{Lang_2014}
{Lang}, D. 2014, \texttt{arXiv:1405.0308}, \aj, 147, 108

\bibitem[{{Le Brun} et~al.(2015){Le Brun}, {McCarthy}, \&
  {Melin}}]{LeBrun_2015}
{Le Brun}, A.~M.~C., {McCarthy}, I.~G., \& {Melin}, J.-B. 2015,
  \texttt{arXiv:1501.05666}, \mnras, 451, 3868

\bibitem[{{Lewis} et~al.(2000){Lewis}, {Challinor}, \& {Lasenby}}]{Lewis_2000}
{Lewis}, A., {Challinor}, A., \& {Lasenby}, A. 2000, \texttt{astro-ph/9911177},
  \apj, 538, 473

\bibitem[{{Lidman} et~al.(2013)}]{Lidman_2013}
{Lidman}, C., et~al. 2013, \texttt{arXiv:1305.0882}, \mnras, 433, 825

\bibitem[{{Liu} et~al.(2015)}]{Liu_2015}
{Liu}, T., et~al. 2015, \texttt{arXiv:1503.04051}, \apjs, 216, 28

\bibitem[{{Lopes} et~al.(2004){Lopes}, {de Carvalho}, {Gal}, {Djorgovski},
  {Odewahn}, {Mahabal}, \& {Brunner}}]{Lopes_2004}
{Lopes}, P.~A.~A., {de Carvalho}, R.~R., {Gal}, R.~R., {Djorgovski}, S.~G.,
  {Odewahn}, S.~C., {Mahabal}, A.~A., \& {Brunner}, R.~J. 2004, \aj, 128, 1017

\bibitem[{{Louis} et~al.(2017)}]{Louis_2016}
{Louis}, T., et~al. 2017, \texttt{arXiv:1610.02360}, \jcap, 6, 031

\bibitem[{{Madsen} et~al.(2017){Madsen}, {Beardmore}, {Forster}, {Guainazzi},
  {Marshall}, {Miller}, {Page}, \& {Stuhlinger}}]{Madsen_2017}
{Madsen}, K.~K., {Beardmore}, A.~P., {Forster}, K., {Guainazzi}, M.,
  {Marshall}, H.~L., {Miller}, E.~D., {Page}, K.~L., \& {Stuhlinger}, M. 2017,
  \texttt{arXiv:1609.09032}, \aj, 153, 2

\bibitem[{{Mahdavi} et~al.(2013){Mahdavi}, {Hoekstra}, {Babul}, {Bildfell},
  {Jeltema}, \& {Henry}}]{Mahdavi_2013}
{Mahdavi}, A., {Hoekstra}, H., {Babul}, A., {Bildfell}, C., {Jeltema}, T., \&
  {Henry}, J.~P. 2013, \texttt{arXiv:1210.3689}, \apj, 767, 116

\bibitem[{{Mantz} et~al.(2014)}]{Mantz_2014}
{Mantz}, A.~B., et~al. 2014, \texttt{arXiv:1401.2087}, \apj, 794, 157

\bibitem[{{Mantz} et~al.(2016)}]{Mantz_2016}
---. 2016, \texttt{arXiv:1606.03407}, \mnras, 463, 3582

\bibitem[{{Marriage} et~al.(2011)}]{Marriage_2011}
{Marriage}, T.~A. et~al. 2011, \texttt{arXiv:1010.1065}, ApJ, 737, 61

\bibitem[{{Mason} et~al.(2016)}]{Mason_2016}
{Mason}, B.~S., et~al. 2016, in American Astronomical Society Meeting
  Abstracts, Vol. 227, American Astronomical Society Meeting Abstracts,  439.04

\bibitem[{{Mayet} et~al.(2017)}]{Mayet_2017}
{Mayet}, F., et~al. 2017, \texttt{arXiv:1709.01255}, ArXiv e-prints

\bibitem[{{McMahon} et~al.(2013){McMahon}, {Banerji}, {Gonzalez}, {Koposov},
  {Bejar}, {Lodieu}, {Rebolo}, \& {VHS Collaboration}}]{McMahon_2013}
{McMahon}, R.~G., {Banerji}, M., {Gonzalez}, E., {Koposov}, S.~E., {Bejar},
  V.~J., {Lodieu}, N., {Rebolo}, R., \& {VHS Collaboration}. 2013, The
  Messenger, 154, 35

\bibitem[{{Mehrtens} et~al.(2012)}]{Mehrtens_2012}
{Mehrtens}, N., et~al. 2012, \texttt{arXiv:1106.3056}, \mnras, 423, 1024

\bibitem[{{Melchior} et~al.(2017)}]{Melchior_2017}
{Melchior}, P., et~al. 2017, \texttt{arXiv:1610.06890}, \mnras, 469, 4899

\bibitem[{{Menanteau} et~al.(2010)}]{Menanteau_2010}
{Menanteau}, F. et~al. 2010, \texttt{arXiv:1006.5126}, ApJ, 723, 1523

\bibitem[{{Menanteau} et~al.(2013)}]{Menanteau_2013}
---. 2013, \texttt{arXiv:1210.4048}, ApJ, 765, 67

\bibitem[{{Muldrew} et~al.(2012)}]{Muldrew_2012}
{Muldrew}, S.~I., et~al. 2012, \texttt{arXiv:1109.6328}, MNRAS, 419, 2670

\bibitem[{{Murray} et~al.(2013){Murray}, {Power}, \& {Robotham}}]{Murray_2013}
{Murray}, S.~G., {Power}, C., \& {Robotham}, A.~S.~G. 2013,
  \texttt{arXiv:1306.6721}, Astronomy and Computing, 3, 23

\bibitem[{{Muzzin} et~al.(2012)}]{Muzzin_2012}
{Muzzin}, A., et~al. 2012, \texttt{arXiv:1112.3655}, \apj, 746, 188

\bibitem[{{Naess} et~al.(2014)}]{Naess_2014}
{Naess}, S., et~al. 2014, \texttt{arXiv:1405.5524}, \jcap, 10, 007

\bibitem[{{Navarro} et~al.(1997){Navarro}, {Frenk}, \& {White}}]{NFW_1997}
{Navarro}, J.~F., {Frenk}, C.~S., \& {White}, S.~D.~M. 1997,
  \texttt{arXiv:astro-ph/9611107}, \apj, 490, 493

\bibitem[{{Nelson} et~al.(2012){Nelson}, {Rudd}, {Shaw}, \&
  {Nagai}}]{Nelson_2012}
{Nelson}, K., {Rudd}, D.~H., {Shaw}, L., \& {Nagai}, D. 2012,
  \texttt{arXiv:1112.3659}, \apj, 751, 121

\bibitem[{{Newman} et~al.(2014){Newman}, {Ellis}, {Andreon}, {Treu},
  {Raichoor}, \& {Trinchieri}}]{Newman_2014}
{Newman}, A.~B., {Ellis}, R.~S., {Andreon}, S., {Treu}, T., {Raichoor}, A., \&
  {Trinchieri}, G. 2014, \texttt{arXiv:1310.6754}, \apj, 788, 51

\bibitem[{{Oguri}(2014)}]{Oguri_2014}
{Oguri}, M. 2014, \texttt{arXiv:1407.4693}, \mnras, 444, 147

\bibitem[{{Oguri} et~al.(2017)}]{Oguri_2017}
{Oguri}, M., et~al. 2017, \texttt{arXiv:1701.00818}, PASJ submitted
  (arXiv:1701.00818)

\bibitem[{{Oke}(1974)}]{Oke_1974}
{Oke}, J.~B. 1974, ApJS, 27, 21

\bibitem[{{Oliver} et~al.(2012)}]{Oliver_2012}
{Oliver}, S.~J., et~al. 2012, \texttt{arXiv:1203.2562}, \mnras, 424, 1614

\bibitem[{{Pacaud} et~al.(2016)}]{Pacaud_2016}
{Pacaud}, F., et~al. 2016, \texttt{arXiv:1512.04264}, \aap, 592, A2

\bibitem[{{Papovich} et~al.(2010)}]{Papovich_2010}
{Papovich}, C., et~al. 2010, \texttt{arXiv:1002.3158}, \apj, 716, 1503

\bibitem[{{Penna-Lima} et~al.(2017){Penna-Lima}, {Bartlett}, {Rozo}, {Melin},
  {Merten}, {Evrard}, {Postman}, \& {Rykoff}}]{Penna-Lima_2017}
{Penna-Lima}, M., {Bartlett}, J.~G., {Rozo}, E., {Melin}, J.-B., {Merten}, J.,
  {Evrard}, A.~E., {Postman}, M., \& {Rykoff}, E. 2017,
  \texttt{arXiv:1608.05356}, \aap, 604, A89

\bibitem[{{Pierre} et~al.(2012){Pierre}, {Clerc}, {Maughan}, {Pacaud},
  {Papovich}, \& {Willmer}}]{Pierre_2012}
{Pierre}, M., {Clerc}, N., {Maughan}, B., {Pacaud}, F., {Papovich}, C., \&
  {Willmer}, C.~N.~A. 2012, \texttt{arXiv:1109.6194}, \aap, 540, A4

\bibitem[{{Piffaretti} et~al.(2011){Piffaretti}, {Arnaud}, {Pratt},
  {Pointecouteau}, \& {Melin}}]{Piffaretti_2011}
{Piffaretti}, R., {Arnaud}, M., {Pratt}, G.~W., {Pointecouteau}, E., \&
  {Melin}, J.-B. 2011, \texttt{arXiv:1007.1916}, \aap, 534, A109

\bibitem[{{Planck Collaboration} et~al.(2014{\natexlab{a}})}]{Planck_XX_2013}
{Planck Collaboration}, et~al. 2014{\natexlab{a}}, \texttt{arXiv:1303.5080},
  \aap, 571, A20

\bibitem[{{Planck Collaboration} et~al.(2014{\natexlab{b}})}]{Planck_XXIX_2013}
---. 2014{\natexlab{b}}, \texttt{arXiv:1303.5089}, \aap, 571, A29

\bibitem[{{Planck Collaboration}
  et~al.(2016{\natexlab{a}})}]{Planck2015Overview_2016}
---. 2016{\natexlab{a}}, \texttt{arXiv:1502.01582}, \aap, 594, A1

\bibitem[{{Planck Collaboration} et~al.(2016{\natexlab{b}})}]{Planck2015_XIII}
---. 2016{\natexlab{b}}, \texttt{arXiv:1502.01589}, \aap, 594, A13

\bibitem[{{Planck Collaboration} et~al.(2016{\natexlab{c}})}]{Planck2015_XXIV}
---. 2016{\natexlab{c}}, \texttt{arXiv:1502.01597}, \aap, 594, A24

\bibitem[{{Planck Collaboration} et~al.(2016{\natexlab{d}})}]{Planck2015_XXVII}
---. 2016{\natexlab{d}}, \texttt{arXiv:1502.01598}, \aap, 594, A27

\bibitem[{{Poole} et~al.(2007){Poole}, {Babul}, {McCarthy}, {Fardal},
  {Bildfell}, {Quinn}, \& {Mahdavi}}]{Poole_2007}
{Poole}, G.~B., {Babul}, A., {McCarthy}, I.~G., {Fardal}, M.~A., {Bildfell},
  C.~J., {Quinn}, T., \& {Mahdavi}, A. 2007, \texttt{astro-ph/0701586}, \mnras,
  380, 437

\bibitem[{{Popesso} et~al.(2005){Popesso}, {B{\"o}hringer}, {Romaniello}, \&
  {Voges}}]{Popesso_2005}
{Popesso}, P., {B{\"o}hringer}, H., {Romaniello}, M., \& {Voges}, W. 2005,
  \texttt{arXiv:astro-ph/0410011}, A\&A, 433, 415

\bibitem[{{Reichardt} et~al.(2013)}]{Reichardt_2013}
{Reichardt}, C.~L., et~al. 2013, \texttt{arXiv:1203.5775}, \apj, 763, 127

\bibitem[{{Rines} et~al.(2016){Rines}, {Geller}, {Diaferio}, \&
  {Hwang}}]{Rines_2016}
{Rines}, K.~J., {Geller}, M.~J., {Diaferio}, A., \& {Hwang}, H.~S. 2016,
  \texttt{arXiv:1507.08289}, \apj, 819, 63

\bibitem[{{Rykoff} et~al.(2014)}]{Rykoff_2014}
{Rykoff}, E.~S., et~al. 2014, \texttt{arXiv:1303.3562}, \apj, 785, 104

\bibitem[{{Rykoff} et~al.(2016)}]{Rykoff_2016}
---. 2016, \texttt{arXiv:1601.00621}, \apjs, 224, 1

\bibitem[{{Sayers} et~al.(2016)}]{Sayers_2016}
{Sayers}, J., et~al. 2016, \texttt{arXiv:1509.02950}, \apj, 820, 101

\bibitem[{{Schellenberger} \& {Reiprich}(2017)}]{Schellenberger_2017}
{Schellenberger}, G. \& {Reiprich}, T.~H. 2017, \texttt{arXiv:1705.05842},
  \mnras, 469, 3738

\bibitem[{{Schlegel} et~al.(1998){Schlegel}, {Finkbeiner}, \&
  {Davis}}]{Schlegel_1998}
{Schlegel}, D.~J., {Finkbeiner}, D.~P., \& {Davis}, M. 1998,
  \texttt{arXiv:astro-ph/9710327}, ApJ, 500, 525

\bibitem[{{Scodeggio} et~al.(2016)}]{Scodeggio_2016}
{Scodeggio}, M., et~al. 2016, \texttt{arXiv:1611.07048}, A\&A submitted
  (arXiv:1611.07048)

\bibitem[{{SDSS Collaboration} et~al.(2016)}]{Albareti_2016}
{SDSS Collaboration}, et~al. 2016, \texttt{arXiv:1608.02013}, ApJS submitted
  (arXiv:1608.02013)

\bibitem[{{Sehgal} et~al.(2011)}]{Sehgal_2011}
{Sehgal}, N. et~al. 2011, \texttt{arXiv:1010.1025}, ApJ, 732, 44

\bibitem[{{Sereno}(2015)}]{Sereno_2015}
{Sereno}, M. 2015, \texttt{arXiv:1409.5435}, \mnras, 450, 3665

\bibitem[{{Sif{\'o}n} et~al.(2013){Sif{\'o}n}, {Menanteau}, {Hasselfield}
  et~al.}]{Sifon_2013}
{Sif{\'o}n}, C., {Menanteau}, F., {Hasselfield}, M., et~al. 2013,
  \texttt{arXiv:1201.0991}, ApJ, 772, 25

\bibitem[{{Sif{\'o}n} et~al.(2016)}]{Sifon_2016}
{Sif{\'o}n}, C., et~al. 2016, \texttt{arXiv:1512.00910}, \mnras, 461, 248

\bibitem[{{Simet} et~al.(2017){Simet}, {McClintock}, {Mandelbaum}, {Rozo},
  {Rykoff}, {Sheldon}, \& {Wechsler}}]{Simet_2017}
{Simet}, M., {McClintock}, T., {Mandelbaum}, R., {Rozo}, E., {Rykoff}, E.,
  {Sheldon}, E., \& {Wechsler}, R.~H. 2017, \texttt{arXiv:1603.06953}, \mnras,
  466, 3103

\bibitem[{{Skrutskie} et~al.(2006)}]{Skrutskie_2006}
{Skrutskie}, M.~F., et~al. 2006, \aj, 131, 1163

\bibitem[{{Smith} et~al.(2016)}]{Smith_2016}
{Smith}, G.~P., et~al. 2016, \texttt{arXiv:1511.01919}, \mnras, 456, L74

\bibitem[{{Staniszewski} et~al.(2009)}]{Staniszewski_2009}
{Staniszewski}, Z. et~al. 2009, \texttt{arXiv:0810.1578}, ApJ, 701, 32

\bibitem[{{Struble} \& {Rood}(1999)}]{Struble_1999}
{Struble}, M.~F. \& {Rood}, H.~J. 1999, \apjs, 125, 35

\bibitem[{{Sunyaev} \& {Zeldovich}(1972)}]{SZ_1972}
{Sunyaev}, R.~A. \& {Zeldovich}, Y.~B. 1972, Comments on Astrophysics and Space
  Physics, 4, 173

\bibitem[{{Swetz} et~al.(2011)}]{Swetz_2011}
{Swetz}, D.~S. et~al. 2011, \texttt{arXiv:1007.0290}, ApJS, 194, 41

\bibitem[{{Szabo} et~al.(2011){Szabo}, {Pierpaoli}, {Dong}, {Pipino}, \&
  {Gunn}}]{Szabo_2011}
{Szabo}, T., {Pierpaoli}, E., {Dong}, F., {Pipino}, A., \& {Gunn}, J. 2011,
  \texttt{arXiv:1011.0249}, \apj, 736, 21

\bibitem[{{Tanaka} et~al.(2010){Tanaka}, {Finoguenov}, \& {Ueda}}]{Tanaka_2010}
{Tanaka}, M., {Finoguenov}, A., \& {Ueda}, Y. 2010, \texttt{arXiv:1004.3606},
  \apjl, 716, L152

\bibitem[{{Thornton} et~al.(2016)}]{Thornton_2016}
{Thornton}, R.~J., et~al. 2016, \texttt{arXiv:1605.06569}, \apjs, 227, 21

\bibitem[{{Tinker} et~al.(2008){Tinker}, {Kravtsov}, {Klypin}, {Abazajian},
  {Warren}, {Yepes}, {Gottl{\"o}ber}, \& {Holz}}]{Tinker_2008}
{Tinker}, J., {Kravtsov}, A.~V., {Klypin}, A., {Abazajian}, K., {Warren}, M.,
  {Yepes}, G., {Gottl{\"o}ber}, S., \& {Holz}, D.~E. 2008,
  \texttt{arXiv:0803.2706}, \apj, 688, 709

\bibitem[{{Tokunaga} \& {Vacca}(2005)}]{TokunagaVacca_2005}
{Tokunaga}, A.~T. \& {Vacca}, W.~D. 2005, \texttt{arXiv:astro-ph/0502120},
  PASP, 117, 421

\bibitem[{{Tundo} et~al.(2012){Tundo}, {Moretti}, {Tozzi}, {Teng}, {Rosati},
  {Tagliaferri}, \& {Campana}}]{Tundo_2012}
{Tundo}, E., {Moretti}, A., {Tozzi}, P., {Teng}, L., {Rosati}, P.,
  {Tagliaferri}, G., \& {Campana}, S. 2012, \texttt{arXiv:1208.2272}, \aap,
  547, A57

\bibitem[{{Valdes}(1998)}]{Valdes_1998}
{Valdes}, F.~G. 1998, in Astronomical Society of the Pacific Conference Series,
  Vol. 145, Astronomical Data Analysis Software and Systems VII, ed.
  R.~{Albrecht}, R.~N. {Hook}, \& H.~A. {Bushouse}, ~53

\bibitem[{{Valtchanov} et~al.(2004)}]{Valtchanov_2004}
{Valtchanov}, I., et~al. 2004, \texttt{astro-ph/0305192}, \aap, 423, 75

\bibitem[{{Vanderlinde} et~al.(2010)}]{Vanderlinde_2010}
{Vanderlinde}, K. et~al. 2010, \texttt{arXiv:1003.0003}, ApJ, 722, 1180

\bibitem[{{Viero} et~al.(2014)}]{Viero_2014}
{Viero}, M.~P., et~al. 2014, \texttt{arXiv:1308.4399}, \apjs, 210, 22

\bibitem[{{von der Linden} et~al.(2014)}]{vonDerLinden_2014}
{von der Linden}, A., et~al. 2014, \texttt{arXiv:1402.2670}, \mnras, 443, 1973

\bibitem[{{Walker} et~al.(2003)}]{Walker_2003}
{Walker}, A.~R., et~al. 2003, in \procspie, Vol. 4841, Instrument Design and
  Performance for Optical/Infrared Ground-based Telescopes, ed. M.~{Iye} \&
  A.~F.~M. {Moorwood},  286--294

\bibitem[{{Wen} \& {Han}(2015)}]{WH_2015}
{Wen}, Z.~L. \& {Han}, J.~L. 2015, \texttt{arXiv:1506.04503}, \apj, 807, 178

\bibitem[{{Wen} et~al.(2009){Wen}, {Han}, \& {Liu}}]{WHL_2009}
{Wen}, Z.~L., {Han}, J.~L., \& {Liu}, F.~S. 2009, \texttt{arXiv:0906.0803},
  \apjs, 183, 197

\bibitem[{{Wen} et~al.(2012){Wen}, {Han}, \& {Liu}}]{WHL_2012}
---. 2012, \texttt{arXiv:1202.6424}, \apjs, 199, 34

\bibitem[{{Wik} et~al.(2008){Wik}, {Sarazin}, {Ricker}, \&
  {Randall}}]{Wik_2008}
{Wik}, D.~R., {Sarazin}, C.~L., {Ricker}, P.~M., \& {Randall}, S.~W. 2008,
  \texttt{arXiv:0802.3695}, \apj, 680, 17

\bibitem[{{Willis} et~al.(2013)}]{Willis_2013}
{Willis}, J.~P., et~al. 2013, \texttt{arXiv:1212.4185}, \mnras, 430, 134

\bibitem[{{Yang} et~al.(2010){Yang}, {Bhattacharya}, \& {Ricker}}]{Yang_2010}
{Yang}, H.-Y.~K., {Bhattacharya}, S., \& {Ricker}, P.~M. 2010,
  \texttt{arXiv:1010.0249}, \apj, 725, 1124

\end{thebibliography}
\end{document}